\documentclass[final,suppldata]{gOMS2e}

\usepackage[numbers]{natbib}
\usepackage{amsmath}
\usepackage{hyperref}
\hypersetup{hidelinks,colorlinks,linkcolor={blue},citecolor={blue},urlcolor={blue}}
\usepackage{MnSymbol}
\usepackage{graphicx}
\usepackage{listings}
\usepackage{color}
\usepackage{qtree}
\usepackage{rotating}
\usepackage{booktabs}
\usepackage{multirow}
\usepackage{wrapfig}
\usepackage[nosort]{cleveref}% load last
\usepackage{mobi}

\newlength{\arrowbump}
\newlength{\arrowlbump}
\newcommand{\FORWARDPROC}[1]{\overrightarrow{#1}}

\newcommand{\REVERSEPROC}[1]{\overleftarrow{#1}}

\newcommand{\IF}{\textbf{if}\;}
\newcommand{\THEN}{\textbf{then}\;}
\newcommand{\ELSE}{\textbf{else}\;}

\newcommand{\LET}{\textbf{let}\;}

\newcommand{\IN}{\textbf{in}\;}

\newcommand{\CLOSURE}[2]{\langle(#1),#2\rangle}

\newcommand{\BUNDLEF}{\rhd}
\newcommand{\BUNDLE}[2]{#1\BUNDLEF#2}
\newcommand{\TAPEF}{\lhd}
\newcommand{\TAPE}[2]{#1\TAPEF#2}

\newcommand{\EVAL}{\mathcal{E}}
\newcommand{\APPLY}{\mathcal{A}}

\newcommand{\JACOBIAN}{\mathcal{J}}

\newcommand{\FORWARDJ}{\FORWARDPROC{\JACOBIAN}}

\newcommand{\REVERSEJ}{\REVERSEPROC{\JACOBIAN}}

\renewcommand{\Re}{\mathbb{R}}

\newcommand{\defoccur}[1]{\textsl{#1}}

\newcommand{\CAPSULE}[2]{\lsem#1,#2\rsem}

\newcommand{\CHECKPOINTREVERSEJ}{\begin{array}[b]{@{}l@{}}
\!\text{\small$\checkmark$}\\*[-6pt]
\mathcal{J}\end{array}}

\newcommand{\COMPILE}{\mathcal{S}}
\newcommand{\NAME}{\mathcal{N}}

\newcommand{\eg}{\emph{e.g.},}

\newcommand{\ie}{\emph{i.e.},}
\newcommand{\Ie}{\emph{I.e.},}

\newcommand{\vs}{\emph{vs.}}

\newcommand{\Tapenade}{\textsc{Tapenade}}
\newcommand{\Scheme}{\textsc{Scheme}}
\newcommand{\SMLNJ}{\textsc{sml/nj}}
\newcommand{\SML}{\textsc{sml}}
\newcommand{\MLton}{\textsc{MLton}}
\newcommand{\Clang}{\textsc{c}}
\newcommand{\POSIX}{\textsc{posix}}
\newcommand{\Fortran}{\textsc{Fortran}}
\newcommand{\Prolog}{\textsc{Prolog}}
\newcommand{\ADOLC}{\textsc{adol-c}}
\newcommand{\VLAD}{\textsc{vlad}}
\newcommand{\checkpointVLAD}{\textsc{checkpointVLAD}}
\newcommand{\FADBAD}{\textsc{fadbad}$++$}
\newcommand{\Stalingrad}{\textsc{Stalin}$\nabla$}
\newcommand{\RsixRSAD}{\textsc{r6rs-ad}}
\newcommand{\DiffSharp}{\textsc{DiffSharp}}
\newcommand{\Cplusplus}{\textsc{c}$++$}

\newcommand{\LD}{.}

\newcommand{\wrap}[1]{\begin{tabular}{@{}c@{}}#1\end{tabular}}

\definecolor{darkred}{rgb}{0.7,0,0}
\definecolor{darkgreen}{rgb}{0,0.5,0}
\definecolor{darkblue}{rgb}{0,0,0.7}

\newcommand{\checkpointstart}{\lstinline{c$ad checkpoint-start}}
\newcommand{\checkpointend}{\lstinline{c$ad checkpoint-end}}
\newcommand{\nocheckpoint}{\lstinline{c$ad nocheckpoint}}
\newcommand{\binomialckp}{\lstinline{c$ad binomial-ckp}}

\crefname{equation}{}{}
\crefname{subequation}{}{}
\crefmultiformat{equation}{(#2#1#3}{, #2#1#3)}{, #2#1#3}{, #2#1#3)}
\crefmultiformat{subequation}{(#2#1#3}{, #2#1#3)}{, #2#1#3}{, #2#1#3)}
\crefrangelabelformat{equation}{(#3#1#4--#5\crefstripprefix{#1}{#2}#6)}
\crefrangelabelformat{subequation}{(#3#1#4--#5\crefstripprefix{#1}{#2}#6)}

\begin{document}

\title{Divide-and-Conquer Checkpointing for Arbitrary Programs\\
with No User Annotation}

% website says All authors of a manuscript should include their full names,
% affiliations, postal addresses, telephone numbers and email addresses on the
% cover page of the manuscript.
%\needswork: We don't have telephone numbers. We only have email for the
%            corresponding author.
\author{
\name{Jeffrey Mark Siskind\textsuperscript{a}$^{\ast}$\thanks{$^\ast$Corresponding author. Email: qobi@purdue.edu}
and Barak A. Pearlmutter\textsuperscript{b}}
\affil{\textsuperscript{a} School of Electrical and Computer Engineering,
  Purdue University, 465 Northwestern Avenue, West Lafayette, IN, USA;
\textsuperscript{b} Dept of Computer Science, Maynooth University, Ireland}
% \received{\needswork}
}

\maketitle

\vspace{4ex}

% 200 words
\begin{abstract}
  Classical reverse-mode automatic differentiation (AD) imposes only a small
  constant-factor overhead in operation count over the original computation,
  but has storage requirements that grow, in the worst case, in proportion to
  the time consumed by the original computation.
  This storage blowup can be ameliorated by checkpointing, a process that
  reorders application of classical reverse-mode AD over an execution interval
  to tradeoff space \vs\ time.
  Application of checkpointing in a divide-and-conquer fashion to strategically
  chosen nested execution intervals can break classical reverse-mode AD into
  stages which can reduce the worst-case growth in storage from linear to
  sublinear.
  Doing this has been fully automated only for computations of particularly
  simple form, with checkpoints spanning execution intervals resulting from
  a limited set of program constructs.
  Here we show how the technique can be automated for arbitrary computations.
  The essential innovation is to apply the technique at the level of the
  language implementation itself, thus allowing checkpoints to span any
  execution interval.
\end{abstract}

% Please provide three to six keywords taken from terms used in your manuscript
% website says 3-8

% authors are encouraged to provide two to six 2010 AMS Subject Classification
% codes
% website says 1-5
% http://www.ams.org/mathscinet/msc/msc2010.html
\begin{classcode}
  68N20, % Compilers and interpreters
  68N18, % Functional programming and lambda calculus
  65F50, % Sparse matrices
  65D25, % Numerical differentiation
  46G05, % Derivatives
  58C20  % Differentiation theory (Gateaux, Fréchet, etc.)
\end{classcode}

\section{Introduction}
\label{sec:introduction}

Reverse-mode automatic differentiation (AD) traverses the run-time dataflow
graph of a calculation in reverse order, in a so-called \defoccur{reverse
  sweep}, so as to calculate a Jacobian-transpose-vector product of the
Jacobian of the given original (or \defoccur{primal}) calculation
\citep{speelpenning80}.
Although the number of arithmetic operations involved in this process is only a
constant factor greater than that of the primal calculation, some values
involved in the primal dataflow graph must be saved for use in the reverse
sweep, thus imposing considerable storage overhead.
This is accomplished by replacing the primal computation with a
\defoccur{forward sweep} that performs the primal computation while saving the
requisite values on a data structure known as the \defoccur{tape}.
A technique called \defoccur{checkpointing} \citep{volin1985aco} reorders
portions of the forward and reverse sweeps to reduce the maximal length of the
requisite tape.
Doing so, however, requires (re)computation of portions of the primal and saving
the requisite program state to support such as \defoccur{snapshots}.
Overall space savings result when the space saved by reducing the maximal
length of the requisite tape exceeds the space cost of storing the snapshots.
Such space saving incurs a time cost in (re)computation of portions of the
primal.
Different checkpointing strategies lead to a space-time tradeoff.

\begin{wrapfigure}[10]{r}{0.5\columnwidth}
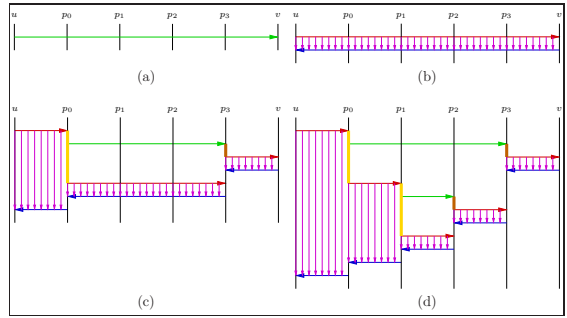

  \vspace*{-25pt}
  \centering
  \resizebox{0.5\columnwidth}{!}{\fbox{\begin{tabular}{@{}cc@{}}
    \vspace*{-25pt}\\
    \begin{tabular}[t]{@{}c@{}}
      \rule{0pt}{0pt}\\
      \input{Xcheckpointing0}
    \end{tabular}&
    \begin{tabular}[t]{@{}c@{}}
      \rule{0pt}{0pt}\\
      \input{Xcheckpointing1}
    \end{tabular}\\
    (a)&(b)\\
    \begin{tabular}[t]{@{}c@{}}
      \rule{0pt}{0pt}\\
      \input{Xcheckpointing2}
    \end{tabular}&
    \begin{tabular}[t]{@{}c@{}}
      \rule{0pt}{0pt}\\
      \input{Xcheckpointing3}
    \end{tabular}\\
    (c)&(d)
    \end{tabular}}}
  \caption{Checkpointing in reverse-mode AD.\@
    See text for description.}
  \label{fig:checkpointing}
\end{wrapfigure}
We introduce some terminology that will be useful in describing checkpointing.
An \defoccur{execution point} is a point in time during the execution of a
program.
A \defoccur{program point} is a location in the program code.
Since program fragments might be \defoccur{invoked} zero or more times during
the execution of a program, each execution point corresponds to exactly one
program point but each program point may correspond to zero or more execution
points.
An \defoccur{execution interval} is a time interval spanning two execution
points.
A \defoccur{program interval} is a fragment of code spanning two program
points.
Program intervals are usually constrained so that they nest, \ie\ they do not
cross one boundary of a syntactic program \defoccur{construct} without crossing
the other.
Each program interval may correspond to zero or more execution intervals,
those execution intervals whose endpoints result from the same invocation of
the program interval.
Each execution interval corresponds to at most one program interval.
An execution interval might not correspond to a program interval because the
endpoints might not result from the same invocation of any program interval.

\begin{wrapfigure}[22]{r}{0.5\columnwidth}
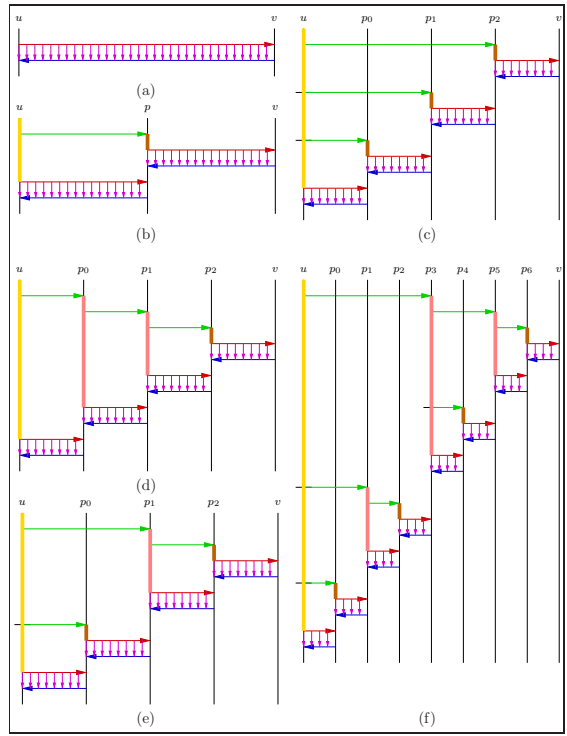

  \vspace*{-25pt}
  \centering
  \resizebox{0.5\columnwidth}{!}{\fbox{\begin{tabular}{@{}cc@{}}
    \vspace*{-25pt}\\
    \begin{tabular}[t]{@{}c@{}}
      \rule{0pt}{0pt}\\
      \input{Xcheckpointing4}\\
      (a)\\
      \input{Xcheckpointing5}
    \end{tabular}&
    \begin{tabular}[t]{@{}c@{}}
      \rule{0pt}{0pt}\\
      \input{Xcheckpointing6}
    \end{tabular}\\
    (b)&(c)\\
    \begin{tabular}[t]{@{}c@{}}
      \rule{0pt}{0pt}\\
      \input{Xcheckpointing7}\\
      (d)\\
      \input{Xcheckpointing8}
    \end{tabular}&
    \begin{tabular}[t]{@{}c@{}}
      \rule{0pt}{0pt}\\
      \input{Xcheckpointing9}
    \end{tabular}\\
    (e)&(f)
    \end{tabular}}}
  \caption{Divide-and-conquer checkpointing in reverse-mode AD.\@
    See text for description.}
  \label{fig:divide-and-conquer-checkpointing}
\end{wrapfigure}
Figs.~\ref{fig:checkpointing} and~\ref{fig:divide-and-conquer-checkpointing}
illustrate the process of performing reverse-mode AD with and without
checkpointing.
Control flows from top to bottom, and along the direction of the arrow within
each row.
The symbols~$u$, $v$, and $p_0,\dotsc,p_6$ denote execution points in the
primal, $u$~being the start of the computation whose derivative is desired,
$v$~being the end of that computation, and each~$p_i$ being an intermediate
execution point in that computation.
Reverse mode involves various sweeps, whose execution intervals are represented
as horizontal green, red, and blue lines.
Green lines denote (re)computation of the primal without taping.
Red lines denote computation of the primal with taping, \ie\ the forward sweep
of reverse mode.
Blue lines denote computation of the Jacobian-transpose-vector product,
\ie\ the reverse sweep of reverse mode.
The vertical black lines denote collections of execution points across the
various sweeps that correspond to execution points in the primal, each
particular execution point being the intersection of a horizontal line and a
vertical line.
In portions of Figs.~\ref{fig:checkpointing}
and~\ref{fig:divide-and-conquer-checkpointing} other than
Fig.~\ref{fig:checkpointing}(a) we refer to execution points for other sweeps
besides the primal in a given collection with the symbols~$u$, $v$, and
$p_0,\dotsc,p_6$ when the intent is clear.
The vertical violet, gold, pink, and brown lines denote execution intervals for
the lifetimes of various saved values.
Violet lines denote the lifetime of a value saved on the tape during the
forward sweep and used during the reverse sweep.
The value is saved at the execution point at the top of the violet line and
used once at the execution point at the bottom of that line.
Gold and pink lines denote the lifetime of a snapshot.\footnote{The distinction
  between gold and pink lines, the meaning of brown lines, and the meaning of
  the black tick marks on the left of the gold and pink lines will be explained
  in Section~\ref{sec:intuition}.}
The snapshot is saved at the execution point at the top of each gold or pink
line and used at various other execution points during its lifetime.
Green lines emanating from a gold or pink line indicate restarting a portion of
the primal computation from a saved snapshot.

Fig.~\ref{fig:checkpointing}(a) depicts the primal computation, $y=f(x)$, which
takes~$t$ time steps, with~$x$ being a portion of the program state at
execution point~$u$ and~$y$ being a portion of the program state at execution
point~$v$ computed from~$x$.
This is performed without taping (green).
Fig.~\ref{fig:checkpointing}(b) depicts classical reverse mode without
checkpointing.
An uninterrupted forward sweep (red) is performed for the entire length of the
primal, then an uninterrupted reverse sweep (blue) is performed for the entire
length.
Since the tape values are consumed in reverse order from which they are saved,
the requisite tape length is $O(t)$.
Fig.~\ref{fig:checkpointing}(c) depicts a \defoccur{checkpoint} introduced
for the execution interval $[p_0,p_3)$.
This interrupts the forward sweep and delays a portion of that sweep until the
reverse sweep.
Execution proceeds by a forward sweep (red) that tapes during the execution
interval $[u,p_0)$, a primal sweep (green) without taping during the execution
interval $[p_0,p_3)$, a taping forward sweep (red) during the execution
interval $[p_3,v)$, a reverse sweep (blue) during the execution interval
$[v,p_3)$, a taping forward sweep (red) during the execution interval
$[p_0,p_3)$, a reverse sweep (blue) during the execution interval $[p_3,p_0)$,
and then a reverse sweep (blue) during the execution interval $[p_0,u)$.
The forward sweep for the execution interval $[p_0,p_3)$ is delayed until after
the reverse sweep for the execution interval $[v,p_3)$.
As a result of this reordering, the tapes required for those sweeps are not
simultaneously live.
Thus the requisite tape length is the maximum of the two tape lengths, not
their sum.
This savings comes at a cost.
To allow such out-of-order execution, a snapshot (gold) must be saved at~$p_0$
and the portion of the primal during the execution interval $[p_0,p_3)$ must be
computed twice, first without taping (green) then with (red).

A checkpoint can be introduced into a portion of the forward sweep that has
been delayed, as shown in Fig.~\ref{fig:checkpointing}(d).
An additional checkpoint can be introduced for the execution interval
$[p_1,p_2)$.
This will delay a portion of the already delayed forward sweep even further.
As a result, the portions of the tape needed for the three execution intervals
$[p_1,p_2)$, $[p_2,p_3)$, and $[p_3,v)$ are not simultaneously live,
thus further reducing the requisite tape length, but requiring more
(re)computation of the primal (green).
The execution intervals for multiple checkpoints must either be disjoint or
must nest; the execution interval of one checkpoint cannot cross one endpoint of
the execution interval of another checkpoint without crossing the other
endpoint.

Execution intervals for checkpoints can be specified in a variety of ways.
\begin{description}
\item[program interval]~\\ Execution intervals of specified program intervals
  constitute checkpoints.
\item[subroutine call site]~\\ Execution intervals of specified subroutine
  call sites constitute checkpoints.
\item[subroutine body]~\\ Execution intervals of specified subroutine bodies
  constitute checkpoints \citep{volin1985aco}.
\end{description}
Nominally, these have the same power; with any one, one could achieve the
effect of the other two.
Specifying a subroutine body could be accomplished by specifying all call
sites to that subroutine.
Specifying some call sites but not others could be accomplished by having two
variants of the subroutine, one whose body is specified and one whose is not,
and calling the appropriate one at each call site.
Specifying a program interval could be accomplished by extracting that interval
as a subroutine.

\Tapenade\ \citep{hascoet2004tug} allows the user to specify program intervals
for checkpoints with the \checkpointstart\ and \checkpointend\ pragmas.
\Tapenade, by default, checkpoints all subroutine calls
\citep{dauvergne2006tdf}.
This default can be overridden for named subroutines with the
\texttt{-nocheckpoint} command-line option and for both named subroutines and
specific call sites with the \nocheckpoint\ pragma.

\begin{wrapfigure}[2]{r}{0.12\columnwidth}
\vspace*{-28pt}
\centering
\leaf{$[u,p)$}
\leaf{$[p,v)$}
\branch{2}{$[u,v)$}
\resizebox{0.12\columnwidth}{!}{\qobitree}
\vspace*{-10pt}
\caption{}
\label{fig:tree1}
\end{wrapfigure}
Recursive application of checkpointing in a divide-and-conquer fashion,
\ie\ ``treeverse,'' can divide the forward and reverse sweeps into stages
run sequentially \citep{griewank1992alg}.
The key idea is that only one stage is live at a time, thus requiring a shorter
tape.
However, the state of the primal computation at various intermediate execution
points needs to be saved as snapshots, in order to (re)run the requisite
portion of the primal to allow the forward and reverse sweeps for each stage to
run in turn.
This process is illustrated in Fig.~\ref{fig:divide-and-conquer-checkpointing}.
Consider a \defoccur{root execution interval} $[u,v)$ of the derivative
calculation.
Without checkpointing, the forward and reverse sweeps span the entire root
execution interval, as shown in
Fig.~\ref{fig:divide-and-conquer-checkpointing}(a).
One can divide the root execution interval $[u,v)$ into two subintervals
$[u,p)$ and $[p,v)$ at the \defoccur{split point} $p$ and checkpoint the
first subinterval $[u,v)$.
This divides the forward (red) and reverse (blue) sweeps into two
\defoccur{stages}.
These two stages are not simultaneously live.
If the two subintervals are the same length, this halves the storage needed for
the tape at the expense of running the primal computation for $[u,p)$ twice,
first without taping (green), then with taping (red).
This requires a single snapshot (gold) at~$u$.
This process can be viewed as constructing a \defoccur{binary checkpoint tree}
(Fig.~\ref{fig:tree1}) whose nodes are labeled with execution intervals, the
intervals of the children of a node are adjacent, the interval of a node is the
disjoint union of the intervals of its children, and left children are
checkpointed.

\begin{wrapfigure}[7]{r}{0.27\columnwidth}
\vspace*{-25pt}
\centering
\begin{tabular}{@{}ll@{}}
\raisebox{-20pt}{{\scriptsize (a)}}&
\leaf{$[u,p_0)$}
\leaf{$[p_0,p_1)$}
\branch{2}{$[u,p_1)$}
\leaf{$[p_1,p_2)$}
\branch{2}{$[u,p_2)$}
\leaf{$[p_2,v)$}
\branch{2}{$[u,v)$}
\hspace*{-25pt}\resizebox{0.2\columnwidth}{!}{\qobitree}\\
\raisebox{-20pt}{{\scriptsize (b)}}&
\leaf{$[u,p_0)$}
\leaf{$[p_0,p_1)$}
\leaf{$[p_1,p_2)$}
\leaf{$[p_2,v)$}
\branch{4}{$[u,v)$}
\hspace*{-25pt}\resizebox{0.2\columnwidth}{!}{\qobitree}
\end{tabular}
\caption{}
\label{fig:tree2}
\end{wrapfigure}
One can construct a left-branching binary checkpoint tree over the same root
execution interval $[u,v)$ with the split points~$p_0$, $p_1$, and~$p_2$
(Fig.~\ref{fig:tree2}a).
This can also be viewed as constructing an \defoccur{n-ary checkpoint tree}
where all children but the rightmost are checkpointed
(Fig.~\ref{fig:tree2}b).
This leads to \emph{nested} checkpoints for the execution intervals
$[u,p_0)$, $[u,p_1)$, and $[u,p_2)$ as shown in
Fig.~\ref{fig:divide-and-conquer-checkpointing}(c).
Since the starting execution point~$u$ is the same for these intervals, a
single snapshot (gold) with longer lifetime suffices.
These checkpoints divide the forward (red) and reverse (blue) sweeps into four
stages.
This allows the storage needed for the tape to be reduced arbitrarily (\ie\ the
red and blue segments can be made arbitrarily short), by rerunning successively
shorter prefixes of the primal computation (green), without taping, running
only short segments (red) with taping.
This requires an $O(t)$ increase in time for (re)computation of the primal
(green).

\begin{wrapfigure}[5]{r}{0.27\columnwidth}
\vspace*{-27pt}
\centering
\leaf{$[u,p_0)$}
\leaf{$[p_0,p_1)$}
\leaf{$[p_1,p_2)$}
\leaf{$[p_2,v)$}
\branch{2}{$[p_1,v)$}
\branch{2}{$[p_0,v)$}
\branch{2}{$[u,v)$}
\resizebox{0.27\columnwidth}{!}{\qobitree}
\vspace*{-10pt}
\caption{}
\label{fig:tree3}
\end{wrapfigure}
Alternatively, one can construct a right-branching binary checkpoint tree
over the same root execution interval $[u,v)$ with the same split
points~$p_0$, $p_1$, and~$p_2$ (Fig.~\ref{fig:tree3}).
This also divides the forward (red) and reverse (blue) sweeps into four stages.
With this, the requisite tape length (the maximal length of the red and blue
segments) can be reduced arbitrarily while running the primal (green) just
once, by saving more snapshots (gold and pink), as shown in
Fig.~\ref{fig:divide-and-conquer-checkpointing}(d),
This requires an $O(t)$ increase in space for storage of the live snapshots
(gold and pink).

Thus we see that divide-and-conquer checkpointing can make the requisite tape
arbitrarily small with either left- or right-branching binary checkpoint trees.
This involves a space-time tradeoff.
The left-branching binary checkpoint trees require a single snapshot but an
$O(t)$ increase in time for (re)computation of the primal (green).
The right-branching binary checkpoint trees require an $O(t)$ increase in space
for storage of the live snapshots (gold and pink) but (re)run the primal only
once.

\begin{wrapfigure}[4]{r}{0.27\columnwidth}
\vspace*{-27pt}
\centering
\leaf{$[u,p_0)$}
\leaf{$[p_0,p_1)$}
\branch{2}{$[u,p_1)$}
\leaf{$[p_1,p_2)$}
\leaf{$[p_2,v)$}
\branch{2}{$[p_1,v)$}
\branch{2}{$[u,v)$}
\resizebox{0.27\columnwidth}{!}{\qobitree}
\vspace*{-10pt}
\caption{}
\label{fig:tree4}
\end{wrapfigure}
One can also construct a complete binary checkpoint tree over the same root
execution interval $[u,v)$ with the same split points~$p_0$, $p_1$, and~$p_2$
(Fig.~\ref{fig:tree4}).
This constitutes application of the approach from
Fig.~\ref{fig:divide-and-conquer-checkpointing}(b) in a divide-and-conquer
fashion as shown in Fig.~\ref{fig:divide-and-conquer-checkpointing}(e).
This also divides the forward (red) and reverse (blue) sweeps into four stages.
One can continue this divide-and-conquer process further, with more split
points, more snapshots, and more but shorter stages, as shown in
Fig.~\ref{fig:divide-and-conquer-checkpointing}(f).
This leads to an $O(\log t)$ increase in space for storage of the live snapshots
(gold and pink) and an $O(\log t)$ increase in time for (re)computation of the
primal (green).
Variations of this technique can tradeoff between different improvements in
space and/or time complexity, leading to overhead in a variety of sublinear
asymptotic complexity classes in one or both.
In order to apply this technique, we must be able to construct a checkpoint
tree of the desired shape with appropriate split points.
This in turn requires the ability to interrupt the primal computation at
appropriate execution points, save the interrupted execution state as a
\defoccur{capsule}, and restart the computation from the capsules, sometimes
repeatedly.\footnote{The correspondence between capsules and snapshots will be
  discussed in Section~\ref{sec:intuition}.}

\begin{wrapfigure}[13]{r}{0.3\columnwidth}
\vspace*{-20pt}
\centering
\begin{tabular}{@{}ll@{}}
\raisebox{-20pt}{{\scriptsize (a)}}&
\leaf{$[u,p_0)$}
\leaf{$[p_0,p_1)$}
\branch{2}{$[u,p_1)$}
\leaf{$[p_1,p_2)$}
\leaf{$[p_2,v)$}
\branch{2}{$[p_1,v)$}
\branch{2}{$[u,v)$}
\hspace*{-25pt}\resizebox{0.2\columnwidth}{!}{\qobitree}\\
\raisebox{-20pt}{{\scriptsize (b)}}&
\leaf{$[u,p_0)$}
\leaf{$[p_0,p_1)$}
\leaf{$[p_1,p_2)$}
\leaf{$[p_2,v)$}
\branch{2}{$[p_1,v)$}
\branch{3}{$[u,v)$}
\hspace*{-25pt}\resizebox{0.20\columnwidth}{!}{\qobitree}\\
\raisebox{-20pt}{{\scriptsize (c)}}&
\leaf{$[u,p_0)$}
\leaf{$[p_0,p_1)$}
\branch{2}{$[u,p_1)$}
\leaf{$[p_1,p_2)$}
\leaf{$[p_2,p_3)$}
\branch{2}{$[p_1,p_3)$}
\branch{2}{$[u,p_3)$}
\leaf{$[p_3,p_4)$}
\leaf{$[p_4,p_5)$}
\branch{2}{$[p_3,p_5)$}
\leaf{$[p_5,p_6)$}
\leaf{$[p_6,v)$}
\branch{2}{$[p_5,v)$}
\branch{2}{$[p_3,v)$}
\branch{2}{$[u,v)$}
\hspace*{-60pt}\resizebox{0.25\columnwidth}{!}{\qobitree}\\
\raisebox{-20pt}{{\scriptsize (d)}}&
\leaf{$[u,p_0)$}
\leaf{$[p_0,p_1)$}
\leaf{$[p_1,p_2)$}
\leaf{$[p_2,p_3)$}
\branch{2}{$[p_1,p_3)$}
\faketreewidth{\hspace*{20pt}}
\leaf{$[p_3,p_4)$}
\leaf{$[p_4,p_5)$}
\leaf{$[p_5,p_6)$}
\leaf{$[p_6,v)$}
\branch{2}{$[p_5,v)$}
\faketreewidth{\hspace*{20pt}}
\branch{3}{$[p_3,v)$}
\faketreewidth{\hspace*{150pt}}
\branch{4}{$[u,v)$}
\hspace*{-60pt}\resizebox{0.25\columnwidth}{!}{\qobitree}
\end{tabular}
\vspace*{-10pt}
\caption{}
\label{fig:tree5}
\end{wrapfigure}
Any given divide-and-conquer decomposition of the same root execution interval
with the same split points can be viewed as either a binary checkpoint tree or
an n-ary checkpoint tree.
Thus Fig.~\ref{fig:divide-and-conquer-checkpointing}(e) can be viewed as either
Fig.~\ref{fig:tree5}(a) or Fig.~\ref{fig:tree5}(b).
Similarly, Fig.~\ref{fig:divide-and-conquer-checkpointing}(f) can be viewed as
either Fig.~\ref{fig:tree5}(c) or Fig.~\ref{fig:tree5}(d).
Thus we distinguish between two algorithms to perform divide-and-conquer
checkpointing.
\begin{description}
\item[binary] An algorithm that constructs a binary checkpoint tree.
\item[treeverse] The algorithm from \cite[Figs.~2 and~3]{griewank1992alg}
  that constructs an n-ary checkpoint tree.
\end{description}
There is, however, a simple correspondence between associated binary and n-ary
checkpoint trees.
The n-ary checkpoint tree is derived from the binary checkpoint tree by
coalescing each maximal sequence of left branches into a single node.
Thus we will see, in Section~\ref{sec:binomial}, that these two algorithms
exhibit the same properties.

Note that (divide-and-conquer) checkpointing does not incur any space or time
overhead in the forward or reverse sweeps themselves (\ie\ the number of violet
lines and the total length of red and blue lines).
Any space overhead results from the snapshots (gold and pink) and any time
overhead results from (re)computation of the primal (green).

Several design choices arise in the application of divide-and-conquer
checkpointing in addition to the choice of binary \vs\ n-ary checkpoint trees.
\begin{itemize}
\item What root execution interval(s) should be subject to divide-and-conquer
  checkpointing?
\item Which execution points are candidate split points?
  The divide-and-conquer process of constructing the checkpoint tree will
  select actual split points from these candidates.
\item What is the shape or depth of the checkpoint tree, \ie\ what is the
  termination criterion for the divide-and-conquer process?
\end{itemize}
Since the leaf nodes of the checkpoint tree correspond to stages, the
termination criterion and the number of evaluation steps in the stage at each
leaf node (the length of a pair of red and blue lines) are mutually constrained.
The number of live snapshots at a leaf (how many gold and pink lines are
crossed by a horizontal line drawn leftward from that stage, the pair of red
and blue lines, to the root) depends on the depth of the leaf and its position
in the checkpoint tree.
Different checkpoint trees, with different shapes resulting from different
termination criteria and split points, can lead to a different maximal number
of live snapshots, resulting in different storage requirements.
The amount of (re)computation of the primal (the total length of the green
lines) can also depend on the shape of the checkpoint tree, thus different
checkpoint trees, with different shapes resulting from different termination
criteria and split points, can lead to different compute-time requirements.
Thus different strategies for specifying the termination criterion and the
split points can influence the space-time tradeoff.

We make a distinction between several different approaches to selecting
root execution intervals subject to divide-and-conquer checkpointing.
\begin{description}
\item[loop] Execution intervals resulting from invocations of specified
  DO loops are subject to divide-and-conquer checkpointing.
\item[entire derivative calculation] The execution interval for an
  entire specified derivative calculation is subject to divide-and-conquer
  checkpointing.
\end{description}
We further make a distinction between several different approaches to selecting
candidate split points.
\begin{description}
\item[iteration boundary] Iteration boundaries of the DO loop specified as
  the root execution interval are taken as candidate split points.
\item[arbitrary] Any execution point inside the root execution interval can
  be taken as a candidate split point.
\end{description}
We further make a distinction between several different approaches to specifying
the termination criterion and deciding which candidate split points to select as
actual split points.
\begin{description}
\item[bisection] Split points are selected so as to divide the
  computation dominated by a node in half as one progresses successively from
  right to left among children \citep[equation (12)]{griewank1992alg}.
  One can employ a variety of termination criteria, including that from
  \citep[p.~46]{griewank1992alg}.
  If the termination criterion is such that the total number of leaves is a
  power of two, one obtains a complete binary checkpoint tree.
  A termination criterion that bounds the number of evaluation steps in a
  leaf limits the size of the tape and achieves logarithmic overhead in both
  asymptotic space and time complexity compared with the primal.
\item[binomial] Split points are selected using the criterion from
  \citep[equation (16)]{griewank1992alg}.
  The termination criterion from \citep[p.~46]{griewank1992alg} is usually
  adopted to achieve the desired properties discussed in
  \citep{griewank1992alg}.
  Different termination criteria can be selected to control space-time
  tradeoffs.
  \begin{description}
  \item[fixed space overhead]
    One can bound the size of the tape and the number of snapshots to obtain
    sublinear but superlogarithmic overhead in asymptotic time complexity
    compared with the primal.
  \item[fixed time overhead]
    One can bound the size of the tape and the (re)computation of the primal
    to obtain sublinear but superlogarithmic overhead in asymptotic space
    complexity compared with the primal.
  \item[logarithmic space and time overhead]
    One can bound the size of the tape and obtain logarithmic overhead in both
    asymptotic space and time complexity compared with the primal.
    The constant factor is less than that of bisection checkpointing.
  \end{description}
\end{description}
We elaborate on the strategies for selecting actual split points from candidate
split points and the associated termination criteria in
Section~\ref{sec:binomial}.

Divide-and-conquer checkpointing has only been provided to date in AD systems
in special cases.
For example, \Tapenade\ allows the user to select invocations of a specified DO
loop as the root execution interval for divide-and-conquer checkpointing with
the \binomialckp\ pragma, taking iteration boundaries of that loop as candidate
split points.
\Tapenade\ employs binomial selection of split points and a fixed space
overhead termination criterion.
Note, however, that \Tapenade\ only guarantees this fixed space overhead
property for DO loop bodies that take constant time.
Similarly \ADOLC\ \citep{griewank1996apf} contains a nested taping mechanism
for time-integration processes \citep{kowarz2006ocf} that also performs
divide-and-conquer checkpointing.
This only applies to code formulated as a time-integration process.

Here, we present a framework for applying divide-and-conquer checkpointing to
arbitrary code with no special annotation or refactoring required.
An entire specified derivative calculation is taken as the root execution
interval, rather than invocations of a specified DO loop.
Arbitrary execution points are taken as candidate split points, rather than
iteration boundaries.
As discussed below in Section~\ref{sec:binomial}, both binary and
n-ary (treeverse) checkpoint trees are supported.
Furthermore, as discussed below in Section~\ref{sec:binomial}, both bisection
and binomial checkpointing are supported.
Additionally, all of the above termination criteria are supported: fixed space
overhead, fixed time overhead, and logarithmic space and time overhead.
Any combination of the above checkpoint-tree generation algorithms, split-point
selection methods, and termination criteria are supported.
In order to apply this framework, we must be able to interrupt the primal
computation at appropriate execution points, save the interrupted execution
state as a capsule, and restart the computation from the capsules, sometimes
repeatedly.
This is accomplished by building divide-and-conquer checkpointing on top of
a general-purpose mechanism for interrupting and resuming computation.
This mechanism is similar to \emph{engines} \citep{haynes1984engines} and is
orthogonal to AD.\@
We present several implementations of our framework which we call
\checkpointVLAD.
In Section~\ref{sec:example}, we compare the space and time usage of our
framework with that of \Tapenade\ on an example.

Note that one cannot generally achieve the space and time guarantees of
divide-and-conquer checkpointing with program-interval, subroutine-call-site,
or subroutine-body checkpointing unless the call tree has the same shape as the
requisite checkpoint tree.
Furthermore, one cannot generally achieve the space and time guarantees of
divide-and-conquer checkpointing for DO loops by specifying the loop body as a
program-interval checkpoint, as that would lead to a right-branching
checkpoint tree and behavior analogous to
Fig.~\ref{fig:divide-and-conquer-checkpointing}(d).
Moreover, if one allows split points at arbitrary execution points, the
resulting checkpoint execution intervals may not correspond to program
intervals.

Some form of divide-and-conquer checkpointing is \emph{necessary}.
One may wish to take the gradient of a long-running computation, even if it has
low asymptotic time complexity.
The length of the tape required by reverse mode without divide-and-conquer
checkpointing increases with increasing run time.
Modern computers can execute several billion floating point operations per
second, even without GPUs and multiple cores, which only exacerbate the problem.
If each such operation required storage of a single eight-byte double precision
number, modern terabyte RAM sizes would fill up after a few seconds of
computation.
Thus without some form of divide-and-conquer checkpointing, it would not be
possible to efficiently take the gradient of a computation that takes more than
a few seconds.

Machine learning methods in general, and deep learning methods in particular,
require taking gradients of long-running high-dimension computations,
particularly when training deep neural networks in general or recurrent
neural networks over long time series.
Thus variants of divide-and-conquer checkpointing have been rediscovered
and deployed by the machine learning community in this context
\citep{chen-etal-2016a, gruslys-etal-2016a}.
These implementations are far from automatic, and depend on compile-time
analysis of the static primal flow graphs.

The general strategy of divide-and-conquer checkpointing, the n-ary treeverse
algorithm, the bisection and binomial strategies for selecting split points,
and the termination criteria that provide fixed space overhead, fixed time
overhead, and logarithmic space and time overhead were all presented in
\citep{griewank1992alg}.
Furthermore, \Tapenade\ has implemented divide-and-conquer checkpointing with
the n-ary treeverse algorithm, the binomial strategy for selecting split
points, and the termination criterion that provides fixed space overhead, but
only for root execution intervals corresponding to invocations of specified DO
loops that meet certain criteria with split points restricted to iteration
boundaries of those loops.
To our knowledge, the binary checkpoint-tree algorithm presented here and the
framework for allowing it to achieve all of the same guarantees as the n-ary
treeverse algorithm is new.
However, our central novel contribution here is providing a framework for
supporting either the binary checkpoint-tree algorithm or the n-ary treeverse
algorithm, either bisection or binomial split point selection, and any of the
termination criteria of fixed space overhead, fixed time overhead, or
logarithmic space and time overhead in a way that supports taking the entire
derivative calculation as the root execution interval and taking arbitrary
execution points as candidate split points, by integrating the framework into
the language implementation.

Some earlier work \cite{heller1998checkpointing, stovboun2000tool,
  kang2003implementation} prophetically presaged the work here.
This work seems to have received far less exposure and attention than deserved.
Perhaps because the ideas therein were so advanced and intricate that it was
difficult to communicate those ideas clearly.
Moreover, the authors report difficulties in getting their implementations to
be fully functional.
Our work here formulates the requisite ideas and mechanisms carefully and
precisely, using methods from the programming-language community, like
formulation of divide-and-conquer checkpointing of a function as
divide-and-conquer application of reverse mode to two functions whose
composition is the original function, formulation of the requisite
decomposition as a precise and abstract interruption and resumption interface,
formulation of semantics precisely through specification of evaluators, use of
CPS evaluators to specify an implementation of the interruption and resumption
interface, and systematic derivation of a compiler from that evaluator via CPS
conversion, to allow complete, correct, comprehensible, and fully general
implementation.

\section{The Limitations of Divide-and-Conquer Checkpointing with Split Points
  at Fixed Syntactic Program Points like Loop Iteration Boundaries}
\label{sec:limitations}

Consider the example in Fig.~\ref{fig:example-fortran}.
This example, $y=f(x;l,\phi)$, while contrived, is a simple caricature of a
situation that arises commonly in practice: modeling a physical system with an
adaptive grid.
An initial state vector $x:\Re^n$ is repeatedly transformed by a state update
process $\Re^n\rightarrow\Re^n$ and, upon termination, an aggregate
property~$y$ of the final state is computed by a function
$\Re^n\rightarrow\Re$.
We wish to compute the gradient of that property~$y$ relative to the initial
state~$x$.
Here, the state update process first rotates the value pairs at adjacent
odd-even coordinates of the state~$x$ by an angle~$\theta$ and then rotates
those at adjacent even-odd coordinates.
The rotation~$\theta$ is taken to be proportional to the magnitude of~$x$.
The adaptive grid manifests in two nested update loops.
The outer loop has duration~$l$, specified as an input hyperparameter.
The duration~$m$ of the inner loop varies wildly as some function
of another input hyperparameter~$\phi$ and the outer loop index~$i$, perhaps
$2^{\lfloor\lg(l)\rfloor-\lfloor\lg(1+(1013\lfloor3^x\rfloor i\mod l))\rfloor}$,
that is small on most iterations of the outer loop but $O(l)$ on a few
iterations.
If the split points were limited to iteration boundaries of the outer loop, as
would be common in existing implementations, the increase in space or time
requirements would grow larger than sublinearly.
The issue is that for the desired sublinear growth properties to hold, it must
be possible to select arbitrary execution points as split points.
In other words, the granularity of the divide-and-conquer decomposition
must be primitive atomic computations, not loop iterations.
The distribution of run time across the program is not modularly reflected in
the static syntactic structure of the source code, in this case the loop
structure.
Often the user is unaware of, or even unconcerned with, the micro-level structure
of atomic computations, and does not wish to break the modularity of the source
code to expose it.
Yet the user may still wish to reap the sublinear space or time overhead
benefits of divide-and-conquer checkpointing.
Moreover, the relative duration of different paths through a program may vary
from loop iteration to loop iteration in a fashion that is data dependent, as
shown by the above example, and not even statically determinable.
We will now proceed to discuss an implementation strategy for
divide-and-conquer checkpointing that does not constrain split points to loop
iteration boundaries or other syntactic program constructs and does not
constrain checkpoints to program intervals or other syntactic program
constructs.
Instead, it can take any arbitrary execution point as a split point and
introduce checkpoints at any resulting execution interval.

\begin{figure}
  \centering
  \resizebox{\columnwidth}{!}{\begin{tabular}{@{}ll@{}}
  \wrap{
\begin{lstlisting}[lastline=40]
      function ilog2(l)
      ilog2 = dlog(real(l, 8))/dlog(2.0d0)
      end

      subroutine rotate(theta, x1, x2, x1p, x2p)
      double precision theta, x1, x2, x1p, x2p, c, s
      c = cos(theta)
      s = sin(theta)
      x1p = c*x1-s*x2
      x2p = s*x1+c*x2
      end

      subroutine rot1(theta, n, x)
      double precision theta, x, x1p, x2p
      dimension x(n)
      do k = 1, n-1, 2
         call rotate(theta, x(k), x(k+1), x1p, x2p)
         x(k) = x1p
         x(k+1) = x2p
      end do
      end

      subroutine rot2(theta, n, x)
      double precision theta, x, x1p, x2p
      dimension x(n)
      do k = 2, n-2, 2
         call rotate(theta, x(k), x(k+1), x1p, x2p)
         x(k) = x1p
         x(k+1) = x2p
      end do
      end

      subroutine magsqr(n, x, y)
      double precision x, y
      dimension x(n)
      y = 0.0d0
      do k = 1, n
         y = y+x(k)*x(k)
      end do
      end
\end{lstlisting}}&
  \wrap{
\begin{lstlisting}[firstline=42]
      subroutine f(n, x, l, phi, y)
      double precision phi, x, y, x1
      dimension x(n), x1(1000)
      do k = 1, n
         x1(k) = x(k)
      end do
c$ad binomial-ckp l+1 40 1
      do i = 1, l
         m = 2**(ilog2(l)-
     +           ilog2(1+int(mod(1013.0d0*3.0d0**phi*real(i, 8),
     +                           real(l, 8)))))
         do j = 1, m
            call magsqr(n, x1, y)
            y = sqrt(y)
            call rot1(1.2d0*y, n, x1)
            call rot2(1.4d0*y, n, x1)
         end do
      end do
      call magsqr(n, x1, y)
      y = y/2.0d0
      end

      program main
      double precision phi, x, xb, y, yb
      dimension x(1000), xb(1000)
      read *, n
      read *, l
      read *, phi
      do k = 1, n
         x(k) = n+1-k
         xb(k) = 0.0d0
      end do
      yb = 1.0d0
      call f(n, x, l, phi, y)
      print *, y
      call f_b(n, x, xb, l, phi, y, yb)
      do k = 1, n
         print *, xb(k)
      end do
      end
\end{lstlisting}}
  \end{tabular}}
  \caption{\Fortran\ example.
    This example is rendered in \checkpointVLAD\ in Fig.~\ref{fig:example-vlad}.
    Space and time overhead of two variants of this example when run under
    \Tapenade\ are presented in Fig.~\ref{fig:example-plots}.
    The pragma used for the variant with divide-and-conquer checkpointing of
    the outer DO loop is shown.
    This pragma is removed for the variant with no checkpointing.}
  \label{fig:example-fortran}
\end{figure}

\section{Technical Details of our Method}
\label{sec:method}

Implementing divide-and-conquer checkpointing requires the capacity to
\begin{enumerate}
\item measure the length of the primal computation,
\item interrupt the primal computation at a portion of the measured length,
\item save the state of the interrupted computation as a capsule, and
\item resume an interrupted computation from a capsule.
\end{enumerate}
For our purposes, the second and third operations are always performed together
and can be fused into a single operation.
These can be difficult to implement efficiently as library routines in an
existing language implementation (see Section~\ref{sec:implementation}).
Thus we design a new language implementation, \checkpointVLAD, with efficient
support for these low-level operations.

\subsection{Core Language}

\checkpointVLAD\ adds builtin AD operators to a functional pre-AD core
language.
In the actual implementation, this core language is provided with a
\Scheme-like surface syntax.\footnote{The surface syntax employed differs
  slightly from \Scheme\ in ways that are irrelevant to the issue at hand.}
But nothing turns on this; the core language can be exposed with any surface
syntax.
For expository purposes, we present the core language here in a simple, more
traditional, math-like notation.

\checkpointVLAD\ employs the same functional core language as our earlier
\VLAD\ system \citep{ad2016b}.
Support for AD in general, and divide-and-conquer checkpointing in particular,
is simplified in a functional programming language (see
Section~\ref{sec:advantages}).
Except for this simplification, which can be eliminated with well-known
techniques (\eg\ monads \citep{wadler90comprehending} and uniqueness types
\citep{achten1993high}) for supporting mutation in functional languages,
nothing turns on our choice of core language.
We intend our core language as a simple expository vehicle for the ideas
presented here; they could be implemented in other core languages (see
Section~\ref{sec:implementation}).

Our core language contains the following constructs:
\begin{equation}
  e::= c\mid
       x\mid
       \lambda x\LD{}e\mid
       e_1\;e_2\mid
       \IF e_1\;\THEN e_2\;\ELSE e_3\mid
       \diamond e\mid
       e_1\bullet e_2
\label{eq:core}
\end{equation}
Here, $e$~denotes expressions, $c$~denotes constants, $x$~denotes variables,
$e_1\;e_2$ denotes function application, $\diamond$~denotes builtin unary
operators, and $\bullet$~denotes builtin binary operators.
For expository simplicity, the discussion of the core language here omits
many vagaries such as support for recursion and functions of multiple
arguments; the actual implementation supports these using standard mechanisms
that are well known within the programming-language community (\eg\ tupling or
Currying).

\subsection{Direct-Style Evaluator for the Core Language}

We start by formulating a simple evaluator for this core language
(Fig.~\ref{fig:direct-core}) and extend it to perform AD and ultimately
divide-and-conquer checkpointing.
This evaluator is written in what is known in the programming-language
community as \defoccur{direct style}, where functions (in this case~$\EVAL$,
denoting `eval' and~$\APPLY$, denoting `apply') take inputs as function-call
arguments and yield outputs as function-call return values
\citep{reynolds1993discoveries}.
While this evaluator can be viewed as an interpreter, it is intended more as a
description of the evaluation mechanism; this mechanism could be the underlying
hardware as exposed via a compiler.
Indeed, as described below in Section~\ref{sec:implementations}, we have
written three implementations, one an interpreter, one a hybrid
compiler/interpreter, and one a compiler.

\begin{figure}
  % no macro expansion
  % is il:restrict necessary or just an optimization?
  % no parameters (destructuring)
  % no wrapping primitives
  % no recursive closures and letrec
  % no checkpoint and resume expressions
  % no continuation and converted lambda expressions and applications
  % no let expressions
  % no il:increment, il:cons, il:limit-check, and il:make-checkpoint
  \centering
  \fbox{\hspace*{-11pt}
    \begin{minipage}{\textwidth}
      \par\vspace*{-15pt}
      \begin{subequations}
      \begin{align}
        \APPLY\;\CLOSURE{\lambda x\LD{}e}{\rho}\;v&=\EVAL\;\rho[x\mapsto v]\;e
        \label{eq:direct-apply}\\
        \EVAL\;\rho\;c&=c\label{eq:direct-constant}\\
        \EVAL\;\rho\;x&=\rho\;x\label{eq:direct-variable}\\
        \EVAL\;\rho\;(\lambda x\LD{}e)&=\CLOSURE{\lambda x\LD{}e}{\rho}
        \label{eq:direct-lambda}\\
        \EVAL\;\rho\;(e_1\;e_2)&=
        \APPLY\;(\EVAL\;\rho\;e_1)\;(\EVAL\;\rho\;e_2)
        \label{eq:direct-application}\\
        \EVAL\;\rho\;(\IF e_1\;\THEN e_2\;\ELSE e_3)&=
        \IF(\EVAL\;\rho\;e_1)\;
        \THEN(\EVAL\;\rho\;e_2)\;
        \ELSE(\EVAL\;\rho\;e_3)
        \label{eq:direct-conditional}\\
        \EVAL\;\rho\;(\diamond e)&=\diamond(\EVAL\;\rho\;e)
        \label{eq:direct-unary}\\
        \EVAL\;\rho\;(e_1\bullet e_2)&=
        (\EVAL\;\rho\;e_1)\bullet(\EVAL\;\rho\;e_2)\label{eq:direct-binary}
      \end{align}
      \end{subequations}
    \end{minipage}}
  \caption{Direct-style evaluator for the core \checkpointVLAD\ language.}
  \label{fig:direct-core}
\end{figure}

With any evaluator, one distinguishes between two language evaluation strata:
the \defoccur{target}, the language being implemented and the process of
evaluating programs in that language, and the \defoccur{host}, the language in
which the evaluator is written and the process of evaluating the evaluator
itself.
In our case, the target is \checkpointVLAD, while the host varies among our
three implementations; for the first two it is \Scheme\ while for the third it
is the underlying hardware, achieved by compilation to machine code via
\Clang.

In the evaluator in Fig.~\ref{fig:direct-core}, $\rho$~denotes an
\defoccur{environment}, a mapping from variables to their values,
$\rho_0$~denotes the empty environment that does not map any variables,
$\rho\;x$ denotes looking up the variable~$x$ in the environment~$\rho$ to
obtain its value, $\rho[x\mapsto v]$ denotes augmenting an environment~$\rho$
to map the variable~$x$ to the value~$v$, and $\EVAL\;\rho\;e$ denotes
evaluating the expression~$e$ in the context of the environment~$\rho$.
There is a clause for $\EVAL$ in Fig.~\ref{fig:direct-core},
\crefrange{eq:direct-constant}{eq:direct-binary}, for each construct
in~\eqref{eq:core}.
Clause~\eqref{eq:direct-constant} says that one evaluates a constant by
returning that constant.
Clause~\eqref{eq:direct-variable} says that one evaluates a variable by
returning its value in the environment.
The notation $\langle e,\rho\rangle$ denotes a \defoccur{closure}, a lambda
expression~$e$ together with an environment~$\rho$ containing values for the
free variables in~$e$.
Clause~\eqref{eq:direct-lambda} says that one evaluates a lambda expression by
returning a closure with the environment in the context that the lambda
expression was evaluated in.
Clause~\eqref{eq:direct-application} says that one evaluates an application by
evaluating the callee expression to obtain a closure, evaluating the argument
expression to obtain a value, and then applying the closure to the value
with~$\APPLY$.\@
$\APPLY$, as described in~\eqref{eq:direct-apply}, evaluates the body of the
lambda expression in the callee closure in the environment of that closure
augmented with the formal parameter of that lambda expression bound to the
argument value.
The remaining clauses are all analogous to clause~\eqref{eq:direct-binary},
which says that one evaluates an expression $e_1\bullet e_2$ in the target by
evaluating~$e_1$ and~$e_2$ to obtain values and then applying~$\bullet$ in the
host to these values.

\subsection{Adding AD Operators to the Core Language}

Unlike many AD systems implemented as libraries, we provide support for AD by
augmenting the core language to include builtin AD operators for both forward
and reverse mode \citep{ad2016b}.
This allows seamless integration of AD into the language in a completely
general fashion with no unimplemented or erroneous corner cases.
In particular, it allows nesting \citep{siskindpearlmutter2008a}.
In \checkpointVLAD, we adopt slight variants of the~$\FORWARDJ$ and~$\REVERSEJ$
operators previously incorporated into \VLAD.
(Nothing turns on this.
The variants adopted here are simpler, better suit our expository
purposes, and allow us to focus on the issue at hand.)
In \checkpointVLAD, these operators have the following signatures.
\begin{align}
  \FORWARDJ&:f\;x\;\acute{x}\mapsto (y, \acute{y}) &
  \REVERSEJ&:f\;x\;\grave{y}\mapsto (y, \grave{x})
\end{align}
We use the notation~$\acute{x}$ and~$\grave{x}$ to denote tangent or
cotangent values associated with the primal value~$x$ respectively, and the
notation~$(x,y)$ to denote a pair of values.
Since in \checkpointVLAD, functions can take multiple arguments but only return
a single result, which can be an aggregate like a pair, the AD operators take
the primal and the associated (co)tangent as distinct arguments but return the
primal and the associated (co)tangent as a pair of values.

The~$\FORWARDJ$ operator provides the portal to forward mode and calls a
function~$f$ on a primal~$x$ with a tangent~$\acute{x}$ to yield a primal~$y$
and a tangent~$\acute{y}$.
The~$\REVERSEJ$ operator provides the portal to reverse mode and calls a
function~$f$ on a primal~$x$ with a cotangent~$\grave{y}$ to yield a primal~$y$
and a cotangent~$\grave{x}$.\footnote{In the implementation,
  $\FORWARDJ$~and~$\REVERSEJ$ are named \texttt{j*} and \texttt{*j}
  respectively.}
Unlike the original \VLAD, here, we restrict ourselves to the case where
(co)tangents are ground data values, \ie\ reals and (arbitrary) data structures
containing reals and other scalar values, but not functions (\ie\ closures).
Nothing turns on this; it allows us to focus on the issue at hand.

The implementations of \VLAD\ and \checkpointVLAD\ are disjoint and use
completely different technology.
The \Stalingrad\ \citep{ad2016b} implementation of \VLAD\ is based on
source-code transformation, conceptually applied reflectively at run time but
migrated to compile time through partial evaluation.
The implementation of \checkpointVLAD\ uses something more akin to operator
overloading.
Again, nothing turns on this; this simplification is for expository purposes and
allows us to focus on the issue at hand (see Section~\ref{sec:implementation}).

In \checkpointVLAD, AD is performed by overloading the arithmetic operations
in the host, in a fashion similar to \FADBAD\ \citep{bendtsen1996faf}.
The actual method used is that employed by
\RsixRSAD\footnote{\url{https://github.com/qobi/R6RS-AD} and
\url{https://engineering.purdue.edu/~qobi/stalingrad-examples2009/}} and
\DiffSharp\footnote{\url{http://diffsharp.github.io/DiffSharp/}}.
The key difference is that \FADBAD\ uses \Cplusplus\ templates to encode a
hierarchy of distinct forward-mode types (\eg\ \verb|F<double>|,
\verb|F<F<double> >|, \ldots), distinct reverse-mode types (\eg\
\verb|B<double>|, \verb|B<B<double> >|, \ldots), and mixtures thereof
(\eg\ \verb|F<B<double> >|, \verb|B<F<double> >|, \ldots) while
here, we use a dynamic, run-time approach where numeric values are tagged with
the nesting level \citep{sussmanwm2001, siskindpearlmutter2008a}.
Template instantiation at compile-time specializes code to different nesting
levels.
The dynamic approach allows a single interpreter (host), formulated around
unspecialized code, to interpret different target programs with different
nesting levels.

\subsection{Augmenting the Direct-Style Evaluator to Support the AD Operators}

We add AD into the target language as new constructs.
\begin{equation}
  e::= \FORWARDJ\;e_1\;e_2\;e_3\mid
       \REVERSEJ\;e_1\;e_2\;e_3
 \label{eq:AD}
\end{equation}
We implement this functionality by augmenting the direct-style evaluator with
new clauses for $\EVAL$ (Fig.~\ref{fig:direct-AD}),
clause~\eqref{eq:direct-jstar} for~$\FORWARDJ$ and
clause~\eqref{eq:direct-starj} for~$\REVERSEJ$.
These clauses are all analogous to clause~\eqref{eq:direct-binary}, formulated
around~$\FORWARDJ$ and~$\REVERSEJ$ operators in the host.
These are defined in~\cref{eq:direct-forward,eq:direct-reverse}.
The~$\FORWARDJ$ and~$\REVERSEJ$ operators in the host behave like~$\APPLY$
except that they level shift to perform AD.\@
Just like $(\APPLY\;f\;x)$ applies a target function~$f$ (closure) to a target
value~$x$, $(\FORWARDJ\;f\;x\;\acute{x})$ performs forward mode by applying a
target function~$f$ (closure) to a target primal value~$x$ and a target tangent
value~$\acute{x}$, while $(\REVERSEJ\;f\;x\;\acute{y})$ performs reverse mode by
applying a target function~$f$ (closure) to a target primal value~$x$ and a
target cotangent value~$\grave{y}$.

\setcounter{equation}{1}
\begin{figure}
  \centering
  \fbox{\hspace*{-11pt}
    \begin{minipage}{\textwidth}
      \par\vspace*{-15pt}
      \begin{subequations}
      \setcounter{equation}{8}
      \begin{align}
        \FORWARDJ\;v_1\;v_2\;\acute{v}_3&=
        \LET(\BUNDLE{v_4}{\acute{v}_5})=
        (\APPLY\;v_1\;(\BUNDLE{v_2}{\acute{v}_3}))
        \;\IN(v_4,\acute{v}_5)\label{eq:direct-forward}\\
        \REVERSEJ\;v_1\;v_2\;\grave{v}_3&=
        \LET(\TAPE{v_4}{\grave{v}_5})=(\TAPE{(\APPLY\;v_1\;v_2)}{\grave{v}_3})
        \;\IN(v_4,\grave{v}_5)\label{eq:direct-reverse}\\
        \EVAL\;\rho\;(\FORWARDJ\;e_1\;e_2\;e_3)&=
        \FORWARDJ\;(\EVAL\;\rho\;e_1)\;(\EVAL\;\rho\;e_2)\;(\EVAL\;\rho\;e_3)
        \label{eq:direct-jstar}\\
        \EVAL\;\rho\;(\REVERSEJ\;e_1\;e_2\;e_3)&=
        \REVERSEJ\;(\EVAL\;\rho\;e_1)\;(\EVAL\;\rho\;e_2)\;(\EVAL\;\rho\;e_3)
        \label{eq:direct-starj}
      \end{align}
      \end{subequations}
    \end{minipage}}
  \caption{Additions to the direct-style evaluator for \checkpointVLAD\ to
    support AD.\@}
  \label{fig:direct-AD}
\end{figure}

As described in~\eqref{eq:direct-forward}, $\FORWARDJ$ operates by recursively
walking~$v_2$, a data structure containing primals, in tandem
with~$\acute{v}_3$, a data structure containing tangents, to yield a single
data structure where each numeric leaf value is a dual number, a numeric
primal value associated with a numeric tangent value.
This recursive walk is denoted as $\BUNDLE{v_2}{\acute{v}_3}$.
$\APPLY$~is then used to apply the function (closure)~$v_1$ to the data
structure produced by $\BUNDLE{v_2}{\acute{v}_3}$.
Since the input argument is level shifted and contains dual numbers instead of
ordinary reals, the underlying arithmetic operators invoked during the
application perform forward mode by dispatching on the tags at run time.
The call to $\APPLY$ yields a result data structure where each numeric leaf
value is a dual number.
This is then recursively walked to separate out two data structures, one,
$v_4$, containing the numeric primal result values, and the other,
$\acute{v}_5$, containing the numeric tangent result values, which are returned
as a pair $(v_4,\acute{v}_5)$
This recursive walk is denoted as
$\LET(\BUNDLE{v_4}{\acute{v}_5})=\ldots\;\IN\ldots$.

As described in~\eqref{eq:direct-reverse}, $\REVERSEJ$ operates by recursively
walking~$v_2$, a data structure containing primals, to replace each numeric
value with a tape node.
$\APPLY$~is then used to apply the function (closure)~$v_1$ to this
modified~$v_2$.
Since the input argument is level shifted and contains tape nodes instead of
ordinary reals, the underlying arithmetic operators invoked during the
application perform the forward sweep of reverse mode by dispatching on the
tags at run time.
The call to $\APPLY$ yields a result data structure where each numeric leaf
value is a tape node.
A recursive walk is performed on this result data structure, in tandem with a
data structure~$\grave{v}_3$ of associated cotangent values, to initiate the
reverse sweep of reverse mode.
This combined operation is denoted as
$(\TAPE{(\APPLY\;v_1\;v_2)}{\grave{v}_3})$.
The result of the forward sweep is then recursively walked to replace each tape
node with its numeric primal value and the input value is recursively walked to
replace each tape node with the cotangent computed by the reverse sweep.
These are returned as a pair $(v_4,\grave{v}_5)$.
This combined operation is denoted as
$\LET(\TAPE{v_4}{\grave{v}_5})=\ldots\;\IN\ldots$.

\subsection{An Operator to Perform Divide-and-Conquer Checkpointing in
  Reverse-Mode AD}

We introduce a new AD operator~$\CHECKPOINTREVERSEJ$ to perform
divide-and-conquer checkpointing.\footnote{In the implementation,
  $\CHECKPOINTREVERSEJ$~is named \texttt{checkpoint-*j}.}
(For expository simplicity, we focus for now on binary bisection checkpointing.
In Section~\ref{sec:binomial} below, we provide alternate implementations
of~$\CHECKPOINTREVERSEJ$ that perform treeverse and/or binomial checkpointing.)
The crucial aspect of the design is that the signature (and semantics)
of~$\CHECKPOINTREVERSEJ$ is \emph{identical} to~$\REVERSEJ$; they are
\emph{completely interchangeable}, differing only in the space/time complexity
tradeoffs.
This means that code \emph{need not be modified} to switch back and forth
between ordinary reverse mode and various forms of divide-and-conquer
checkpointing, save interchanging calls to~$\REVERSEJ$
and~$\CHECKPOINTREVERSEJ$.\@

\begin{figure}
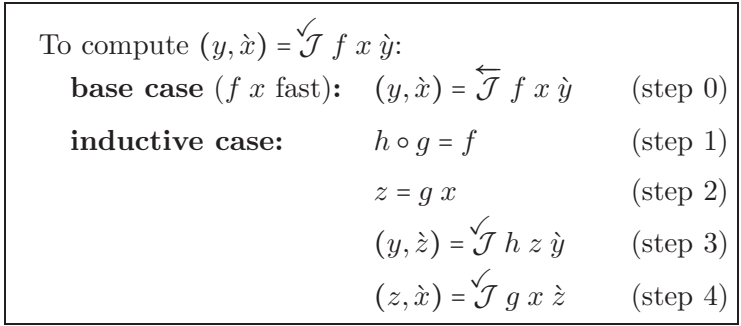

  \centering
  \fbox{
    \begin{tabular}{lll@{\hspace{2em}}l@{}}
      \multicolumn{4}{l}{To compute $(y,\grave{x})=
        \CHECKPOINTREVERSEJ\;f\;x\;\grave{y}$:}\\
      & \textbf{base case} ($f\;x$ fast)\textbf{:}
      & $(y,\grave{x})=\REVERSEJ\;f\;x\;\grave{y}$ & (step~0)\\[1.5ex]
      & \textbf{inductive case:}
      & $h\circ g=f$ & (step~1)\\[1.2ex]
      && $z=g\;x$ & (step~2)\\
      && $(y,\grave{z})=\CHECKPOINTREVERSEJ\;h\;z\;\grave{y}$ & (step~3)\\
      && $(z,\grave{x})=\CHECKPOINTREVERSEJ\;g\;x\;\grave{z}$ & (step~4)
    \end{tabular}}
  \caption{Algorithm for binary checkpointing.}
  \label{fig:binary}
\end{figure}

Conceptually, the behavior of~$\CHECKPOINTREVERSEJ$ is shown in
Fig.~\ref{fig:binary}.
In this inductive definition, a function~$f$ is split into the composition of
two functions~$g$ and~$h$ in step~(1), the capsule~$z$ is computed by
applying~$g$ to the input~$x$ in step~(2), and the cotangent is computed by
recursively applying~$\CHECKPOINTREVERSEJ$ to~$h$ and~$g$ in steps~3 and~4.
This divide-and-conquer behavior is terminated in a base case, when the
function~$f$ is small, at which point the cotangent is computed
with~$\REVERSEJ$, in step~(0).
If step~(1) splits a function~$f$ into two functions~$g$ and~$h$ that take the
same number of evaluation steps, and we terminate the recursion when~$f$
takes a bounded number of steps, the recursive divide-and-conquer process
yields logarithmic asymptotic space/time overhead complexity.

The central difficulty in implementing the above is performing step~(1), namely
splitting a function~$f$ into two functions~$g$ and~$h$, such that $f=h\circ
g$, ideally where we can specify the split point, the number of evaluation
steps through~$f$ where~$g$ transitions into~$h$.
A sophisticated user can manually rewrite a subprogram~$f$ into two
subprograms~$g$ and~$h$.
A sufficiently powerful compiler or source transformation tool might also be
able to do so, with access to nonlocal program text.
But an overloading system, with access only to local information, would not be
able to.

\subsection{General-Purpose Interruption and Resumption Mechanism}

We solve this problem by providing an interface to a general-purpose
interruption and resumption mechanism that is orthogonal to AD
(Fig.~\ref{fig:interface}).
This interface allows (a)~determining the number of evaluation steps of a
computation, (b)~interrupting a computation after a specified number of steps,
usually half the number of steps determined by the mechanism in~(a), and
(c)~resuming an interrupted computation to completion.
A variety of implementation strategies for this interface are possible.
We present two in detail below, in Sections~\ref{sec:implementing}
and~\ref{sec:augmenting}, and briefly discuss another in
Section~\ref{sec:implementation}.

\begin{figure}
  \centering
  \fbox{
    \begin{tabular}{lp{0.65\textwidth}}
      $\textsc{primops}\;f\;x\mapsto l$&Return the number~$l$ of evaluation
      steps needed to compute $y=f(x)$.\\[1ex]
      $\textsc{interrupt}\;f\;x\;l\mapsto z$&Run the first~$l$ steps of the
      computation of $f(x)$ and return a capsule $z$.\\[1ex]
      $\textsc{resume}\;z\mapsto y$&
      If $z=(\textsc{interrupt}\;f\;x\;l)$, return $y=f(x)$.
    \end{tabular}}
  \caption{General-purpose interruption and resumption interface.}
  \label{fig:interface}
\end{figure}

Irrespective of how one implements the general-purpose interruption and
resumption interface, one can use it to implement the binary bisection variant
of~$\CHECKPOINTREVERSEJ$ in the host, as shown in
Fig.~\ref{fig:implementation}.
The function~$f$ is split into the composition of two functions~$g$ and~$h$ by
taking~$g$ as $(\lambda x\LD{}\textsc{interrupt}\;f\;x\;l)$, where~$l$ is half
the number of steps determined by $(\textsc{primops}\;f\;x)$, and~$h$ as
$(\lambda z\LD{}\textsc{resume}\;z)$.

\begin{figure}
  \centering
  \fbox{
    \begin{tabular}{lll@{\hspace{2em}}l@{}}
      \multicolumn{4}{l}{To compute $(y,\grave{x})=
        \CHECKPOINTREVERSEJ\;f\;x\;\grave{y}$:}\\
      & \textbf{base case:}
      & $(y,\grave{x})=\REVERSEJ\;f\;x\;\grave{y}$ & (step~0)\\[1.5ex]
      & \textbf{inductive case:}
      & $l=\textsc{primops}\;f\;x$ & (step~1)\\[1ex]
      && $z=\textsc{interrupt}\;f\;x\;\lfloor\frac{l}{2}\rfloor$ & (step~2)\\
      && $(y,\grave{z})=
      \CHECKPOINTREVERSEJ\;(\lambda z\LD{}\textsc{resume}\;z)\;z\;\grave{y}$ &
      (step~3)\\
      && $(z,\grave{x})=
      \CHECKPOINTREVERSEJ\;
      (\lambda x\LD{}\textsc{interrupt}\;f\;x\;\lfloor\frac{l}{2}\rfloor)
      \;x\;\grave{z}$ & (step~4)
  \end{tabular}}
  \caption{Binary bisection checkpointing via the general-purpose interruption
    and resumption interface.
    Step~(1) need only be performed once at the beginning of the recursion,
    with steps~(2) and~(4) taking~$l$ at the next recursion level to be
    $\lfloor\frac{l}{2}\rfloor$ and step~(3) taking~$l$ at the next recursion
    level to be $\lceil\frac{l}{2}\rceil$.
    As discussed in the text, this implementation is not quite correct, because
    $(\lambda z\LD{}\textsc{resume}\;z)$ in step~(3) and
    $(\lambda x\LD{}\textsc{interrupt}\;f\;x\;\lfloor\frac{l}{2}\rfloor)$ in
    step~(4) are host closures but need to be target closures.
    A proper implementation is given in Fig.~\ref{fig:proper-implementation}.}
  \label{fig:implementation}
\end{figure}

\subsection{Continuation-Passing-Style Evaluator}

One way of implementing the general-purpose interruption and resumption
interface is to convert the evaluator from direct style to what is known in the
programming-language community as \defoccur{continuation-passing style} (CPS)
\citep{reynolds1993discoveries}, where functions (in
this case~$\EVAL$, $\APPLY$, $\FORWARDJ$, and~$\REVERSEJ$ in the host) take an
additional continuation input~$k$ and instead of yielding outputs via
function-call return, do so by calling the continuation with said output as
arguments (Figs.~\ref{fig:cps-core} and~\ref{fig:cps-AD}).
In CPS, functions never return: they just call their continuation.
With tail-call merging, this corresponds to a computed \texttt{go to} and does
not incur stack growth.
This crucially allows an interruption to actually return a capsule containing
the saved state of the evaluator, including its continuation, allowing the
evaluation to be resumed by calling the evaluator with this saved state.
This `level shift' of return to calling a continuation, allowing an actual
return to constitute interruption, is analogous to the way backtracking is
classically implemented in \Prolog, with success implemented as calling a
continuation and failure implemented as actual return.
In our case, we further instrument the evaluator to thread two values as inputs
and outputs: the count~$n$ of the number of evaluation steps, which is
incremented at each call to~$\EVAL$, and the limit~$l$ of the number of steps,
after which an interrupt is triggered.

\setcounter{equation}{4}
\begin{subequations}

\begin{figure}
  % no macro expansion
  % is il:restrict necessary or just an optimization?
  % no parameters (destructuring)
  % no wrapping primitives
  % no recursive closures and letrec
  % vlad.ss doesn't yet have conditional and let expressions
  % n+1 and the implicit = in (4)
  % This doesn't pack values into the continuation.
  % I don't know if that is actually necessary. Actually, it is.
  % This doesn't do il:environment->values and then il:values->environment.
  \centering
  \fbox{\hspace*{-11pt}
    \begin{minipage}{\textwidth}
      \par\vspace*{-15pt}
      \begin{align}
        \APPLY\;k\;n\;l\;\CLOSURE{\lambda x\LD{}e}{\rho}\;v&=
        \EVAL\;k\;n\;l\;\rho[x\mapsto v]\;e\label{eq:cps-apply}\\
        \EVAL\;k\;l\;l\;\rho\;e&=
        \CAPSULE{k}{\CLOSURE{\lambda\_\LD{}e}{\rho}}
        \label{eq:cps-interruption}\\
        \EVAL\;k\;n\;l\;\rho\;c&=k\;(n+1)\;l\;c\label{eq:cps-constant}\\
        \EVAL\;k\;n\;l\;\rho\;x&=k\;(n+1)\;l\;(\rho\;x)\label{eq:cps-variable}\\
        \EVAL\;k\;n\;l\;\rho\;(\lambda x\LD{}e)&=
        k\;(n+1)\;l\;\CLOSURE{\lambda x\LD{}e}{\rho}\label{eq:cps-lambda}\\
        \EVAL\;k\;n\;l\;\rho\;(e_1\;e_2)&=
        \EVAL\;
        \begin{array}[t]{@{}l@{}}
        (\begin{array}[t]{@{}l@{}}
        \lambda n\;l\;v_1\LD{}\\
        (\EVAL\;
        \begin{array}[t]{@{}l@{}}
        (\begin{array}[t]{@{}l@{}}
        \lambda n\;l\;v_2\LD{}\\
        (\APPLY\;k\;n\;l\;v_1\;v_2))
        \end{array}\\
        n\;l\;\rho\;e_2))
        \end{array}
        \end{array}\\
        (n+1)\;l\;\rho\;e_1
        \end{array}\label{eq:cps-application}\\
        \EVAL\;k\;n\;l\;\rho\;(\IF e_1\;\THEN e_2\;\ELSE e_3)&=
        \EVAL\;
        \begin{array}[t]{@{}l@{}}
        (\begin{array}[t]{@{}l@{}}
        \lambda n\;l\;v_1\LD{}\\
        (\begin{array}[t]{@{}l@{}}
        \IF v_1\\
        \THEN(\EVAL\;k\;n\;l\;\rho\;e_2)\\
        \ELSE(\EVAL\;k\;n\;l\;\rho\;e_3)))
        \end{array}
        \end{array}\\
        (n+1)\;l\;\rho\;e_1
        \end{array}\label{eq:cps-conditional}\\
        \EVAL\;k\;n\;l\;\rho\;(\diamond e)&=
        \EVAL\;
        \begin{array}[t]{@{}l@{}}
        (\begin{array}[t]{@{}l@{}}
        \lambda n\;l\;v\LD{}\\
        (k\;n\;l\;(\diamond v)))
        \end{array}\\
        (n+1)\;l\;\rho\;e
        \end{array}\label{eq:cps-unary}\\
        \EVAL\;k\;n\;l\;\rho\;(e_1\bullet e_2)&=
        \EVAL\;
        \begin{array}[t]{@{}l@{}}
        (\begin{array}[t]{@{}l@{}}
        \lambda n\;l\;v_1\LD{}\\
        (\EVAL\;
        \begin{array}[t]{@{}l@{}}
        (\begin{array}[t]{@{}l@{}}
        \lambda n\;l\;v_2\LD{}\\
        (k\;n\;l\;(v_1\bullet v_2)))
        \end{array}\\
        n\;l\;\rho\;e_2))
        \end{array}
        \end{array}\\
        (n+1)\;l\;\rho\;e_1
        \end{array}
        \label{eq:cps-binary}
      \end{align}
    \end{minipage}}
  \caption{CPS evaluator for the core \checkpointVLAD\ language.}
  \label{fig:cps-core}
\end{figure}

\begin{figure}
  \centering
  \fbox{\hspace*{-11pt}
    \begin{minipage}{\textwidth}
      \par\vspace*{-15pt}
      \begin{align}
        \FORWARDJ\;v_1\;v_2\;\acute{v}_3&=
        \APPLY\;
        \begin{array}[t]{@{}l@{}}
        (\begin{array}[t]{@{}l@{}}
        \lambda n\;l\;(\BUNDLE{v_4}{\acute{v}_5})\LD{}\\
        (v_4,\acute{v}_5))
        \end{array}\\
        0\;\infty\;v_1\;(\BUNDLE{v_2}{\acute{v}_3})
        \end{array}\label{eq:cps-forward}\\
        \REVERSEJ\;v_1\;v_2\;\grave{v}_3&=
        \APPLY\;
        \begin{array}[t]{@{}l@{}}
        (\begin{array}[t]{@{}l@{}}
        \lambda n\;l\;v\LD{}\\
        \LET(\TAPE{v_4}{\grave{v}_5})=\TAPE{v}{\grave{v}_3}\\
        \IN(v_4,\grave{v}_5))
        \end{array}\\
        0\;\infty\;v_1\;v_2
        \end{array}\label{eq:cps-reverse}\\
        \EVAL\;k\;n\;l\;\rho\;(\FORWARDJ\;e_1\;e_2\;e_3)&=
        \EVAL\;
        \begin{array}[t]{@{}l@{}}
        (\begin{array}[t]{@{}l@{}}
        \lambda n\;l\;v_1\LD{}\\
        (\EVAL\;
        \begin{array}[t]{@{}l@{}}
        (\begin{array}[t]{@{}l@{}}
        \lambda n\;l\;v_2\LD{}\\
        (\EVAL\;
        \begin{array}[t]{@{}l@{}}
        (\lambda n\;l\;v_3\LD{}(k\;n\;l\;(\FORWARDJ\;v_1\;v_2\;v_3)))\\
        n\;l\;\rho\;e_3))
        \end{array}
        \end{array}\\
        n\;l\;\rho\;e_2))
        \end{array}
        \end{array}\\
        (n+1)\;l\;\;\rho\;e_1
        \end{array}\label{eq:cps-jstar}\\
        \EVAL\;k\;n\;l\;\rho\;(\REVERSEJ\;e_1\;e_2\;e_3)&=
        \EVAL\;
        \begin{array}[t]{@{}l@{}}
        (\begin{array}[t]{@{}l@{}}
        \lambda n\;l\;v_1\LD{}\\
        (\EVAL\;
        \begin{array}[t]{@{}l@{}}
        (\begin{array}[t]{@{}l@{}}
        \lambda n\;l\;v_2\LD{}\\
        (\EVAL\;
        \begin{array}[t]{@{}l@{}}
        (\lambda n\;l\;v_3\LD{}(k\;n\;l\;(\REVERSEJ\;v_1\;v_2\;v_3)))\\
        n\;l\;\rho\;e_3))
        \end{array}
        \end{array}\\
        n\;l\;\rho\;e_2))
        \end{array}
        \end{array}\\
        (n+1)\;l\;\rho\;e_1
        \end{array}\label{eq:cps-starj}
      \end{align}
    \end{minipage}}
  \caption{Additions to the CPS evaluator for \checkpointVLAD\ to support AD.\@}
  \label{fig:cps-AD}
\end{figure}

\end{subequations}

Fig.~\ref{fig:cps-core} contains the portion of the CPS evaluator for the core
language corresponding to Fig.~\ref{fig:direct-core}, while
Fig.~\ref{fig:cps-AD} contains the portion of the CPS evaluator for the AD
constructs corresponding to Fig.~\ref{fig:direct-AD}.
Except for~\eqref{eq:cps-interruption}, the equations in
Figs.~\ref{fig:direct-core} and~\ref{fig:direct-AD} are in one-to-one
correspondence to those in Figs.~\ref{fig:cps-core} and~\ref{fig:cps-AD}, in
order.
Clauses \crefrange{eq:cps-constant}{eq:cps-lambda} are analogous to the
corresponding clauses \crefrange{eq:direct-constant}{eq:direct-lambda} except
that they call the continuation~$k$ with the result, instead of returning that
result.
The remaining clauses for~$\EVAL$ in the CPS evaluator are all variants of
\setcounter{equation}{5}
\begin{equation}
  \EVAL\;
  \begin{array}[t]{@{}l@{}}
  (\begin{array}[t]{@{}l@{}}
  \lambda n\;l\;v_1\LD{}
  (\EVAL\;
  \begin{array}[t]{@{}l@{}}
  (\begin{array}[t]{@{}l@{}}
  \lambda n\;l\;v_2\LD{}
  (k\;n\;l\;\ldots))
  \end{array}\;
  n\;l\;\rho\;e_2))
  \end{array}
  \end{array}\;
  (n+1)\;l\;\rho\;e_1
  \end{array}
  \label{eq:cps-template}
\end{equation}
for one-, two-, or three-argument constructs.
This evaluates the first argument~$e_1$ and calls the continuation
$(\lambda n\;l\;v_1\LD{}\ldots)$ with its value~$v_1$.
This continuation then evaluates the second argument~$e_2$ and calls the
continuation $(\lambda n\;l\;v_2\LD{}\ldots)$ with its value~$v_2$.
This continuation computes something, denoted by $\ldots$, and calls the
continuation~$k$ with the resulting value.

The CPS evaluator threads a step count~$n$ and a step limit~$l$ through the
evaluation process.
Each clause of~$\EVAL$ increments the step count exactly once to provide a
coherent fine-grained measurement of the execution time.
Clause~\eqref{eq:cps-interruption} of~$\EVAL$ implements interruption.
When the step count reaches the step limit, a capsule containing the saved
state of the evaluator, denoted $\CAPSULE{k}{f}$, is returned.
Here, $f$~is a closure $\CLOSURE{\lambda\_\LD{}e}{\rho}$ containing the
environment~$\rho$ and the expression~$e$ at the time of interruption.
This closure takes an argument that is not used.
The step count~$n$ must equal the step limit~$l$ at the time of interruption.
As will be discussed below, in Section~\ref{sec:implementing}, neither the step
count nor the step limit need to be saved in the capsule, as the computation
is always resumed with different step count and limit values.

Several things about this CPS evaluator are of note.
First, all builtin unary and binary operators are assumed to take unit time.
This follows from the fact that all clauses for~$\EVAL$, as typified
by~\eqref{eq:cps-template}, increment the step count by one.
Second, the builtin unary and binary operators in the host are implemented in
direct style and are not passed a continuation.
This means that clauses~\cref{eq:cps-unary,eq:cps-binary}, as typified
by~\eqref{eq:cps-template}, must call the continuation~$k$ on the result of the
unary and binary operators.
Third, like all builtin operators, invocations of the~$\FORWARDJ$
and~$\REVERSEJ$ operators, including the application of~$v_1$, are assumed to
take unit time.
This follows from the fact that clauses~\cref{eq:cps-jstar,eq:cps-starj}, again
as typified by~\eqref{eq:cps-template}, increment the step count by one.
Fourth, like all builtin operators, $\FORWARDJ$~and~$\REVERSEJ$ in the host,
in~\cref{eq:cps-forward,eq:cps-reverse}, are implemented in direct style and
are not passed a continuation.
This means that clauses~\cref{eq:cps-jstar,eq:cps-starj}, as typified
by~\eqref{eq:cps-template}, must call the continuation~$k$ on the result
of~$\FORWARDJ$ and~$\REVERSEJ$.\@
Finally, since~$\FORWARDJ$ and~$\REVERSEJ$ receive target functions (closures)
for~$v_1$, they must apply these to their arguments with~$\APPLY$.\@
Since~$\APPLY$ is written in CPS in the CPS evaluator, these calls to~$\APPLY$
in~\cref{eq:cps-forward,eq:cps-reverse} must be provided with a
continuation~$k$, a step count~$n$, and a step limit~$l$ as arguments.
The continuation argument simply returns the result.
The step count, however, is restarted at zero, and the step limit is set to
$\infty$.
This means that invocations of~$\FORWARDJ$ and~$\REVERSEJ$ are atomic and
cannot be interrupted internally.
We discuss this further below in Section~\ref{sec:nesting}.

\subsection{Implementing the General-Purpose Interruption and Resumption
  Interface with the CPS Evaluator}
\label{sec:implementing}

With this CPS evaluator, it is possible to implement the general-purpose
interruption and resumption interface (Fig.~\ref{fig:interface-implementation}).
The implementation of $\textsc{primops}$~\eqref{eq:cps-primops} calls the
evaluator with no step limit and simply counts the number of steps to
completion.
The implementation of $\textsc{interrupt}$~\eqref{eq:cps-interrupt} calls the
evaluator with a step limit that must be smaller than that needed to complete
so an interrupt is forced and the capsule
$\CAPSULE{k}{\CLOSURE{\lambda\_\LD{}e}{\rho}}$ is returned.
The implementation of $\textsc{resume}$~\eqref{eq:cps-resume} calls the
evaluator with arguments from the saved capsule.
Since the closure in the capsule does not use its argument, an arbitrary
value~$\bot$ is passed as that argument.

Note that the calls to~$\APPLY$ in~$\FORWARDJ$~\eqref{eq:cps-forward},
$\REVERSEJ$~\eqref{eq:cps-reverse}, $\textsc{primops}$~\eqref{eq:cps-primops},
$\textsc{interrupt}$~\eqref{eq:cps-interrupt}, and
$\textsc{resume}$~\eqref{eq:cps-resume} are the only portals into the CPS
evaluator.
The only additional call to~$\APPLY$ is in the evaluator itself,
clause~\eqref{eq:cps-application} of~$\EVAL$.\@
All of the portals restart the step count at zero.
Except for the call in $\textsc{interrupt}$~\eqref{eq:cps-interrupt}, none
of the portals call the evaluator with a step limit.
In particular, $\textsc{resume}$~\eqref{eq:cps-resume} does not provide a step
limit; other mechanisms detailed below provide for interrupting a resumed
capsule.

\setcounter{equation}{6}
\begin{figure}
  \centering
  \fbox{\hspace*{-11pt}
    \begin{minipage}{\textwidth}
      \par\vspace*{-15pt}
      \begin{subequations}
      \setcounter{equation}{0}
      \begin{align}
        \textsc{primops}\;f\;x&=\APPLY\;(\lambda n\;l\;v\LD{}n)\;0\;\infty\;f\;x
        \label{eq:cps-primops}\\
        \textsc{interrupt}\;f\;x\;l&=\APPLY\;(\lambda n\;l\;v\LD{}v)\;0\;l\;f\;x
        \label{eq:cps-interrupt}\\
        \textsc{resume}\;\CAPSULE{k}{f}&=\APPLY\;k\;0\;\infty\;f\;\bot
        \label{eq:cps-resume}
      \end{align}
      \end{subequations}
    \end{minipage}}
  \caption{Implementation of the general-purpose interruption and resumption
    interface using the CPS evaluator.}
  \label{fig:interface-implementation}
\end{figure}

This implementation of the general-purpose interruption and resumption
interface cannot be used to fully implement $\CHECKPOINTREVERSEJ$ in the host
as depicted in Fig.~\ref{fig:implementation}.
The reason is that the calls to $\REVERSEJ$ in the base case, step~(0), and
$\textsc{interrupt}$ in step~(2), must take a target function (closure)
for~$f$, because that is what is invoked by the calls to~$\APPLY$ in
$\REVERSEJ$~\eqref{eq:cps-reverse} and
$\textsc{interrupt}$~\eqref{eq:cps-interrupt}.\@
As written in Fig.~\ref{fig:implementation}, the recursive calls to
$\CHECKPOINTREVERSEJ$, namely steps~(3) and~(4), pass
$(\lambda z\LD{}\textsc{resume}\;z)$ and
$(\lambda x\LD{}\textsc{interrupt}\;f\;x\;\lfloor\frac{l}{2}\rfloor)$ for~$f$.
There are two problems with this.
First, these are host closures produced by host lambda expressions, not target
closures.
Second, these call the host functions $\textsc{resume}$ and
$\textsc{interrupt}$ that are not available in the target.
Thus it is not possible to formulate these as target closures without
additional machinery.

Examination of Fig.~\ref{fig:implementation} reveals that the general-purpose
interruption and resumption interface is invoked four times in the
implementation of $\CHECKPOINTREVERSEJ$.
$\textsc{primops}$ is invoked in step~(1), $\textsc{interrupt}$ is invoked in
steps~(2) and~(4), and $\textsc{resume}$ is invoked in step~(3).
Of these, $\textsc{primops}$ is invoked only in the host, $\textsc{resume}$ is
invoked only in the target, and $\textsc{interrupt}$ is invoked in both the
host and the target.
Thus we need to expose $\textsc{interrupt}$ and $\textsc{resume}$ to the
target.
We do not need to expose $\textsc{primops}$ to the target; the implementation in
Fig.~\ref{fig:implementation} only uses it in the host.
For $\textsc{interrupt}$, the call in step~(2) can use the host
implementation~\eqref{eq:cps-interrupt} in
Fig.~\ref{fig:interface-implementation} but the call in step~(4) must use a new
variant exposed to the target.
For $\textsc{resume}$, the call in step~(3) must also use a new variant exposed
to the target.
The host implementation~\eqref{eq:cps-resume} in
Fig.~\ref{fig:interface-implementation} is never used since $\textsc{resume}$
is never invoked in the host.

We expose $\textsc{interrupt}$ and $\textsc{resume}$ to the target by adding
them to the target language as new constructs.
\setcounter{equation}{7}
\begin{equation}
  e::= \textbf{interrupt}\;e_1\;e_2\;e_3\mid
       \textbf{resume}\;e
  \label{eq:interface}
\end{equation}
We implement this functionality by augmenting the CPS evaluator with new
clauses for $\EVAL$ (Fig.~\ref{fig:cps-interface}),
clause~\eqref{eq:cps-interrupt-expression} for $\textbf{interrupt}$ and
clause~\eqref{eq:cps-resume-expression} for $\textbf{resume}$.
We discuss the implementation of these below.
But we first address several other issues.

\setcounter{equation}{5}
\begin{figure}
  %\needswork: v_3 needs primal*
  \centering
  \fbox{\hspace*{-11pt}
    \begin{minipage}{\textwidth}
      \par\vspace*{-15pt}
      \begin{subequations}
      \setcounter{equation}{13}
      \begin{align}
        \mathcal{I}\;f\;l&=
        \CLOSURE{\lambda x\LD{}(\textbf{interrupt}\;f\;x\;l)}
                {\rho_0[f\mapsto f][l\mapsto l]}
        \label{eq:I}\\
        \mathcal{R}&=\CLOSURE{\lambda z\LD{}(\textbf{resume}\;z)}{\rho_0}
        \label{eq:R}\\
        \EVAL\;k\;n\;l\;\rho\;(\textbf{interrupt}\;e_1\;e_2\;e_3)&=
        \EVAL\;
        \begin{array}[t]{@{}l@{}}
        (\begin{array}[t]{@{}l@{}}
        \lambda n\;l\;v_1\LD{}\\
        (\EVAL\;
        \begin{array}[t]{@{}l@{}}
        (\begin{array}[t]{@{}l@{}}
        \lambda n\;l\;v_2\LD{}\\
        (\EVAL\;
        \begin{array}[t]{@{}l@{}}
        (\begin{array}[t]{@{}l@{}}
        \lambda n\;l\;v_3\LD{}\\
        \IF\;l=\infty\\
        \THEN(\APPLY\;k\;0\;v_3\;v_1\;v_2)\\
        \ELSE\begin{array}[t]{@{}l@{}}
        \LET\CAPSULE{k}{f}=(\APPLY\;k\;0\;l\;v_1\;v_2)\\
        \IN\CAPSULE{k}{(\mathcal{I}\;f\;(v_3-l))})
        \end{array}
        \end{array}\\
        n\;l\;\rho\;e_3))
        \end{array}
        \end{array}\\
        n\;l\;\rho\;e_2))
        \end{array}
        \end{array}\\
        (n+1)\;l\;\rho\;e_1
        \end{array}\label{eq:cps-interrupt-expression}\\
        \EVAL\;k\;n\;l\;\rho\;(\textbf{resume}\;e)&=
        \EVAL\;
        (\lambda n\;l\;\CAPSULE{k'}{f}\LD{}
        (\APPLY\;k'\;0\;l\;f\;\bot))\;
        (n+1)\;l\;\rho\;e
        \label{eq:cps-resume-expression}
      \end{align}
      \end{subequations}
    \end{minipage}}
  \caption{Additions to the CPS evaluator for \checkpointVLAD\ to expose the
    general-purpose interruption and resumption interface to the target.}
  \label{fig:cps-interface}
\end{figure}

With appropriate implementations of $\textbf{interrupt}$ and
$\textbf{resume}$ expressions in the target language, one can create target
closures for the expressions
$(\lambda z\LD{}\textsc{resume}\;z)$ and
$(\lambda x\LD{}\textsc{interrupt}\;f\;x\;\lfloor\frac{l}{2}\rfloor)$, and use
these to formulate a proper implementation of~$\CHECKPOINTREVERSEJ$ in the host.
We formulate a target closure to correspond to
$(\lambda z\LD{}\textsc{resume}\;z)$ and denote this as~$\mathcal{R}$.
The definition is given in~\eqref{eq:R} in Fig.~\ref{fig:cps-interface}.
Note that since $(\lambda z\LD{}\textsc{resume}\;z)$ does not contain any free
variables, the closure created by~$\mathcal{R}$ is constructed from the empty
environment~$\rho_0$.
Thus there is a single constant~$\mathcal{R}$.
We similarly formulate a target closure to correspond to
$(\lambda x\LD{}\textsc{interrupt}\;f\;x\;l)$ and denote this as~$\mathcal{I}$.
The definition is given in~\eqref{eq:I} in Fig.~\ref{fig:cps-interface}.
Here, however, $(\lambda x\LD{}\textsc{interrupt}\;f\;x\;l)$ contains two free
variables: $f$ and $l$.
Thus the closure created by~$\mathcal{I}$ contains a nonempty environment with
values for these two variables.
To provide these values, $\mathcal{I}$ is formulated as a function that takes
these values as arguments.

With $(\mathcal{I}\;f\;l)$ and~$\mathcal{R}$, it is now possible to reformulate
the definition of~$\CHECKPOINTREVERSEJ$ in the host from
Fig.~\ref{fig:implementation}, replacing the host closure
$(\lambda z\LD{}\textsc{resume}\;z)$ in step~(3) with the target
closure~$\mathcal{R}$ and the host closure
$(\lambda x\LD{}\textsc{interrupt}\;f\;x\;\lfloor\frac{l}{2}\rfloor)$
in step~(4)
with the target closure~$(\mathcal{I}\;f\;\lfloor\frac{l}{2}\rfloor)$.
This new, proper, definition of~$\CHECKPOINTREVERSEJ$ in the host is given in
Fig.~\ref{fig:proper-implementation}.

\begin{figure}
  \centering
  \fbox{
    \begin{tabular}{lll@{\hspace{2em}}l@{}}
      \multicolumn{4}{l}{To compute $(y,\grave{x})=
        \CHECKPOINTREVERSEJ\;f\;x\;\grave{y}$:}\\
      & \textbf{base case:}
      & $(y,\grave{x})=\REVERSEJ\;f\;x\;\grave{y}$ & (step~0)\\[1.5ex]
      & \textbf{inductive case:}
      & $l=\textsc{primops}\;f\;x$ & (step~1)\\[1ex]
      && $z=\textsc{interrupt}\;f\;x\;\lfloor\frac{l}{2}\rfloor$ & (step~2)\\
      && $(y,\grave{z})=\CHECKPOINTREVERSEJ\;\mathcal{R}\;z\;\grave{y}$ &
      (step~3)\\
      && $(z,\grave{x})=
      \CHECKPOINTREVERSEJ\;
      (\mathcal{I}\;f\;\lfloor\frac{l}{2}\rfloor)
      \;x\;\grave{z}$ & (step~4)
  \end{tabular}}
  \caption{Binary bisection checkpointing in the CPS evaluator.
    This is a proper implementation of the algorithm in
    Fig.~\ref{fig:implementation} where the host closure
    $(\lambda z\LD{}\textsc{resume}\;z)$ in step~(3) is replaced with the target
    closure~$\mathcal{R}$ and the host closure
    $(\lambda x\LD{}\textsc{interrupt}\;f\;x\;\lfloor\frac{l}{2}\rfloor)$ in
    step~(4) is replaced with the target
    closure~$(\mathcal{I}\;f\;\lfloor\frac{l}{2}\rfloor)$.}
  \label{fig:proper-implementation}
\end{figure}

In this proper implementation of~$\CHECKPOINTREVERSEJ$ in the host, the
$\textbf{interrupt}$ and $\textbf{resume}$ operations need to be able to nest,
even without nesting of calls to~$\CHECKPOINTREVERSEJ$ in the target.
The recursive calls to~$\CHECKPOINTREVERSEJ$ in the inductive case of
Fig.~\ref{fig:proper-implementation} imply that it must be possible to
interrupt a resumed capsule.
This happens when passing~$\mathcal{R}$ for~$f$ in step~(3) and then passing
$(\mathcal{I}\;f\;\ldots)$ for~$f$ in step~(4), \ie\ the left branch of a right
branch in the checkpoint tree.
The resulting function $f=(\mathcal{I}\;\mathcal{R}\;\ldots)$ will interrupt
when applied to some capsule.
It also happens when passing $(\mathcal{I}\;f\;\ldots)$ for~$f$ twice in
succession in step~(4), \ie\ the left branch of a left branch in the checkpoint
tree.
The resulting function $f=(\mathcal{I}\;(\mathcal{I}\;f\;\ldots)\;\ldots)$ will
interrupt and the capsule produced will interrupt when resumed.

Consider all the ways that evaluations of $\textbf{interrupt}$ and
$\textbf{resume}$ expressions can nest.
User code will never contain $\textbf{interrupt}$ and
$\textbf{resume}$ expressions; they are created only by invocations
of~$\mathcal{I}$ and~$\mathcal{R}$.\@
$\mathcal{R}$ is only invoked by step~(3) of~$\CHECKPOINTREVERSEJ$ in
Fig.~\ref{fig:proper-implementation}.
$\mathcal{I}$ is invoked two ways: step~(4) of~$\CHECKPOINTREVERSEJ$ in
Fig.~\ref{fig:proper-implementation} and a way that we have not yet
encountered, evaluation of nested $\textbf{interrupt}$ expressions in
the $\textbf{else}$ branch of clause~\eqref{eq:cps-interrupt-expression} in
Fig.~\ref{fig:cps-interface}.

Consider all the ways that evaluations of~$\mathcal{I}$ and~$\mathcal{R}$ can
be invoked in~$\CHECKPOINTREVERSEJ$ in Fig.~\ref{fig:proper-implementation}.
$\CHECKPOINTREVERSEJ$~is invoked with some user code for~$f$, \ie\ code that
does not contain $\textbf{interrupt}$ and $\textbf{resume}$ expressions.
The inductive cases for~$\CHECKPOINTREVERSEJ$ create a binary checkpoint tree of
invocations.
The leaf nodes of this binary checkpoint tree correspond to the base case in
step~(0) where the host~$\REVERSEJ$ is invoked.
At internal nodes, the host $\textsc{interrupt}$ is invoked in step~(2).
The target closure values that can be passed to the host~$\REVERSEJ$ and
$\textsc{interrupt}$ are constructed from~$f$, $\mathcal{I}$, and $\mathcal{R}$
in steps~(3) and~(4).
What is the space of all possible constructed target closures?
The constructed target closures invoked along the left spine of the binary
checkpoint tree look like the following
\setcounter{equation}{8}
\begin{equation}
  (\mathcal{I}\;(\mathcal{I}\;\ldots\;(\mathcal{I}\;(\mathcal{I}\;f\;l_0)\;
  l_1)\;\ldots\;l_{i-1})\;l_i)
  \label{eq:case-one}
\end{equation}
with zero or more nested calls to~$\mathcal{I}$.
In this case $l_i<l_{i-1}<\dotsb<l_1<l_0$, because the recursive calls
to~$\CHECKPOINTREVERSEJ$ in step~(4) of Fig.~\ref{fig:proper-implementation}
always reduce~$l$.
The constructed target closures invoked in any other node in the binary
checkpoint tree look like the following
\begin{equation}
  (\mathcal{I}\;(\mathcal{I}\;\ldots\;(\mathcal{I}\;(\mathcal{I}\;\mathcal{R}
  \;l_0)\;l_1)\;\ldots\;l_{i-1})\;l_i)
  \label{eq:case-two}
\end{equation}
with zero or more nested calls to~$\mathcal{I}$.
In this case, again, $l_i<l_{i-1}<\dotsb<l_1<l_0$, for the same reason.
These are the possible target closures~$f$ passed to~$\REVERSEJ$ in step~(0)
or $\textsc{interrupt}$ in step~(2) of~$\CHECKPOINTREVERSEJ$ in
Fig.~\ref{fig:proper-implementation}.
(We assume that the call to $\textsc{primops}$ in step~(1) is hoisted out of
the recursion.)

A string of calls to~$\mathcal{I}$ as in~\eqref{eq:case-one} will result in
a nested closure structure whose invocation will lead to nested invocations of
$\textbf{interrupt}$ expressions.
\begin{equation}
  \langle
  \begin{array}[t]{@{}l@{}}
  (\lambda x\LD{}(\textbf{interrupt}\;f\;x\;l)),\\
  \rho_0
  \begin{array}[t]{@{}l@{}}
  \lbrack f\mapsto
  \langle
  \begin{array}[t]{@{}l@{}}
  (\lambda x\LD{}(\textbf{interrupt}\;f\;x\;l)),\\
  \rho_0
  \begin{array}[t]{@{}l@{}}
  \lbrack f\mapsto
  \langle
  \begin{array}[t]{@{}l@{}}
  \ldots\\
  (\lambda x\LD{}(\textbf{interrupt}\;f\;x\;l)),\\
  \rho_0
  \begin{array}[t]{@{}l@{}}
  \lbrack f\mapsto
  \langle
  \begin{array}[t]{@{}l@{}}
  (\lambda x\LD{}(\textbf{interrupt}\;f\;x\;l)),\\
  \rho_0
  \begin{array}[t]{@{}l@{}}
  \lbrack f\mapsto f\rbrack
  \lbrack l\mapsto l_0\rbrack\rangle\rbrack
  \lbrack l\mapsto l_1\rbrack\ldots\rangle\rbrack
  \lbrack l\mapsto l_{i-1}\rbrack\rangle\rbrack
  \lbrack l\mapsto l_i\rbrack\rangle
  \end{array}
  \end{array}
  \end{array}
  \end{array}
  \end{array}
  \end{array}
  \end{array}
  \end{array}
  \label{eq:preamble-one}
\end{equation}
A string of calls to~$\mathcal{I}$ as in~\eqref{eq:case-two} will also result in
a nested closure structure whose invocation will lead to nested invocations of
$\textbf{interrupt}$ expressions.
\begin{equation}
  \begin{small}
  \langle
  \begin{array}[t]{@{}l@{}}
  (\lambda x\LD{}(\textbf{interrupt}\;f\;x\;l)),\\
  \rho_0
  \begin{array}[t]{@{}l@{}}
  \lbrack f\mapsto
  \langle
  \begin{array}[t]{@{}l@{}}
  (\lambda x\LD{}(\textbf{interrupt}\;f\;x\;l)),\\
  \rho_0
  \begin{array}[t]{@{}l@{}}
  \lbrack f\mapsto
  \langle
  \begin{array}[t]{@{}l@{}}
  \ldots\\
  (\lambda x\LD{}(\textbf{interrupt}\;f\;x\;l)),\\
  \rho_0
  \begin{array}[t]{@{}l@{}}
  \lbrack f\mapsto
  \langle
  \begin{array}[t]{@{}l@{}}
  (\lambda x\LD{}(\textbf{interrupt}\;f\;x\;l)),\\
  \rho_0
  \begin{array}[t]{@{}l@{}}
  \lbrack f\mapsto\CLOSURE{\lambda z\LD{}(\textbf{resume}\;z)}{\rho_0}\rbrack\\
  \lbrack l\mapsto l_0\rbrack\rangle\rbrack
  \lbrack l\mapsto l_1\rbrack\ldots\rangle\rbrack
  \lbrack l\mapsto l_{i-1}\rbrack\rangle\rbrack
  \lbrack l\mapsto l_i\rbrack\rangle
  \end{array}
  \end{array}
  \end{array}
  \end{array}
  \end{array}
  \end{array}
  \end{array}
  \end{array}
  \end{small}
  \label{eq:preamble-two}
\end{equation}
In both of these, $l_i<l_{i-1}<\dotsb<l_1<l_0$, so the outermost
$\textbf{interrupt}$ expression will interrupt first.
Since the CPS evaluator only maintains a single step limit, $l_i$~will be that
step limit during the execution of the innermost content of these nested
closures, namely~$f$ in~\eqref{eq:preamble-one} and
$\CLOSURE{\lambda z\LD{}(\textbf{resume}\;z)}{\rho_0}$
in~\eqref{eq:preamble-two}.
None of the other intervening $\textbf{interrupt}$ expressions will enforce
their step limits during this execution.
Thus we need to arrange for the capsule created when the step limit~$l_i$
is reached during the execution of~$f$ or
$\CLOSURE{\lambda z\LD{}(\textbf{resume}\;z)}{\rho_0}$ to itself interrupt with
the remaining step limits $l_{i-1},\dotsc,l_1,l_0$.
This is done by rewrapping the closure in a capsule with $\textbf{interrupt}$
expressions.
The interruption of~$f$ or
$\CLOSURE{\lambda z\LD{}(\textbf{resume}\;z)}{\rho_0}$
will produce a capsule that looks like the following
\begin{equation}
  \CAPSULE{k}{f}
  \label{eq:preamble-three}
\end{equation}
where the closure~$f$ contains only user code, \ie\ no $\textbf{interrupt}$ or
$\textbf{resume}$ expressions.
The~$f$ in~\eqref{eq:preamble-three} is wrapped with calls to~$\mathcal{I}$ to
reintroduce the step limits $l_{i-1},\dotsc,l_1,l_0$.
\begin{equation}
  \CAPSULE{k}{(\mathcal{I}\;\ldots\;(\mathcal{I}\;(\mathcal{I}\;f\;
  l_0)\;l_1)\;\ldots\;l_{i-1})}
  \label{eq:preamble-four}
\end{equation}
This will yield a capsule that looks like the following
\begin{equation}
  \lsem k,\langle
  \begin{array}[t]{@{}l@{}}
  (\lambda x\LD{}(\textbf{interrupt}\;f\;x\;l)),\\
  \rho_0
  \begin{array}[t]{@{}l@{}}
  \lbrack f\mapsto
  \langle
  \begin{array}[t]{@{}l@{}}
  \ldots\\
  (\lambda x\LD{}(\textbf{interrupt}\;f\;x\;l)),\\
  \rho_0
  \begin{array}[t]{@{}l@{}}
  \lbrack f\mapsto
  \langle
  \begin{array}[t]{@{}l@{}}
  (\lambda x\LD{}(\textbf{interrupt}\;f\;x\;l)),\\
  \rho_0
  \begin{array}[t]{@{}l@{}}
  \lbrack f\mapsto f\rbrack
  \lbrack l\mapsto l_0\rbrack\rangle\rbrack
  \lbrack l\mapsto l_1\rbrack\ldots\rangle\rbrack
  \lbrack l\mapsto l_{i-1}\rbrack\rangle\rsem
  \end{array}
  \end{array}
  \end{array}
  \end{array}
  \end{array}
  \end{array}
  \label{eq:preamble-five}
\end{equation}
which will interrupt upon resumption.
Each such interruption will peel off one $\textbf{interrupt}$ expression.
Note that since the closure~$f$ in a capsule~\eqref{eq:preamble-three}
contains only user code, it will not contain a $\textbf{resume}$ expression.
Further, since the wrapping process~\eqref{eq:preamble-five} only introduces
$\textbf{interrupt}$ expressions via calls
to~$\mathcal{I}$~\eqref{eq:preamble-four}, and never introduces
$\textbf{resume}$ expressions, the closures in capsules, whether wrapped or
not, will never contain $\textbf{resume}$ expressions.

When there is no contextual step limit, \ie\ when $l=\infty$, the
$\textbf{interrupt}$ expression must introduce~$v_3$, the step limit specified
as the argument to the $\textbf{interrupt}$ expression, as the step limit.
This is handled by the $\textbf{then}$ branch of
clause~\eqref{eq:cps-interrupt-expression} in Fig.~\ref{fig:cps-interface}.
When there is a contextual step limit, \ie\ when $l\not=\infty$, the
$\textbf{interrupt}$ expression must wrap the returned capsule.
This wrapping is handled by the $\textbf{else}$ branch of
clause~\eqref{eq:cps-interrupt-expression} in Fig.~\ref{fig:cps-interface}.
Since capsule resumption restarts the step count at zero, the wrapping that
handles nested step limits is relativized to this restart by the $v_3-l$ in the
$\textbf{else}$ branch in clause~\eqref{eq:cps-interrupt-expression}.

Capsule resumption happens in one place, the call to~$\APPLY$ in
clause~\eqref{eq:cps-resume-expression} in Fig.~\ref{fig:cps-interface} for
a $\textbf{resume}$ expression.
Except for the contextual step limit~$l$, this is the same as the call
to~$\APPLY$ in the implementation of $\textsc{resume}$ in~\eqref{eq:cps-resume}
in Fig.~\ref{fig:interface-implementation}.
Said resumption is performed by applying the capsule closure~$f$, a target
closure, to~$\bot$, since the lambda expression in the capsule closure
ignores its argument.
This call to~$\APPLY$ is passed the capsule continuation~$k'$ as its
continuation.
Unlike the implementation of $\textsc{resume}$ in~\eqref{eq:cps-resume}, the
step limit~$l$ is that which is in effect for the execution of the
$\textbf{resume}$ expression.
This is to allow capsule resumption to itself interrupt.
Because capsules are resumed with a step count of zero and the step limit at
the time of resumption, the step count and limit at the time of the
interruption need not be saved in the capsule.

As a result of this, all $\textbf{interrupt}$ expressions will appear in one
of two places.
The first is a preamble~\eqref{eq:preamble-one} or~\eqref{eq:preamble-two}
wrapped around either a user function~$f$ by~\eqref{eq:case-one} or a
$\textbf{resume}$ expression in~$\mathcal{R}$ by~\eqref{eq:case-two},
respectively.
This will always be invoked either by~$\REVERSEJ$ in the base case, step~(0),
or by $\textsc{interrupt}$ in step~(2), of
Fig.~\ref{fig:proper-implementation}.
The second is a preamble~\eqref{eq:preamble-five} wrapped around the closure of
a capsule by the $\textbf{else}$ branch in
clause~\eqref{eq:cps-interrupt-expression} of
Fig.~\ref{fig:cps-interface}, \ie~\eqref{eq:preamble-four}.
This will always be invoked during capsule resumption,
\ie\ clause~\eqref{eq:cps-resume-expression} of
Fig.~\ref{fig:cps-interface}.
We assume that the step limits are such that an interruption never occurs
during either of these preambles.
This is enforced by ensuring that the termination criterion that triggers the
base case, step~(0), of Fig.~\ref{fig:proper-implementation} is sufficiently
long so that the calls to~$\APPLY$ in~$\REVERSEJ$ in step~(0) and
$\textsc{interrupt}$ in step~(2) won't interrupt before completion of the
preamble.

There is one further requirement to allow the CPS evaluator to support
divide-and-conquer checkpointing.
The base case use of~$\REVERSEJ$ in step~(0) of
Fig.~\ref{fig:proper-implementation} needs to be able to produce
cotangents~$\grave{z}$ of capsules~$z$ in step~(3) and consume them in
step~(4).
A capsule $\CAPSULE{k}{f}$ is the saved state of the evaluator.
The value~$f$ is a target closure $\CLOSURE{\lambda x\LD{}e}{\rho}$ which
contains an environment with saved state.
This state is visible to~$\REVERSEJ$.\@
But the continuation~$k$ is a host continuation, which is opaque.
Any evaluator variables that it closes over are not visible to~$\REVERSEJ$.\@
Thus the implementation of host continuations in the CPS evaluator must employ
a mechanism to expose them.
When we replace the CPS evaluator with a direct-style evaluator applied to
CPS-converted target code, in Sections~\ref{sec:CPS-conversion}
and~\ref{sec:augmenting} below, this will no-longer be necessary since
continuations will be represented as target closures which are visible
to~$\REVERSEJ$.\@

\subsection{Augmenting the CPS Evaluator to Support Divide-and-Conquer
  Checkpointing}

We can now add the~$\CHECKPOINTREVERSEJ$ operator to the target language as a
new construct.
\begin{equation}
  e::= \CHECKPOINTREVERSEJ\;e_1\;e_2\;e_3
  \label{eq:checkpointreversej}
\end{equation}
We implement this functionality by augmenting the CPS evaluator with a new
clause~\eqref{eq:cps-checkpoint-starj} for $\EVAL$
(Fig.~\ref{fig:checkpointreversej}).

\setcounter{equation}{5}
\begin{figure}
  \centering
  \fbox{\hspace*{-11pt}
    \begin{minipage}{\textwidth}
      \par\vspace*{-15pt}
      \begin{subequations}
      \setcounter{equation}{17}
      \begin{align}
        \EVAL\;k\;n\;l\;\rho\;(\CHECKPOINTREVERSEJ\;e_1\;e_2\;e_3)&=
        \EVAL\;
        \begin{array}[t]{@{}l@{}}
        (\begin{array}[t]{@{}l@{}}
        \lambda n\;l\;v_1\LD{}\\
        (\EVAL\;
        \begin{array}[t]{@{}l@{}}
        (\begin{array}[t]{@{}l@{}}
        \lambda n\;l\;v_2\LD{}\\
        (\EVAL\;
        \begin{array}[t]{@{}l@{}}
        (\begin{array}[t]{@{}l@{}}
        \lambda n\;l\;v_3\LD{}\\
        (k\;n\;l\;(\CHECKPOINTREVERSEJ\;v_1\;v_2\;v_3)))
        \end{array}\\
        n\;l\;\rho\;e_3))
        \end{array}
        \end{array}\\
        n\;l\;\rho\;e_2))
        \end{array}
        \end{array}\\
        (n+1)\;l\;\rho\;e_1
        \end{array}\label{eq:cps-checkpoint-starj}
      \end{align}
      \end{subequations}
    \end{minipage}}
  \caption{Addition to the CPS evaluator for \checkpointVLAD\ to support
    divide-and-conquer checkpointing.}
  \label{fig:checkpointreversej}
\end{figure}

With this addition, target programs can perform divide-and-conquer checkpointing
simply by calling $\CHECKPOINTREVERSEJ$ instead of $\REVERSEJ$.
Note that it is not possible to add the~$\CHECKPOINTREVERSEJ$ operator to the
direct-style evaluator because the implementation of binary bisection
checkpointing is built on the general-purpose interruption and resumption
interface which is, in turn, built on the CPS evaluator.
We remove this limitation below in Section~\ref{sec:augmenting}.
Also note that since the implementation of binary bisection checkpointing is
built on the general-purpose interruption and resumption interface which is, in
turn, built on an evaluator, it is only available for programs that are
evaluated, \ie\ for programs in the target, but not for programs in the host.
We remove this limitation below as well, in Section~\ref{sec:compiling}.

\subsection{Some Intuition}
\label{sec:intuition}

The algorithm in Fig.~\ref{fig:proper-implementation} corresponds to
Fig.~\ref{fig:divide-and-conquer-checkpointing}(b).
The start of the computation of~$f$ in Fig.~\ref{fig:proper-implementation}
corresponds to~$u$ in Fig.~\ref{fig:divide-and-conquer-checkpointing}(b).
The computation state at~$u$ is~$x$ in Fig.~\ref{fig:proper-implementation}.
Collectively, the combination of~$f$ and~$x$ in
Fig.~\ref{fig:proper-implementation} comprises a snapshot, the gold
line in Fig.~\ref{fig:divide-and-conquer-checkpointing}(b).
The end of the computation of~$f$ in Fig.~\ref{fig:proper-implementation}
corresponds to~$v$ in Fig.~\ref{fig:divide-and-conquer-checkpointing}(b).
The computation state at~$v$ is~$y$ in Fig.~\ref{fig:proper-implementation}.
Step~(1) computes~$\lfloor\frac{l}{2}\rfloor$ which corresponds to the split
point~$p$ in Fig.~\ref{fig:divide-and-conquer-checkpointing}(b).
Step~(2) corresponds to the green line in
Fig.~\ref{fig:divide-and-conquer-checkpointing}(b), \ie\ running the primal
without taping from the snapshot~$f$ and~$x$ at~$u$ until the split point~$p$
which is~$\lfloor\frac{l}{2}\rfloor$.
The capsule~$z$ in Fig.~\ref{fig:proper-implementation} corresponds to the
computation state at~$p$ in Fig.~\ref{fig:divide-and-conquer-checkpointing}(b).
Brown and pink lines in Fig.~\ref{fig:divide-and-conquer-checkpointing} denote
capsules.
If step~(3) would incur the base case, step~(0), in the recursive call, it
would correspond to the right stage (pair of red and blue lines) in
Fig.~\ref{fig:divide-and-conquer-checkpointing}(b).
If step~(4) would incur the base case, step~(0), in the recursive call, it
would correspond to the left stage (pair of red and blue lines) in
Fig.~\ref{fig:divide-and-conquer-checkpointing}(b).
Note that~$f$ and~$x$ is used both in steps~(2) and~(4).
Referring to this as a snapshot is meant to convey that the information must be
saved across the execution of step~(3).
And it must be possible to apply~$f$ to~$x$ twice, once in step~(2) and once in
step~(4).
In some implementations, such a snapshot involves saving mutable state that
must be restored.
In our formulation in a functional framework (Section~\ref{sec:advantages}), we
need not explicitly save and restore state; we simply apply a function twice.
Nonetheless, the storage required for the snapshot is implicit in the extended
lifetime of the values~$f$ and~$x$ which extends from the entry into
$\CHECKPOINTREVERSEJ$, over step~(3), until step~(4).

Note that recursive calls to~$\CHECKPOINTREVERSEJ$ in step~(4) extend the
lifetime of a snapshot.
These are denoted as the black tick marks on the left of the gold and pink
lines.
In the treeverse algorithm from \cite[Figs.~2 and~3]{griewank1992alg}, the
lifetime of one snapshot ends at a tick mark by a call to \textbf{retrieve} in
one recursive call to \textbf{treeverse} in the \textbf{while} loop of the
parent and the lifetime of a new snapshot begins by a call to \textbf{snapshot}
in the next recursive call to \textbf{treeverse} in the \textbf{while} loop of
the parent.
But since the state retrieved and then immediately saved again as a new
snapshot is the same, these adjacent snapshot execution intervals can
conceptually be merged.

Also note that recursive calls to~$\CHECKPOINTREVERSEJ$ in step~(3)
pass~$\mathcal{R}$ and a capsule~$z$ as the~$f$ and~$x$ of the recursive call.
Thus capsules from one level of the recursion become snapshots at the next
level, for all but the base case step~(0).
Pink lines in Fig.~\ref{fig:divide-and-conquer-checkpointing} denote values
that are capsules at one level but snapshots at lower levels.
Some, but not all, capsules are snapshots.
Some, but not all, snapshots are capsules.
Gold lines in Fig.~\ref{fig:divide-and-conquer-checkpointing} denote snapshots
that are not capsules.
Brown lines in Fig.~\ref{fig:divide-and-conquer-checkpointing} denote capsules
that are not snapshots.
Pink lines in Fig.~\ref{fig:divide-and-conquer-checkpointing} denote values
that are both snapshots and capsules.

\setcounter{figure}{20}
\begin{wrapfigure}[3]{r}{0.27\columnwidth}
\vspace*{-27pt}
\centering
\leaf{\rule[-6pt]{0pt}{18pt}$(z,\grave{x})=(\CHECKPOINTREVERSEJ\;(\mathcal{I}\;f\;\lfloor\frac{l}{2}\rfloor)\;x\;\grave{z})$}
\leaf{\rule[-6pt]{0pt}{18pt}$(y,\grave{z})=(\CHECKPOINTREVERSEJ\;\mathcal{R}\;z\;\grave{y})$}
\branch{2}{\rule[-6pt]{0pt}{18pt}$z=(\textsc{interrupt}\;f\;x\;\lfloor\frac{l}{2}\rfloor)$}
\resizebox{0.27\columnwidth}{!}{\qobitree}
\vspace*{-10pt}
\caption{}
\label{fig:tree6}
\end{wrapfigure}
It is now easy to see that the recursive call tree of the algorithm in
Fig.~\ref{fig:proper-implementation} is isomorphic to a binary checkpoint tree.
The binary checkpoint tree in Fig.~\ref{fig:tree1} corresponds to the call tree
in Fig.~\ref{fig:tree6} produced by the algorithm in
Fig.~\ref{fig:proper-implementation}.
This depicts just one level of the recursion.
If one unrolls this call tree, one obtains a binary checkpoint tree.

\subsection{CPS Conversion}
\label{sec:CPS-conversion}

So far, we have formulated divide-and-conquer checkpointing via a CPS evaluator.
This can be---and has been---used to construct an interpreter.
A compiler can be---and has been---constructed by generating target code in CPS
that is instrumented with step counting, step limits, and limit checks that
lead to interrupts.
Code in direct style can be automatically converted to CPS using a program
transformation known in the programming language community as
\defoccur{CPS conversion}.
Many existing compilers, such as \SMLNJ\ for \SML, perform CPS conversion as
part of the compilation process \citep{appel2006compiling}.

We illustrate CPS conversion for the untyped lambda calculus
(Fig.~\ref{fig:cps-conversion}).
\setcounter{equation}{17}
\begin{equation}
  e::= x\mid
       \lambda x\LD{}e\mid
       e_1\;e_2
  \label{eq:lambda}
\end{equation}
The notation $\ullcorner e|k\ulrcorner$ denotes the transformation of the
expression~$e$ to CPS so that it calls the continuation~$k$ with the result.
There is a clause for $\ullcorner e|k\ulrcorner$ in
Fig.~\ref{fig:cps-conversion},
\crefrange{eq:convert-variable}{eq:convert-application}, for each construct
in~\eqref{eq:lambda}.
Clause~\eqref{eq:convert-variable} says that one converts a variable~$x$ by
calling the continuation~$k$ with the value of that variable.
Clause~\eqref{eq:convert-lambda} says that one converts a lambda expression
$(\lambda x\LD{}e)$ by adding a continuation variable~$k'$ to the lambda binder,
converting the body relative to that variable, and then calling the
continuation~$k$ with that lambda expression.
Clause~\eqref{eq:convert-application} says that one converts an
application $(e_1\;e_2)$ by converting~$e_1$ with a continuation that receives
the value~$x_1$ of~$e_1$, then converts~$e_2$ with a continuation that receives
the value~$x_2$ of~$e_2$, and then calls~$x_1$ with the continuation~$k$
and~$x_2$.
Clause~\eqref{eq:convert-top} says that the top level expression~$e_0$ can be
converted with the identity function as the continuation.

\setcounter{figure}{19}
\setcounter{equation}{16}
\begin{figure}
  \centering
  \fbox{\hspace*{-11pt}
    \begin{minipage}{\textwidth}
      \par\vspace*{-15pt}
      \begin{subequations}
      \begin{align}
        \ullcorner x|k\ulrcorner&\leadsto k\;x\label{eq:convert-variable}\\
        \ullcorner(\lambda x\LD{}e)|k\ulrcorner&
        \leadsto k\;(\lambda k'\;x\LD{}\ullcorner e|k'\ulrcorner)
        \label{eq:convert-lambda}\\
        \ullcorner(e_1\;e_2)|k\ulrcorner&\leadsto{}
        \ullcorner e_1|
        (\lambda x_1\LD{}\ullcorner e_2|(\lambda x_2\LD{}(x_1\;k\;x_2))
        \ulrcorner)\ulrcorner
        \label{eq:convert-application}\\
        e_0&\leadsto{}\ullcorner e_0|(\lambda x\LD{}x)\ulrcorner
        \label{eq:convert-top}
      \end{align}
      \end{subequations}
    \end{minipage}}
  \caption{CPS conversion for the untyped lambda calculus.}
  \label{fig:cps-conversion}
\end{figure}

This technique can be extended to thread a step count~$n$ and a step limit~$l$
through the computation along with the continuation~$k$, and to arrange for the
step count to be incremented appropriately.
Further, this technique can be applied to the entire target language
(Fig.~\ref{fig:threaded-cps-conversion}).
Clauses
\crefrange{eq:threaded-convert-constant}{eq:threaded-convert-checkpoint-starj}
correspond one-to-one to the \checkpointVLAD\ constructs in~\eqref{eq:core},
\eqref{eq:AD}, and~\eqref{eq:checkpointreversej}.
Since CPS conversion is only applied once at the beginning of compilation, to
the user program, and the user program does not contain $\textbf{interrupt}$
and $\textbf{resume}$ expressions, since these only appear internally in the
target closures created by~$\mathcal{I}$ and~$\mathcal{R}$, CPS conversion need
not handle these constructs.
Finally, $\llangle e\rrangle_{k,n,l}$ denotes a limit check that interrupts and
returns a capsule when the step count~$n$ reaches the step limit~$l$.
The implementation of this limit check is given
in~\eqref{eq:threaded-convert-limit-check}.
Each of the clauses
\crefrange{eq:threaded-convert-constant}{eq:threaded-convert-checkpoint-starj}
is wrapped in a limit check.

\setcounter{figure}{21}
\setcounter{equation}{19}
\begin{figure}
  %\needswork: there is a residual white lie here in that we don't factor e in
  %            the limit check
  \centering
  \fbox{\hspace*{-11pt}
    \begin{minipage}{\textwidth}
      \par\vspace*{-15pt}
      \begin{subequations}
      \begin{align}
        \ullcorner c|k,n,l\ulrcorner&\leadsto
        \llangle k\;(n+1)\;l\;c\rrangle_{k,n,l}
        \label{eq:threaded-convert-constant}\\
        \ullcorner x|k,n,l\ulrcorner&\leadsto
        \llangle k\;(n+1)\;l\;x\rrangle_{k,n,l}
        \label{eq:threaded-convert-variable}\\
        \ullcorner(\lambda x\LD{}e)|k,n,l\ulrcorner&\leadsto
        \llangle k\;(n+1)\;l\;
        (\lambda_4 k\;n\;l\;x\LD{}\ullcorner e|k,n,l\ulrcorner)
        \rrangle_{k,n,l}
        \label{eq:threaded-convert-lambda}\\
        \ullcorner(e_1\;e_2)|k,n,l\ulrcorner&\leadsto
        \llangle\ullcorner e_1|
        \begin{array}[t]{@{}l@{}}
        (\lambda_3
        \begin{array}[t]{@{}l@{}}
        n\;l\;x_1\LD{}\\
        \ullcorner e_2|
        \begin{array}[t]{@{}l@{}}
        (\lambda_3
        \begin{array}[t]{@{}l@{}}
        n\;l\;x_2\LD{}\\
        (x_1\;k\;n\;l\;x_2)),
        \end{array}\\
        n,l\ulrcorner),
        \end{array}
        \end{array}\\
        (n+1),l\ulrcorner\rrangle_{k,n,l}
        \end{array}
        \label{eq:threaded-convert-application}\\
        \ullcorner(\IF e_1\;\THEN e_2\;\ELSE e_3)|k,n,l\ulrcorner&\leadsto
        \llangle\ullcorner e_1|
        \begin{array}[t]{@{}l@{}}
        (\lambda_3
        \begin{array}[t]{@{}l@{}}
        n\;l\;x_1\LD{}\\
        \begin{array}[t]{@{}l@{}}
        (
        \begin{array}[t]{@{}l@{}}
        \IF x_1\\
        \THEN\ullcorner e_2|k,n,l\ulrcorner\\
        \ELSE\ullcorner e_3|k,n,l\ulrcorner)),
        \end{array}
        \end{array}
        \end{array}\\
        (n+1),l\ulrcorner\rrangle_{k,n,l}
        \end{array}
        \label{eq:threaded-convert-conditional}\\
        \ullcorner(\diamond e)|k,n,l\ulrcorner&\leadsto
        \llangle\ullcorner e|
        \begin{array}[t]{@{}l@{}}
        (\lambda_3
        \begin{array}[t]{@{}l@{}}
        n\;l\;x\LD{}\\
        (k\;n\;l\;(\diamond x))),
        \end{array}\\
        (n+1),l\ulrcorner\rrangle_{k,n,l}
        \end{array}
        \label{eq:threaded-convert-unary}\\
        \ullcorner(e_1\bullet e_2)|k,n,l\ulrcorner&\leadsto
        \llangle\ullcorner e_1|
        \begin{array}[t]{@{}l@{}}
        (\lambda_3
        \begin{array}[t]{@{}l@{}}
        n\;l\;x_1\LD{}\\
        \ullcorner e_2|
        \begin{array}[t]{@{}l@{}}
        (\lambda_3
        \begin{array}[t]{@{}l@{}}
        n\;l\;x_2\LD{}\\
        (k\;n\;l\;(x_1\bullet x_2))),
        \end{array}\\
        n,l\ulrcorner),
        \end{array}
        \end{array}\\
        (n+1),l\ulrcorner\rrangle_{k,n,l}
        \end{array}
        \label{eq:threaded-convert-binary}\\
        \ullcorner(\FORWARDJ\;e_1\;e_2\;e_3)|k,n,l\ulrcorner&\leadsto
        \llangle\ullcorner e_1|
        \begin{array}[t]{@{}l@{}}
        (\lambda_3
        \begin{array}[t]{@{}l@{}}
        n\;l\;x_1\LD{}\\
        \ullcorner e_2|
        \begin{array}[t]{@{}l@{}}
        (\lambda_3
        \begin{array}[t]{@{}l@{}}
        n\;l\;x_2\LD{}\\
        \ullcorner e_3|
        \begin{array}[t]{@{}l@{}}
        (\lambda_3
        \begin{array}[t]{@{}l@{}}
        n\;l\;x_3\LD{}\\
        (k\;n\;l\;(\FORWARDJ\;x_1\;x_2\;x_3))),
        \end{array}\\
        n,l\ulrcorner),
        \end{array}
        \end{array}\\
        n,l\ulrcorner),
        \end{array}
        \end{array}\\
        (n+1),l\ulrcorner\rrangle_{k,n,l}
        \end{array}
        \label{eq:threaded-convert-jstar}\\
        \ullcorner(\REVERSEJ\;e_1\;e_2\;e_3)|k,n,l\ulrcorner&\leadsto
        \llangle\ullcorner e_1|
        \begin{array}[t]{@{}l@{}}
        (\lambda_3
        \begin{array}[t]{@{}l@{}}
        n\;l\;x_1\LD{}\\
        \ullcorner e_2|
        \begin{array}[t]{@{}l@{}}
        (\lambda_3
        \begin{array}[t]{@{}l@{}}
        n\;l\;x_2\LD{}\\
        \ullcorner e_3|
        \begin{array}[t]{@{}l@{}}
        (\lambda_3
        \begin{array}[t]{@{}l@{}}
        n\;l\;x_3\LD{}\\
        (k\;n\;l\;(\REVERSEJ\;x_1\;x_2\;x_3))),
        \end{array}\\
        n,l\ulrcorner),
        \end{array}
        \end{array}\\
        n,l\ulrcorner),
        \end{array}
        \end{array}\\
        (n+1),l\ulrcorner\rrangle_{k,n,l}
        \end{array}
        \label{eq:threaded-convert-starj}\\
        \ullcorner(\CHECKPOINTREVERSEJ\;e_1\;e_2\;e_3)|k,n,l\ulrcorner&\leadsto
        \llangle\ullcorner e_1|
        \begin{array}[t]{@{}l@{}}
        (\lambda_3
        \begin{array}[t]{@{}l@{}}
        n\;l\;x_1\LD{}\\
        \ullcorner e_2|
        \begin{array}[t]{@{}l@{}}
        (\lambda_3
        \begin{array}[t]{@{}l@{}}
        n\;l\;x_2\LD{}\\
        \ullcorner e_3|
        \begin{array}[t]{@{}l@{}}
        (\lambda_3
        \begin{array}[t]{@{}l@{}}
        n\;l\;x_3\LD{}\\
        (k\;n\;l\;(\CHECKPOINTREVERSEJ\;x_1\;x_2\;x_3))),
        \end{array}\\
        n,l\ulrcorner),
        \end{array}
        \end{array}\\
        n,l\ulrcorner),
        \end{array}
        \end{array}\\
        (n+1),l\ulrcorner\rrangle_{k,n,l}
        \end{array}
        \label{eq:threaded-convert-checkpoint-starj}\\
        \llangle e\rrangle_{k,n,l}&
        \leadsto\IF n=l\;\THEN\CAPSULE{k}{\lambda_4 k\;n\;l\;\_\LD{}e}\;\ELSE e
        \label{eq:threaded-convert-limit-check}
      \end{align}
      \end{subequations}
    \end{minipage}}
  \caption{CPS conversion for the \checkpointVLAD\ language that threads
    step counts and limits.}
  \label{fig:threaded-cps-conversion}
\end{figure}

\subsection{Augmenting the Direct-Style Evaluator to Support CPS-Converted Code
and Divide-and-Conquer Checkpointing}
\label{sec:augmenting}

The direct-style evaluator must be modified in several ways to support
CPS-converted code and divide-and-conquer checkpointing
(Fig.~\ref{fig:extended}).
First, CPS conversion introduced lambda expressions with multiple arguments and
their corresponding applications.
Continuations have three arguments and converted lambda expressions have four.
Thus we add several new constructs into the target language to replace
the single argument lambda expressions and applications from~\eqref{eq:core}.
\setcounter{equation}{18}
\begin{equation}
  e::= \lambda_3 n\;l\;x\LD{}e\mid
       \lambda_4 k\;n\;l\;x\LD{}e\mid
       e_1\;e_2\;e_3\;e_4\mid
       e_1\;e_2\;e_3\;e_4\;e_5
  \label{eq:extended}
\end{equation}
Second, we need to modify~$\EVAL$ to support these new constructs.
We replace clause~\eqref{eq:direct-apply} with
clauses~\cref{eq:extended-continuation-apply,eq:extended-converted-apply} to
update~$\APPLY$ and clauses~\cref{eq:direct-lambda,eq:direct-application} with
clauses
\crefrange{eq:extended-continuation-lambda}{eq:extended-converted-application}
to update~$\EVAL$.
Third, we need to add support for $\textbf{interrupt}$ and $\textbf{resume}$
expressions, as is done with
clauses~\cref{eq:extended-interrupt-expression,eq:extended-resume-expression}.
These are direct-style variants of
clauses~\cref{eq:cps-interrupt-expression,eq:cps-resume-expression} from the
CPS evaluator and are needed to add support for the general-purpose
interruption and resumption interface to the direct-style evaluator when
evaluating CPS code.
Note that the calls to~$\APPLY$
from~\cref{eq:cps-interrupt-expression,eq:cps-resume-expression} are modified
to use the converted form~$\APPLY_4$
of~$\APPLY$~\eqref{eq:extended-converted-apply}
in~\cref{eq:extended-interrupt-expression,eq:extended-resume-expression}.
Similarly, the calls to continuations
from~\cref{eq:cps-interrupt-expression,eq:cps-resume-expression} are modified
to use the continuation form~$\APPLY_3$
of~$\APPLY$~\eqref{eq:extended-continuation-apply}
in~\cref{eq:extended-interrupt-expression,eq:extended-resume-expression}.
Fourth, the calls to~$\APPLY_4$ must be modified in the host implementations of
the AD operators~$\FORWARDJ$ and~$\REVERSEJ$, as is done
with~\cref{eq:extended-forward,eq:extended-reverse}.
Note that unlike the corresponding~\cref{eq:direct-forward,eq:direct-reverse},
the calls to~$\APPLY_4$ here take target closures instead of host closures.
Fifth, the general-purpose interruption and resumption interface,
\cref{eq:cps-primops,eq:cps-interrupt,eq:I,eq:R}, must be migrated from the CPS
evaluator to the direct-style evaluator as
\crefrange{eq:extended-primops}{eq:extended-R}.
In doing so, the calls to~$\APPLY_4$ in $\textsc{primops}$ and
$\textsc{interrupt}$ are changed to use~\eqref{eq:extended-converted-apply},
the host continuations are modified to be target continuations
in~\cref{eq:extended-primops,eq:extended-interrupt}, and the lambda expressions
in~\cref{eq:extended-I,eq:extended-R} are CPS converted.

\setcounter{equation}{20}
\begin{figure}
  %\needswork: v_3 needs primal*
  \centering
  \resizebox{\textwidth}{!}{\fbox{\hspace*{-11pt}
    \begin{minipage}{1.05\textwidth}
      \par\vspace*{-15pt}
      \begin{subequations}
      \begin{align}
        \APPLY_3\;\CLOSURE{\lambda_3 n\;l\;x\LD{}e}{\rho}\;n'\;l'\;v&=
        \EVAL\;\rho[n\mapsto n'][l\mapsto l'][x\mapsto v]\;e
        \label{eq:extended-continuation-apply}\\
        \APPLY_4\;\CLOSURE{\lambda_4 k\;n\;l\;x\LD{}e}{\rho}\;k'\;n'\;l'\;v&=
        \EVAL\;\rho[k\mapsto k'][n\mapsto n'][l\mapsto l'][x\mapsto v]\;e
        \label{eq:extended-converted-apply}\\
        \EVAL\;\rho\;(\lambda_3 n\;l\;x\LD{}e)&
        =\CLOSURE{\lambda_3 n\;l\;x\LD{}e}{\rho}
        \label{eq:extended-continuation-lambda}\\
        \EVAL\;\rho\;(\lambda_4 k\;n\;l\;x\LD{}e)&=
        \CLOSURE{\lambda_4 k\;n\;l\;x\LD{}e}{\rho}
        \label{eq:extended-converted-lambda}\\
        \EVAL\;\rho\;(e_1\;e_2\;e_3\;e_4)&=
        \APPLY_3\;(\EVAL\;\rho\;e_1)\;(\EVAL\;\rho\;e_2)\;
        (\EVAL\;\rho\;e_3)\;(\EVAL\;\rho\;e_4)
        \label{eq:extended-continuation-application}\\
        \EVAL\;\rho\;(e_1\;e_2\;e_3\;e_4\;e_5)&=
        \APPLY_4\;(\EVAL\;\rho\;e_1)\;(\EVAL\;\rho\;e_2)\;
        (\EVAL\;\rho\;e_3)\;(\EVAL\;\rho\;e_4)\;(\EVAL\;\rho\;e_5)
        \label{eq:extended-converted-application}\\
        \EVAL\rho\;(\textbf{interrupt}\;e_1\;e_2\;e_3)&=
        \begin{array}[t]{@{}l@{}}
        \LET
        \begin{array}[t]{@{}l@{}}
        v_1=(\EVAL\;\rho\;e_1)\\
        v_2=(\EVAL\;\rho\;e_2)\\
        v_3=(\EVAL\;\rho\;e_3)\\
        k=\rho\;\textrm{`k'}\\
        l=\rho\;\textrm{`l'}
        \end{array}\\
        \IN
        \begin{array}[t]{@{}l@{}}
        \IF\;l=\infty\\
        \THEN(\APPLY_4\;v_1\;k\;0\;v_3\;v_2)\\
        \ELSE\begin{array}[t]{@{}l@{}}
        \LET\CAPSULE{k}{f}=(\APPLY_4\;v_1\;k\;0\;l\;v_2)\\
        \IN\CAPSULE{k}{(\mathcal{I}\;f\;(v_3-l))}
        \end{array}
        \end{array}
        \end{array}
        \label{eq:extended-interrupt-expression}\\
        \EVAL\;\rho\;(\textbf{resume}\;e)&=
        \begin{array}[t]{@{}l@{}}
        \LET
        \begin{array}[t]{@{}l@{}}
        \CAPSULE{k'}{f}=(\EVAL\;\rho\;e)\\
        l=\rho\;\textrm{`l'}
        \end{array}\\
        \IN
        (\APPLY_4\;f\;k'\;0\;l\;\bot)
        \end{array}
        \label{eq:extended-resume-expression}\\
        \FORWARDJ\;v_1\;v_2\;\acute{v}_3&=
        \begin{array}[t]{@{}l@{}}
        \LET(\BUNDLE{v_4}{\acute{v}_5})=
        (\APPLY_4\;v_1\;\CLOSURE{\lambda_3 n\;l\;v\LD{}v}{\rho_0}
        \;0\;\infty\;(\BUNDLE{v_2}{\acute{v}_3}))\\
        \IN(v_4,\acute{v}_5)
        \end{array}\label{eq:extended-forward}\\
        \REVERSEJ\;v_1\;v_2\;\grave{v}_3&=
        \begin{array}[t]{@{}l@{}}
        \LET(\TAPE{v_4}{\grave{v}_5})=
        (\TAPE{(\APPLY_4\;v_1\;\CLOSURE{\lambda_3 n\;l\;v\LD{}v}{\rho_0}
        \;0\;\infty\;v_2)}{\grave{v}_3})\\
        \IN(v_4,\grave{v}_5)
        \end{array}\label{eq:extended-reverse}\\
        \textsc{primops}\;f\;x&=
        \APPLY_4\;f\;\CLOSURE{\lambda_3 n\;l\;v\LD{}n}{\rho_0}\;0\;\infty\;x
        \label{eq:extended-primops}\\
        \textsc{interrupt}\;f\;l\;n&=
        \APPLY_4\;f\;\CLOSURE{\lambda_3 n\;l\;v\LD{}v}{\rho_0}\;0\;l\;x
        \label{eq:extended-interrupt}\\
        \mathcal{I}\;f\;l&=\CLOSURE{\lambda_4
        k\;n\;l\;x\LD{}
        (\textbf{interrupt}\;f\;x\;l)}{\rho_0[f\mapsto f][l\mapsto l]}
        \label{eq:extended-I}\\
        \mathcal{R}&=
        \CLOSURE{\lambda_4 k\;n\;l\;z\LD{}(\textbf{resume}\;z)}{\rho_0}
        \label{eq:extended-R}
      \end{align}
      \end{subequations}
    \end{minipage}}}
  \caption{Extensions to the direct-style evaluator and the implementation of
    the general-purpose interruption and resumption interface to support
    divide-and-conquer checkpointing on target code that has been converted to
    CPS.\@}
  \label{fig:extended}
\end{figure}

\subsection{Compiling Direct-Style Code to C}
\label{sec:compiling}

One can compile target \checkpointVLAD\ code, after CPS conversion, to
\Clang\ (Figs.~\ref{fig:compiler-one} and~\ref{fig:compiler-two}).
Modern implementations of \Clang, like GCC, together with modern memory
management technology, like the Boehm-Demers-Weiser garbage collector, allow
the compilation process to be a straightforward mapping of each construct to a
small fragment of \Clang\ code.
In particular, garbage collection, \texttt{GC\_malloc}, eases the
implementation of closures and statement expressions,
\texttt{(\symbol{123}\ldots\symbol{125})}, together with nested functions, ease
the implementation of lambda expressions.
Furthermore, the flow analysis, inlining, and tail-call merging performed by
GCC generates reasonably efficient code.
In Figs.~\ref{fig:compiler-one} and~\ref{fig:compiler-two}, $\COMPILE$~denotes
such a mapping from \checkpointVLAD\ expressions~$e$ to \Clang\ code fragments.
Instead of environments~$\rho$, $\COMPILE$~takes~$\pi$, a mapping from
variables to indices in \texttt{environment}, the run-time environment
data structure.
Here, $\pi\;x$ denotes the index of~$x$, $\pi_i$ denotes the variable for
index~$i$, $\phi\;e$ denotes a mapping for the free variables in~$e$, and
$\NAME$~denotes a mapping from a \checkpointVLAD\ operator to the name of the
\Clang\ function that implements that operator.
This, together with a library containing the \texttt{typedef} for
\texttt{thing}, the \texttt{enum} for \texttt{tag}, definitions for
\texttt{null\_constant}, \texttt{true\_constant}, \texttt{false\_constant},
\texttt{cons}, \texttt{as\_closure}, \texttt{set\_closure},
\texttt{continuation\_apply}, \texttt{converted\_apply}, \texttt{is\_false},
and all of the functions named by~$\NAME$ (essentially a translation of
\RsixRSAD, the general-purpose interruption and resumption interface from
Fig.~\ref{fig:extended}, and the implementation of binary bisection
checkpointing from Fig.~\ref{fig:proper-implementation} into \Clang), allows
arbitrary \checkpointVLAD\ code to be compiled to machine code, via \Clang,
with complete support for AD, including forward mode, reverse mode, and binary
bisection checkpointing.

\begin{subequations}

\begin{figure}
  % cache
  % letrec
  % let
  \centering
  \resizebox{\textwidth}{!}{\fbox{\hspace*{-11pt}
    \begin{minipage}{1.2\textwidth}
      \par\vspace*{-15pt}
      \begin{align}
        \COMPILE\;\pi\;()&=\texttt{null\_constant}
        \label{eq:compile-null}\\
        \COMPILE\;\pi\;\textbf{true}&=\texttt{true\_constant}
        \label{eq:compile-true}\\
        \COMPILE\;\pi\;\textbf{false}&=\texttt{false\_constant}
        \label{eq:compile-false}\\
        \COMPILE\;\pi\;(c_1,c_2)&=
        \texttt{cons(}(\COMPILE\;\pi\;c_1)\texttt{, }(\COMPILE\;\pi\;c_1){)}
        \label{eq:compile-pair}\\
        \COMPILE\;\pi\;n&=n\label{eq:compile-constant}\\
        \COMPILE\;\pi\;\texttt{`k'}&=\texttt{continuation}
        \label{eq:compile-k}\\
        \COMPILE\;\pi\;\texttt{`n'}&=\texttt{count}
        \label{eq:compile-n}\\
        \COMPILE\;\pi\;\texttt{`l'}&=\texttt{limit}
        \label{eq:compile-l}\\
        \COMPILE\;\pi\;\texttt{`x'}&=\texttt{argument}
        \label{eq:compile-x}\\
        \COMPILE\;\pi\;x&=
        \texttt{as\_closure(target)->environment[}\pi\;x\texttt{]}
        \label{eq:compile-variable}\\
        \COMPILE\;\pi\;(\lambda_3 n\;l\;x\LD{}e)&=
        \begin{array}[t]{@{}l@{}}
        \texttt{(\symbol{123}}\\
        \texttt{\ \ thing function(thing target,}\\
        \texttt{\ \ \ \ \ \ \ \ \ \ \ \ \ \ \ \ \ thing count,}\\
        \texttt{\ \ \ \ \ \ \ \ \ \ \ \ \ \ \ \ \ thing limit,}\\
        \texttt{\ \ \ \ \ \ \ \ \ \ \ \ \ \ \ \ \ thing argument) \symbol{123}}\\
        \texttt{\ \ return }(\COMPILE\;(\phi\;e)\;e)\texttt{;}\\
        \texttt{\ \ \symbol{125}}\\
        \texttt{\ \ thing lambda = (thing)GC\_malloc(sizeof(struct \symbol{123}}\\
        \texttt{\ \ \ \ enum tag tag;}\\
        \texttt{\ \ \ \ struct \symbol{123}}\\
        \texttt{\ \ \ \ \ \ thing (*function)();}\\
        \texttt{\ \ \ \ \ \ unsigned n;}\\
        \texttt{\ \ \ \ \ \ thing environment[}\lvert\phi\;e\rvert{];}\\
        \texttt{\ \ \ \ \symbol{125}}))\\
        \texttt{\ \ \symbol{125}}\\
        \texttt{\ \ set\_closure(lambda);}\\
        \texttt{\ \ as\_closure(lambda)->function = \&function;}\\
        \texttt{\ \ as\_closure(lambda)->n = }\lvert\phi\;e\rvert{;}\\
        \texttt{\ \ as\_closure(lambda)->environment[}0\texttt{] = }
        \COMPILE\;\pi\;(\phi\; e)_0\\
        \texttt{\ \ }\vdots\\
        \texttt{\ \ as\_closure(lambda)->environment[}
        \lvert\phi\;e\rvert-1\texttt{] = }
        \COMPILE\;\pi\;(\phi\;e)_{\lvert\phi\;e\rvert-1}\\
        \texttt{\ \ lambda;}\\
        \texttt{\symbol{125})}
        \end{array}
        \label{eq:compile-continuation-lambda}
      \end{align}
    \end{minipage}}}
  \caption{Compiler for the \checkpointVLAD\ language when in CPS.\@
    Part I.\@}
  \label{fig:compiler-one}
\end{figure}

\begin{figure}
  \centering
  \resizebox{\textwidth}{!}{\fbox{\hspace*{-11pt}
    \begin{minipage}{1.2\textwidth}
      \par\vspace*{-15pt}
      \begin{align}
        \COMPILE\;\pi\;(\lambda_4 k\;n\;l\;x\LD{}e)&=
        \begin{array}[t]{@{}l@{}}
        \texttt{(\symbol{123}}\\
        \texttt{\ \ thing function(thing target,}\\
        \texttt{\ \ \ \ \ \ \ \ \ \ \ \ \ \ \ \ \ thing continuation,}\\
        \texttt{\ \ \ \ \ \ \ \ \ \ \ \ \ \ \ \ \ thing count,}\\
        \texttt{\ \ \ \ \ \ \ \ \ \ \ \ \ \ \ \ \ thing limit,}\\
        \texttt{\ \ \ \ \ \ \ \ \ \ \ \ \ \ \ \ \ thing argument) \symbol{123}}\\
        \texttt{\ \ return }(\COMPILE\;(\phi\;e)\;e)\texttt{;}\\
        \texttt{\ \ \symbol{125}}\\
        \texttt{\ \ thing lambda = (thing)GC\_malloc(sizeof(struct \symbol{123}}\\
        \texttt{\ \ \ \ enum tag tag;}\\
        \texttt{\ \ \ \ struct \symbol{123}}\\
        \texttt{\ \ \ \ \ \ thing (*function)();}\\
        \texttt{\ \ \ \ \ \ unsigned n;}\\
        \texttt{\ \ \ \ \ \ thing environment[}\lvert\phi\;e\rvert{];}\\
        \texttt{\ \ \ \ \symbol{125}}))\\
        \texttt{\ \ \symbol{125}}\\
        \texttt{\ \ set\_closure(lambda);}\\
        \texttt{\ \ as\_closure(lambda)->function = \&function;}\\
        \texttt{\ \ as\_closure(lambda)->n = }\lvert\phi\;e\rvert{;}\\
        \texttt{\ \ as\_closure(lambda)->environment[}0\texttt{] = }
        \COMPILE\;\pi\;(\phi\;e)_0\\
        \texttt{\ \ }\vdots\\
        \texttt{\ \ as\_closure(lambda)->environment[}
        \lvert\phi\;e\rvert-1\texttt{] = }
        \COMPILE\;\pi\;(\phi\;e)_{\lvert\phi\;e\rvert-1}\\
        \texttt{\ \ lambda;}\\
        \texttt{\symbol{125})}
        \end{array}
        \label{eq:compile-converted-lambda}\\
        \COMPILE\;\pi\;(e_1\;e_2\;e_3\;e_4)&=
        \begin{array}[t]{@{}l@{}}
        \texttt{continuation\_apply(}
        (\COMPILE\;\pi\;e_1)\texttt{, }\\
        \texttt{\ \ \ \ \ \ \ \ \ \ \ \ \ \ \ \ \ \ \ }
        (\COMPILE\;\pi\;e_2)\texttt{, }\\
        \texttt{\ \ \ \ \ \ \ \ \ \ \ \ \ \ \ \ \ \ \ }
        (\COMPILE\;\pi\;e_3)\texttt{, }\\
        \texttt{\ \ \ \ \ \ \ \ \ \ \ \ \ \ \ \ \ \ \ }
        (\COMPILE\;\pi\;e_4)\texttt{)}
        \end{array}
        \label{eq:compile-continuation-application}\\
        \COMPILE\;\pi\;(e_1\;e_2\;e_3\;e_4\;e_5)&=
        \begin{array}[t]{@{}l@{}}
        \texttt{converted\_apply(}
        (\COMPILE\;\pi\;e_1)\texttt{, }\\
        \texttt{\ \ \ \ \ \ \ \ \ \ \ \ \ \ \ \ }
        (\COMPILE\;\pi\;e_2)\texttt{, }\\
        \texttt{\ \ \ \ \ \ \ \ \ \ \ \ \ \ \ \ }
        (\COMPILE\;\pi\;e_3)\texttt{, }\\
        \texttt{\ \ \ \ \ \ \ \ \ \ \ \ \ \ \ \ }
        (\COMPILE\;\pi\;e_4)\texttt{, }\\
        \texttt{\ \ \ \ \ \ \ \ \ \ \ \ \ \ \ \ }
        (\COMPILE\;\pi\;e_5)\texttt{)}
        \end{array}
        \label{eq:compile-converted-application}\\
        \COMPILE\;\pi\;(\IF e_1\;\THEN e_2\;\ELSE e_3)&=
        \texttt{(!is\_false(}
        (\COMPILE\;\pi\;e_1)
        \texttt{)?}
        (\COMPILE\;\pi\;e_2)
        \texttt{:}
        (\COMPILE\;\pi\;e_3)
        \texttt{)}
        \label{eq:compile-conditional}\\
        \COMPILE\;\pi\;(\diamond e)&=
        (\NAME\;\diamond)\texttt{(}(\COMPILE\;\pi\;e)\texttt{)}
        \label{eq:compile-unary}\\
        \COMPILE\;\pi\;(e_1\bullet e_2)&=
        (\NAME\;\bullet)\texttt{(}(\COMPILE\;\pi\;e_1)\texttt{, }
        (\COMPILE\;\pi\;e_2)\texttt{)}
        \label{eq:compile-binary}\\
        \COMPILE\;\pi\;(\FORWARDJ\;e_1\;e_2\;e_3)&=
        (\NAME\;\FORWARDJ)\texttt{(}(\COMPILE\;\pi\;e_1)\texttt{, }
        (\COMPILE\;\pi\;e_2)\texttt{, }(\COMPILE\;\pi\;e_3)\texttt{)}
        \label{eq:compile-jstar}\\
        \COMPILE\;\pi\;(\REVERSEJ\;e_1\;e_2\;e_3)&=
        (\NAME\;\REVERSEJ)\texttt{(}(\COMPILE\;\pi\;e_1)\texttt{, }
        (\COMPILE\;\pi\;e_2)\texttt{, }(\COMPILE\;\pi\;e_3)\texttt{)}
        \label{eq:compile-starj}\\
        \COMPILE\;\pi\;(\CHECKPOINTREVERSEJ\;e_1\;e_2\;e_3)&=
        (\NAME\;\CHECKPOINTREVERSEJ)\texttt{(}(\COMPILE\;\pi\;e_1)\texttt{, }
        (\COMPILE\;\pi\;e_2)\texttt{, }(\COMPILE\;\pi\;e_3)\texttt{)}
        \label{eq:compile-checkpoint-starj}
      \end{align}
    \end{minipage}}}
  \caption{Compiler for the \checkpointVLAD\ language when in CPS.\@
    Part II.\@}
  \label{fig:compiler-two}
\end{figure}

\end{subequations}

\subsection{Implementations}
\label{sec:implementations}

We have written three complete implementations of \checkpointVLAD.\footnote{The
  code for all three implementations will be released on \texttt{github} upon
  acceptance of this manuscript.}
All three accept exactly the same source language in its entirety and are able
to run the example discussed in Section~\ref{sec:example}.
The first implementation is an interpreter based on the CPS evaluator
(Figs.~\ref{fig:cps-core}, \ref{fig:cps-AD}, \ref{fig:cps-interface},
and~\ref{fig:checkpointreversej}), where the evaluator, the operator
overloading implementation of AD, the general-purpose interruption and
resumption mechanism (Fig.~\ref{fig:interface-implementation}), and the binary
bisection checkpointing driver (Fig.~\ref{fig:proper-implementation}) are
implemented in \Scheme.
The second implementation is a hybrid compiler/interpreter that translates the
\checkpointVLAD\ source program into CPS using CPS conversion
(Fig.~\ref{fig:threaded-cps-conversion}) and then interprets this with an
interpreter based on the direct-style evaluator (Figs.~\ref{fig:direct-core},
\ref{fig:direct-AD}, and~\ref{fig:extended}), where the compiler, the
evaluator, the operator overloading implementation of AD, the general-purpose
interruption and resumption mechanism (Fig.~\ref{fig:extended}), and the binary
bisection checkpointing driver (Fig.~\ref{fig:proper-implementation}) are
implemented in \Scheme.
The third implementation is a compiler that translates the
\checkpointVLAD\ source program into CPS using CPS conversion
(Fig.~\ref{fig:threaded-cps-conversion}) and then compiles this to machine code
via \Clang\ using GCC, where the compiler (Figs.~\ref{fig:compiler-one}
and~\ref{fig:compiler-two}) is implemented in \Scheme, the evaluator is the
underlying hardware, and the operator overloading implementation of AD, the
general-purpose interruption and resumption mechanism
(Fig.~\ref{fig:extended}), and the binary bisection checkpointing driver
(Fig.~\ref{fig:implementation}) are implemented in \Clang.
The first implementation was used to generate the results reported in
\citep{ad2016a} and presented at AD~(2016).
The techniques of Figs.~\ref{fig:cps-conversion} and
\ref{fig:threaded-cps-conversion} were presented at AD~(2016).
The third implementation was used to generate the results reported here.

\section{Complexity}
\label{sec:complexity}

The internal nodes of a binary checkpoint tree correspond to invocations of
$\textsc{interrupt}$ in step~(2).
The right branches of each node correspond to step~(3).
The left branches of each node correspond to step~(4).
The leaf nodes correspond to invocations of~$\REVERSEJ$ in the base case,
step~(0).
Each leaf node corresponds to a stage, the red, blue, and violet lines in
Fig.~\ref{fig:divide-and-conquer-checkpointing}(g).
The checkpoint tree is traversed in depth-first right-to-left preorder.
In our implementation, we terminate the recursion when the step limit~$l$ is
below a fixed constant.
Consider a general primal computation~$f$ that uses maximal live storage~$w$
and that runs for~$t$ steps.
This results in the following space and time complexities for reverse mode
without checkpointing, including our implementation of~$\REVERSEJ$, and for
binary bisection checkpointing, including our implementation
of~$\CHECKPOINTREVERSEJ$.
\begin{center}
\resizebox{\columnwidth}{!}{\begin{tabular}{@{}llllll|lllll@{}}
\toprule
& \multicolumn{10}{c}{\textbf{Complexity}}\\
\multirow{4}{*}[1.5ex]{\textbf{Computation}}
& \multicolumn{5}{c}{\textbf{space}} & \multicolumn{5}{c}{\textbf{time}}\\
\cmidrule{2-11}
& primal & snapshots & tape & total & overhead
& recomputation & forward & reverse & total & overhead\\
& & & & & & & sweep & sweep & &\\
\hline
\rule{0pt}{15pt}without checkpointing  & $O(w)$ & & $O(t)$ & $O(w+t)$ &
$O(\frac{w+t}{w})$ & & $O(t)$ & $O(t)$ & $O(t)$ & $O(1)$\\
binary bisection checkpointing & $O(w)$ & $O(w\log t)$ & $O(1)$ & $O(w\log t)$
& $O(\log t)$ & $O(t\log t)$ & $O(t)$ & $O(t)$ & $O(t\log t)$ & $O(\log t)$\\
\bottomrule
\end{tabular}}
\end{center}
If we assume that $O(t)\geq O(w)$, \ie\ that the computation uses all storage,
the total space requirement without checkpointing becomes $O(t)$ and the
overhead becomes $O(t)$.

\section{Extensions to Support Treeverse and Binomial Checkpointing}
\label{sec:binomial}

The general-purpose interruption and resumption interface allows implementation
of~$\CHECKPOINTREVERSEJ$ using the treeverse algorithm from
\citep[Fig.~4]{griewank1992alg} as shown in Fig.~\ref{fig:treeverse}.
This supports the full functionality of that algorithm with the ability to
select arbitrary execution points as split points.
By selecting the choice of $\textsc{mid}$ as either
\citep[equation (12)]{griewank1992alg} or
\citep[equation (16)]{griewank1992alg}, one can select between bisection and
binomial checkpointing.
By selecting which of~$d$, $t$, and~$\alpha$ the user specifies, computing the
others from the ones specified, together with~$n$, using the methods described
in \citep{griewank1992alg} one can select the termination criterion to be
either fixed space overhead, fixed time overhead, or logarithmic space and time
overhead.\footnote{In Section~\ref{sec:binomial} and Figs.~\ref{fig:treeverse}
  and~\ref{fig:binary-two} we use notation similar to that in
  \citep{griewank1992alg} to facilitate understanding.
  Thus~$n$ and~$t$ here means something different then elsewhere in this
  manuscript.}
All of this functionality has been implemented in all three of our
implementations: the interpreter, the hybrid compiler/interpreter, and the
compiler.

But it turns out that the binary checkpointing algorithm from
Fig.~\ref{fig:proper-implementation} can be easily modified to support all of
the functionality of the treeverse algorithm, including the ability to select
either bisection or binomial checkpointing and the ability to select any of the
termination criteria, including either fixed space overhead, fixed time
overhead, or logarithmic space and time overhead, with exactly the same
guarantees as treeverse.
The idea is simple and follows from the observation that the right branch
introduces a snapshot and the left branch introduces (re)computation of the
primal.
One maintains two counts, a right-branch count~$\delta$ and a left-branch
count~$\tau$, decrementing them as one descends into a right or left branch
respectively, to limit the number of snapshots or the amount of
(re)computation introduced.
The base case is triggered when either gets to zero or a specified constant
bound on the number of steps to be taped is reached.
The binary checkpoint tree so produced corresponds to the associated n-ary
checkpoint tree produced by treeverse, as discussed in
Section~\ref{sec:introduction}.
Again, this supports the full functionality of treeverse with the ability to
select arbitrary execution points as split points.
By selecting the choice of $\textsc{mid}$ as either
\citep[equation (12)]{griewank1992alg} or
\citep[equation (16)]{griewank1992alg}, one can select between bisection and
binomial checkpointing.
By selecting which of~$d$, $t$, and~$\alpha$ the user specifies, computing the
others from the ones specified, together with~$n$, using the methods described
in \citep{griewank1992alg} one can select the termination criterion to be
either fixed space overhead, fixed time overhead, or logarithmic space and time
overhead.
All of this functionality has been implemented in all three of our
implementations: the interpreter, the hybrid compiler/interpreter, and the
compiler.

\begin{figure}
  \centering
  \fbox{\hspace*{-11pt}
    \begin{minipage}{\textwidth}
      \par\vspace*{-15pt}
      \begin{align*}
        \textsc{treeverse}\;
        f\;x\;\grave{y}\;\alpha\;\delta\;\tau\;\beta\;\sigma\;\phi&=
        \begin{array}[t]{@{}l@{}}
        \IF\sigma>\beta\\
        \THEN\begin{array}[t]{@{}l@{}}
        \LET z=\textsc{interrupt}\;f\;x\;(\sigma-\beta)\\
        \IN\textsc{first}\;
        \mathcal{R}\;z\;\grave{y}\;\alpha\;(\delta-1)\;\tau\;\beta\;\sigma\;\phi
        \end{array}\\
        \ELSE\textsc{first}\;
        f\;x\;\grave{y}\;\alpha\;\delta\;\tau\;\beta\;\sigma\;\phi
        \end{array}\\
        \textsc{first}\;
        f\;x\;\grave{y}\;\alpha\;\delta\;\tau\;\beta\;\sigma\;\phi&=
        \begin{array}[t]{@{}l@{}}
        \IF\phi-\sigma>\alpha\wedge\delta\not=0\wedge\tau\not=0\\
        \THEN
        \begin{array}[t]{@{}l@{}}
        \LET\begin{array}[t]{@{}l@{}}
        \kappa=\textsc{mid}\;\delta\;\tau\;\sigma\;\phi\\
        (y,\grave{z})=
        \textsc{treeverse}
        \;f\;x\;\grave{y}\;\alpha\;\delta\;\tau\;\sigma\;\kappa\;\phi
        \end{array}\\
        \IN\textsc{rest}\;
        f\;x\;\grave{z}\;\alpha\;y\;\delta\;(\tau-1)\;\beta\;\sigma\;\kappa
        \end{array}\\
        \ELSE\REVERSEJ\;(\mathcal{I}\;f\;(\phi-\sigma))\;x\;\grave{y}
        \end{array}\\
        \textsc{rest}\;
        f\;x\;\grave{y}\;\alpha\;y\;\delta\;\tau\;\beta\;\sigma\;\phi&=
        \begin{array}[t]{@{}l@{}}
        \IF\phi-\sigma>\alpha\wedge\delta\not=0\wedge\tau\not=0\\
        \THEN\begin{array}[t]{@{}l@{}}
        \LET\begin{array}[t]{@{}l@{}}
        \kappa=\textsc{mid}\;\delta\;\tau\;\sigma\;\phi\\
        (\texttt{\_},\grave{z})=
        \textsc{treeverse}
        \;f\;x\;\grave{y}\;\alpha\;\delta\;\tau\;\sigma\;\kappa\;\phi
        \end{array}\\
        \IN\textsc{rest}\;
        f\;x\;\grave{z}\;\alpha\;y\;\delta\;(\tau-1)\;\beta\;\sigma\;\kappa
        \end{array}\\
        \ELSE\begin{array}[t]{@{}l@{}}
        \LET(\texttt{\_},\grave{x})=
        \REVERSEJ\;(\mathcal{I}\;f\;(\phi-\sigma))\;x\;\grave{y}\\
        \IN(y,\grave{x})
        \end{array}
        \end{array}\\
        \CHECKPOINTREVERSEJ\;f\;x\;\grave{y}&=
        \begin{array}[t]{@{}l@{}}
        \LET\begin{array}[t]{@{}l@{}}
        n=\textsc{primops}\;f\;x\\
        \textsc{pick}\;\alpha\;d\;t
        \end{array}\\
        \IN\textsc{treeverse}\;f\;x\;\grave{y}\;\alpha\;d\;t\;0\;0\;n
        \end{array}
      \end{align*}
    \end{minipage}}
  \caption{Implementation of $\textsc{treeverse}$ from
    \citep[Fig.~4]{griewank1992alg} using the general-purpose interruption and
    resumption interface, written in a functional style with no mutation.
    The variables~$\delta$, $\tau$, $\beta$, $\sigma$, $\phi$, $n$, $d$,
    and~$t$ have the same meaning as in \citep{griewank1992alg}.
    The variables~$f$, $x$, $\grave{x}$, $y$, $\grave{y}$, $z$, and~$\grave{z}$
    have the same meaning as earlier in this manuscript.
    The variable~$\alpha$ denotes an upper bound on the number of evaluation
    steps for a leaf node.
    Different termination criteria allow the user to specify some
    of~$\alpha$, $d$, and~$t$ and compute the remainder as a function of the
    ones specified, together with~$n$.}
  \label{fig:treeverse}
\end{figure}

\begin{figure}
  \centering
  \fbox{\hspace*{-11pt}
    \begin{minipage}{\textwidth}
      \par\vspace*{-15pt}
      \begin{align*}
        \textsc{binary}\;
        f\;x\;\grave{y}\;\alpha\;\delta\;\tau\;\phi&=
        \begin{array}[t]{@{}l@{}}
        \IF\phi\leq\alpha\vee\delta=0\vee\tau=0\\
        \THEN\REVERSEJ\;f\;x\;\grave{y}\\
        \ELSE\begin{array}[t]{@{}l@{}}
        \LET\begin{array}[t]{@{}l@{}}
        \kappa=\textsc{mid}\;\delta\;\tau\;0\;\phi\\
        z=\textsc{interrupt}\;f\;x\;\kappa\\
        (y,\grave{z})=\textsc{binary}\;\mathcal{R}\;z\;\grave{y}\;(\delta-1)\;
        \tau\;(\phi-\kappa)\\
        (z,\grave{x})=\textsc{binary}\;(\mathcal{I}\;f\;\kappa)\;x\;\grave{z}\;
        \delta\;(\tau-1)\;\kappa
        \end{array}\\
        \IN(y,\grave{x})
        \end{array}
        \end{array}\\
        \CHECKPOINTREVERSEJ\;f\;x\;\grave{y}&=
        \begin{array}[t]{@{}l@{}}
        \LET\begin{array}[t]{@{}l@{}}
        n=\textsc{primops}\;f\;x\\
        \textsc{pick}\;\alpha\;d\;t
        \end{array}\\
        \IN\textsc{binary}\;f\;x\;\grave{y}\;\alpha\;d\;t\;n
        \end{array}
      \end{align*}
    \end{minipage}}
  \caption{Implementation of binary checkpointing using the general-purpose
    interruption and resumption interface in a fashion that supports all of the
    functionality of $\textsc{treeverse}$ from \citep[Fig.~4]{griewank1992alg}.
    The variables~$\delta$, $\tau$, $\phi$, $n$, $d$, and~$t$ have the same
    meaning as in \citep{griewank1992alg}.
    The variables~$f$, $x$, $\grave{x}$, $y$, $\grave{y}$, $z$, and~$\grave{z}$
    have the same meaning as earlier in this manuscript.
    The variable~$\alpha$ denotes an upper bound on the number of evaluation
    steps for a leaf node.
    Different termination criteria allow the user to specify some
    of~$\alpha$, $d$, and~$t$ and compute the remainder as a function of the
    ones specified, together with~$n$.}
  \label{fig:binary-two}
\end{figure}

As per \citep{griewank1992alg}, with a binomial strategy for selecting split
points, the termination criteria can be implemented as follows.
Given a measured number~$n$ of evaluation steps, $d$, $t$, and~$\alpha$ are
mutually constrained by a single constraint.
\begin{align}
  n&=\textsc{primops}\;f\;x
  &
  \eta(d,t)&=\left(\begin{array}{c}d+t\\t\end{array}\right)
  &
  \alpha&=\left\lceil\frac{n}{\eta(d,t)}\right\rceil
\end{align}
One can select any two and determine the third.
Selecting~$d$ and~$\alpha$ to determine $t=O\left(\sqrt[d]{n}\right)$ yields the
fixed space overhead termination criterion.
Selecting~$t$ and~$\alpha$ to determine $d=O\left(\sqrt[t]{n}\right)$ yields the
fixed time overhead termination criterion.
Alternatively, one can further constrain $d=t$.
With this, selecting~$\alpha$ to determine~$d$ and~$t$ yields the logarithmic
space and time overhead termination criterion.

\section{An Example}
\label{sec:example}

As discussed in Section~\ref{sec:limitations}, existing implementations of
divide-and-conquer checkpointing, such as \Tapenade, are limited to placing
split points at execution points corresponding to particular syntactic program
points in the source code, \ie\ loop iteration boundaries.
Our approach can place split points at arbitrary execution points.
The example in Fig.~\ref{fig:example-fortran} illustrates a situation where
placing split points only at loop iteration boundaries can fail to yield the
sublinear space overhead of divide-and-conquer checkpointing while placement
of split points at arbitrary execution points will yield the sublinear space
overhead of divide-and-conquer checkpointing.
To illustrate this, we run the \Fortran\ variant of this example two different
ways with \Tapenade:
\begin{enumerate}
\item without checkpointing, by removing all pragmas and
\item with divide-and-conquer checkpointing, particularly the treeverse
  algorithm applied to a root execution interval corresponding to the
  invocations of the outer DO loop, split points selected with the binomial
  criterion from execution points corresponding to iteration boundaries of the
  outer DO loop, and a fixed space overhead termination criterion, by placing
  the \binomialckp\ pragma as shown in Fig.~\ref{fig:example-fortran}.
\end{enumerate}
For comparison, we reformulate this \Fortran\ example in
\checkpointVLAD\ (Fig.~\ref{fig:example-vlad}) and run it two different ways:
\begin{enumerate}
\item without checkpointing, by calling~$\REVERSEJ$, written here as
  \texttt{*j} and
\item with divide-and-conquer checkpointing, particularly the binary
  checkpointing algorithm applied to a root execution interval corresponding to
  the entire derivative calculation, split points selected with the bisection
  criterion from arbitrary execution points, and a logarithmic space and time
  overhead termination criterion, by calling~$\CHECKPOINTREVERSEJ$, written
  here as \texttt{checkpoint-*j}.
\end{enumerate}

\lstset{language=Lisp,
        morekeywords={begin,either,if,lambda,define},
        deletekeywords={remove,floor,log,expt,sqrt,loop,null},
	basicstyle=\ttfamily\small,
	keywordstyle=\color{darkblue}\bfseries\ttfamily\small,
	commentstyle=\color{darkred}\itshape\ttfamily\small}

\begin{figure}
  \centering
  \resizebox{\columnwidth}{!}{\begin{tabular}{@{}ll@{}}
  \wrap{
\begin{lstlisting}[lastline=30]
(define (car (cons car cdr)) car)

(define (cdr (cons car cdr)) cdr)

(define (first x) (car x))

(define (rest x) (cdr x))

(define (second x) (first (rest x)))

(define (iota n) (if (zero? n) '() (cons n (iota (- n 1)))))

(define (ilog2 l) (floor (/ (log l) (log 2))))

(define (rotate theta x1 x2)
 (let ((c (cos theta)) (s (sin theta)))
  (cons (- (* c x1) (* s x2)) (+ (* s x1) (* c x2)))))

(define (rot1 theta x)
 (if (or (null? x) (null? (rest x)))
     x
     (let ((x12 (rotate theta (first x) (second x))))
      (cons (car x12)
	    (cons (cdr x12) (rot1 theta (rest (rest x))))))))

(define (rot2 theta x)
 (if (null? x) x (cons (first x) (rot1 theta (rest x)))))

(define (magsqr x)
 (if (null? x) 0 (+ (* (first x) (first x)) (magsqr (rest x)))))
\end{lstlisting}}
  \wrap{
\begin{lstlisting}[firstline=32]
(define (write-vector v)
 (if (null? v) '() (cons (write-real (first v)) (write-vector (rest v)))))

(define (f x l phi)
 (let outer ((i 1) (x1 x))
  (if (> i l)
      (/ (magsqr x1) 2)
      (let ((m (expt
		2
		(- (ilog2 l)
		   (ilog2
		    (+ 1 (modulo (* (* 1013 (floor (expt 3 phi))) i) l)))))))
       (let inner ((j 1) (x1 x1))
	(if (> j m)
	    (outer (+ i 1) x1)
	    (inner (+ j 1)
		   (let ((y (sqrt (magsqr x1))))
		    (rot2 (* 1.4 y) (rot1 (* 1.2 y) x1))))))))))

(let* ((n (read-real))
       (l (read-real))
       (phi (read-real))
       (x (iota n))
       (result (checkpoint-*j (lambda (x) (f x l phi)) x 1)))
 (cons (write-real (car result)) (write-vector (cdr result))))
\end{lstlisting}}
  \end{tabular}}
  \caption{A rendering of the example from Fig.~\ref{fig:example-fortran} in
    \checkpointVLAD.
    Space and time overhead of two variants of this example when run under
    \checkpointVLAD\ are presented in Fig.~\ref{fig:example-plots}.
    The variant for divide-and-conquer checkpointing is shown.
    The variant with no checkpointing replaces \lstinline{checkpoint-*j} with
    \lstinline{*j}.}
  \label{fig:example-vlad}
\end{figure}

For this example, $n$~is the input dimension and~$l$ is the number of
iterations of the outer loop.
Using the notation from Section~\ref{sec:complexity}, the maximal space usage
of the primal for this example should be $w=O(n)$.
The time required for the primal for this example should be $t=O(l)$, since
there are~$l$ iterations of the outer loop and the inner loop has average case
$O(1)$ iterations per iteration of the outer loop.
The analysis in Sections~\ref{sec:complexity} and~\ref{sec:binomial} predicts
the following asymptotic space and time complexity of the \Tapenade\ and
\checkpointVLAD\ variants that compute gradients on this particular example:
\begin{center}
  \begin{tabular}{lll} \toprule
    \multirow{2}{*}[-0.75ex]{\textbf{Computation}} &
    \multicolumn{2}{l}{\textbf{Complexity}} \\
    \cmidrule{2-3}
    & \textbf{space} & \textbf{time} \\
    \midrule
    primal & $O(n)$ & $O(l)$\\
    \addlinespace
    \Tapenade\ no checkpointing & $O(n+l)$ & $O(l)$\\
    \Tapenade\ divide-and-conquer checkpointing & $O(nl)$\footnotemark &
    $O\left(l\sqrt[d]{l}\right)$\\
    \addlinespace
    \checkpointVLAD\ no checkpointing & $O(n+l)$ & $O(l)$\\
    \checkpointVLAD\ divide-and-conquer checkpointing & $O(n\log l)$ &
    $O(l\log l)$
    \\\bottomrule
  \end{tabular}
\end{center}
\footnotetext{The space complexity of \Tapenade\ with divide-and-conquer
  checkpointing would be $O(n)$ if the inner DO loop would have a constant
  number of iterations.
  The fact that it has average case $O(1)$ iterations but $O(l)$ worst case,
  foils checkpointing and causes the space complexity to increase.}
The efficacy of our method can be seen in the plots
(Fig.~\ref{fig:example-plots}) of the observed space and time usage of the
above two \Fortran\ variants and the above two \checkpointVLAD\ variants with
varying~$l$ and $n=\textrm{1000}$.
We observe that \Tapenade\ space and time usage grows with~$l$ for all cases.
\checkpointVLAD\ space and time usage grows with~$l$ with~$\REVERSEJ$.
\checkpointVLAD\ space usage is sublinear with~$\CHECKPOINTREVERSEJ$.
\checkpointVLAD\ time usage grows with~$l$ with~$\CHECKPOINTREVERSEJ$.

\setcounter{figure}{29}
\begin{figure}
  \centering
  \resizebox{\textwidth}{!}{\begin{tabular}{@{}c@{}c@{}}
      \includegraphics[bb=0 0 360 216]{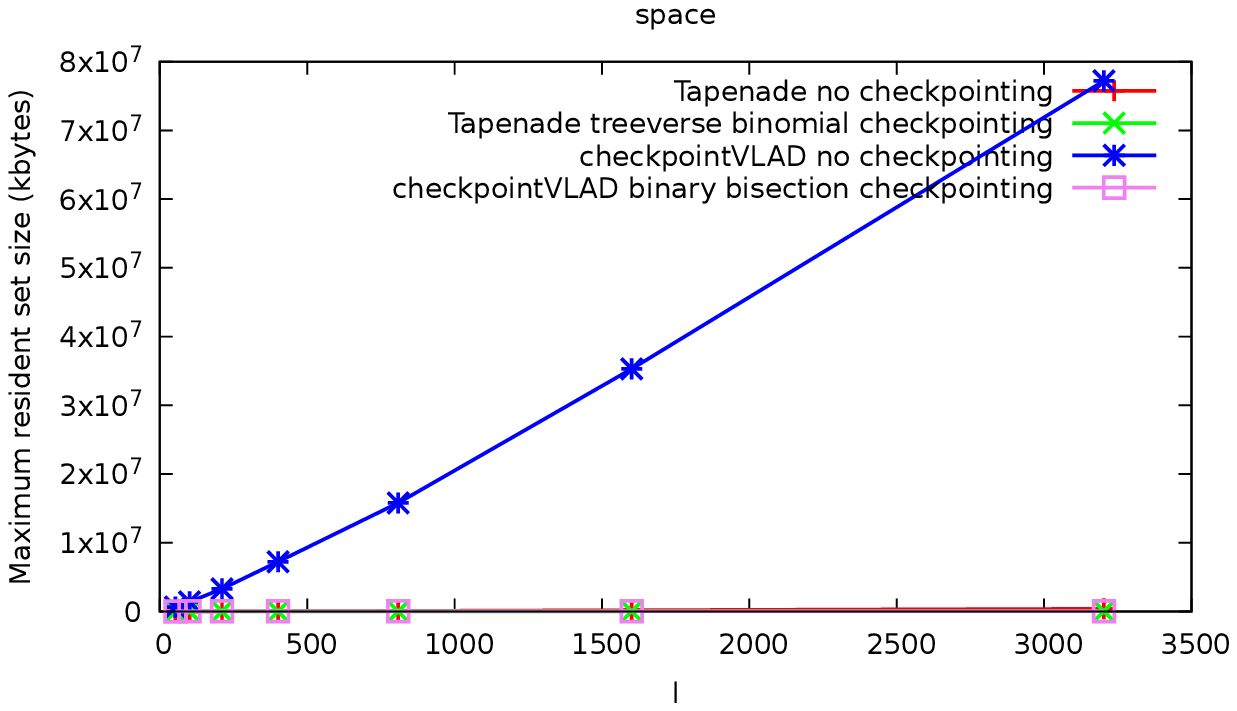}&
      \includegraphics[bb=0 0 360 216]{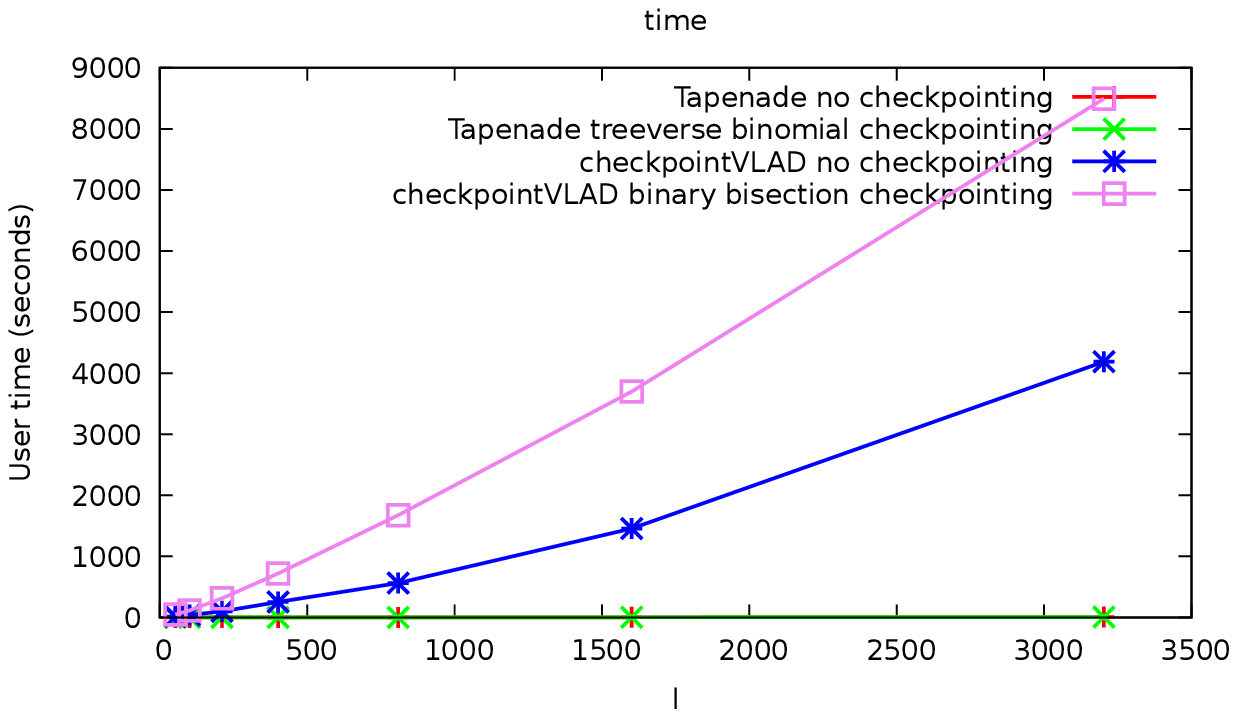}\\
      \includegraphics[bb=0 0 360 216]{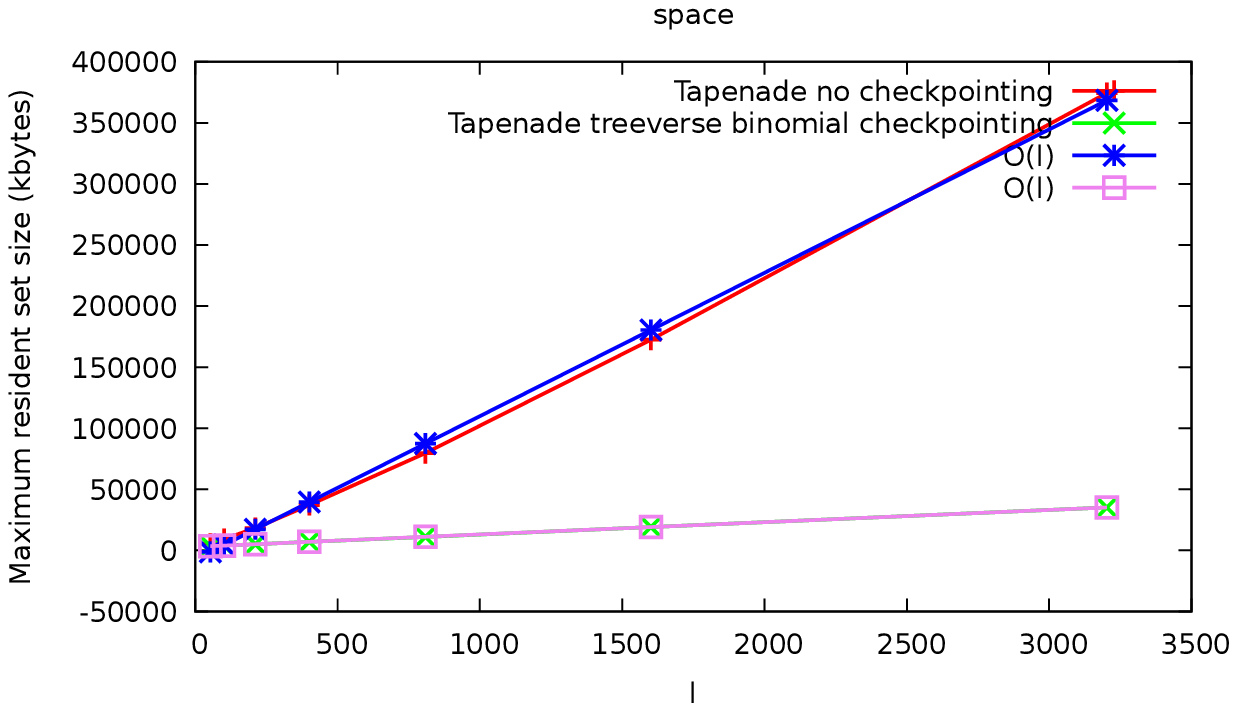}&
      \includegraphics[bb=0 0 360 216]{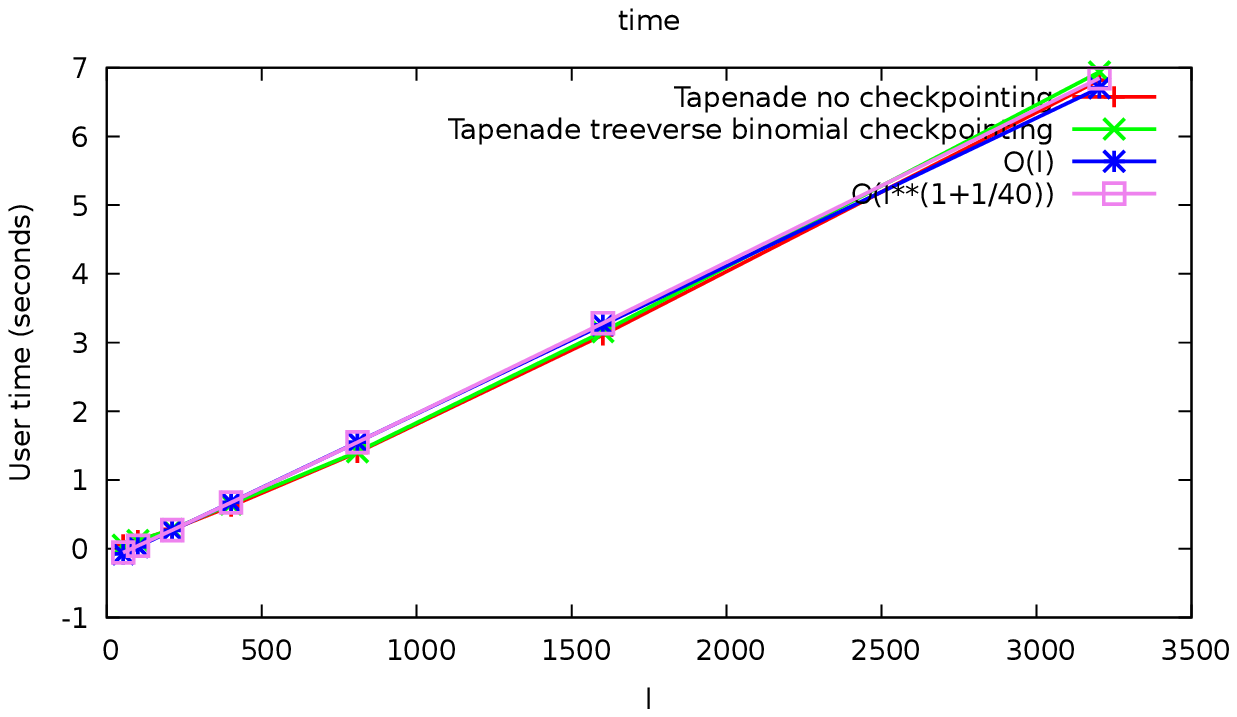}\\
      \includegraphics[bb=0 0 360 216]{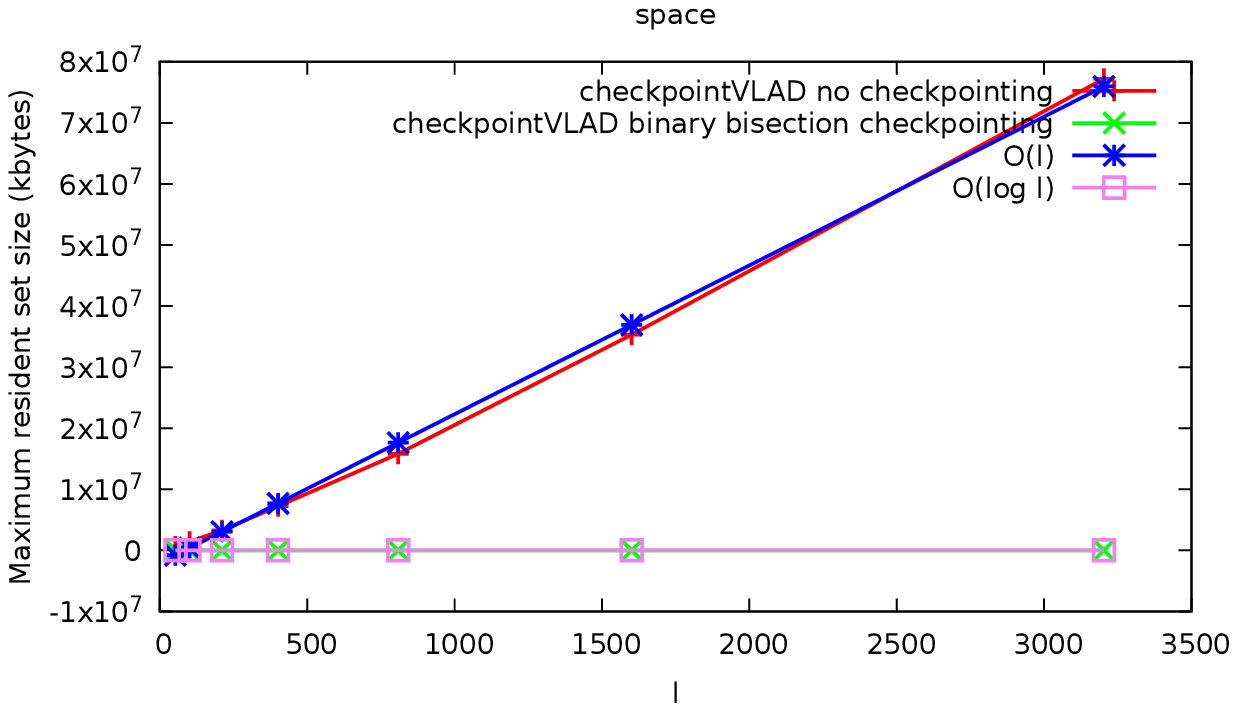}&
      \includegraphics[bb=0 0 360 216]{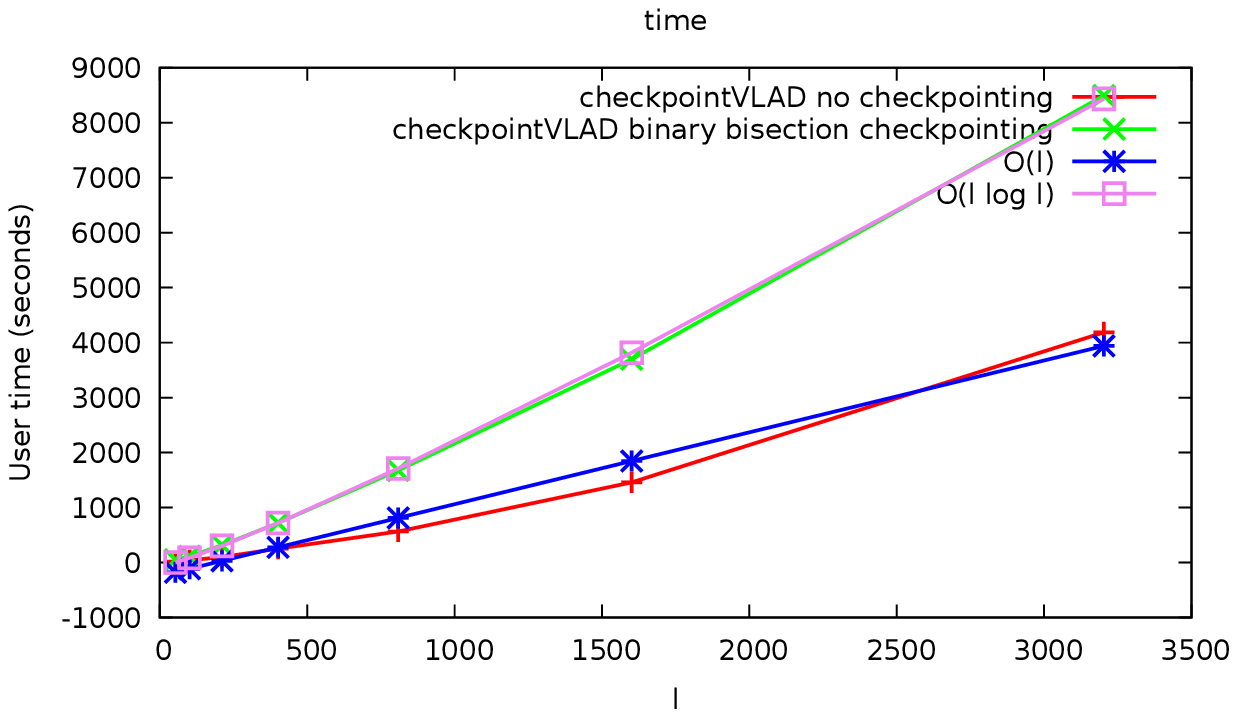}
  \end{tabular}}
  \caption{Space and time usage of reverse-mode AD with and without
    divide-and-conquer checkpointing for the example in
    Figs.~\ref{fig:example-fortran} and~\ref{fig:example-vlad}.
    Space and time usage was measured with \texttt{/usr/bin/time --verbose}
    The center and bottom rows repeat the information from the top row just for
    \Tapenade\ and \checkpointVLAD, overlaid with the theoretical asymptotic
    complexity fit to the actual data by linear regression.}
  \label{fig:example-plots}
\end{figure}

The crucial aspect of this example is that we observe sublinear space usage
overhead with divide-and-conquer checkpointing in \checkpointVLAD\ but not
in \Tapenade.\footnote{Technically, the space and time usage overhead of
  the \checkpointVLAD\ variant of this example with divide-and-conquer
  checkpointing should be logarithmic.
  It is difficult to see that precise overhead in the plots.
  Linear regression does indeed fit logarithmic growth to the time usage better
  than linear growth.
  But the observed space usage appears to be grow in steps.
  This is likely due to the coarse granularity of the measurement techniques
  that are based on kernel memory page allocation and thus fail to measure
  the actual live fraction of the heap data managed by the garbage collector.}
The reason that we fail to observe sublinear space usage overhead with
divide-and-conquer checkpointing in \Tapenade\ is that the space overhead
guarantees only hold when the asymptotic time complexity of the loop body is
constant.
Since the asymptotic time complexity of the loop body is $O(l)$, the requisite
tape size grows with $O(l)$ even though the number of snapshots, and the size
of those snapshots, is bounded by a constant.

\section{Discussion}

\setcounter{figure}{28}
\begin{wrapfigure}[9]{r}{0.50\columnwidth}
\vspace*{-85pt}
\centering
\resizebox{0.50\columnwidth}{!}{\includegraphics[bb=0 0 360 216]{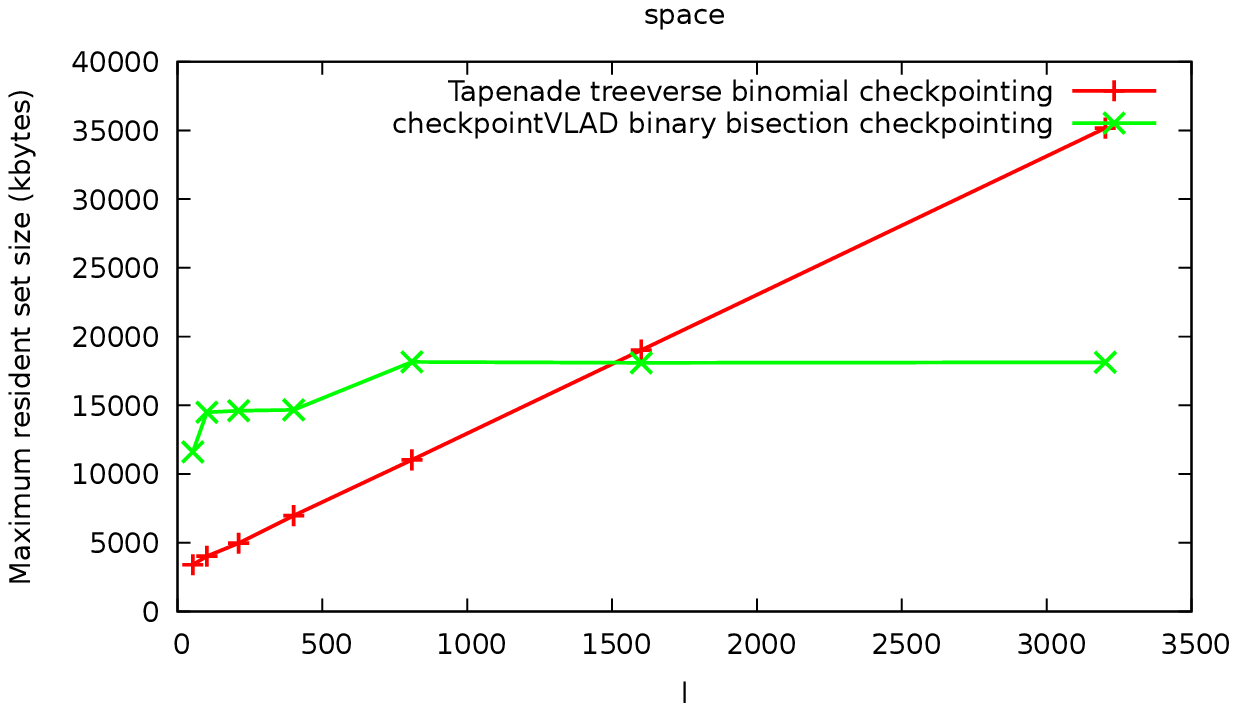}}
\vspace*{-20pt}
\caption{Comparison of space usage of the the example in
    Figs.~\ref{fig:example-fortran} and~\ref{fig:example-vlad} when run with
    divide-and-conquer checkpointing.
    \checkpointVLAD\ achieves sublinear growth while \Tapenade\ does not,
    thus for long enough run times, the space usage of \Tapenade\ exceeds that
    of \checkpointVLAD.}
\label{fig:compare}
\end{wrapfigure}
Our current implementations, as evaluated in Section~\ref{sec:example}, are
expository prototypes and not intended as practical artifacts.
Nonetheless, technology exists that can support construction of large-scale
practical and efficient implementations based on the conceptual ideas presented
here.

\subsection{Implementation Technologies}
\label{sec:implementation}

One can use \POSIX\ \texttt{fork()} to implement the general-purpose
interruption and resumption interface, allowing it to apply in the host, rather
than the target, and thus it could be used to provide an overloaded
implementation of divide-and-conquer checkpointing in a fashion that was
largely transparent to the user \citep{griewank1992alg}.
The last paragraph of \citep{griewank1992alg} states:
\begin{quote}
  For the sake of user convenience and computational efficiency, it would be
  ideal if reverse automatic differentiation were implemented at the compiler
  level.
\end{quote}
We have exhibited such a compiler.
Our implementation, however, is not as efficient as \Tapenade.
There can be a number of reasons for this.
First, \Fortran\ unboxes double precision numbers whereas
\checkpointVLAD\ boxes them.
This introduces storage allocation, reclamation, and access overhead for
arithmetic operations.
Second, array access and update in \Fortran\ take constant time whereas
access and update in \checkpointVLAD\ take linear time.
Array update in \checkpointVLAD\ further involves storage allocation and
reclamation overhead.
Third, \Tapenade\ implements the base case reverse mode of divide-and-conquer
checkpointing using source-code transformation whereas
\checkpointVLAD\ implements it with operator overloading.
In particular, the dynamic method for supporting nesting involves tag dispatch
for every arithmetic operation.

We have exhibited a very aggressive compiler (\Stalingrad) for \VLAD\ that
ameliorates some of these issues.
It unboxes double precision numbers and implements AD via source-code
transformation instead of operator overloading as is done by \checkpointVLAD.
This allows it to have numerical performance rivaling \Fortran.
While it does not support constant-time array access and update, methods that
are well-known in the programming languages community (\eg\ monads and
uniqueness types) can be used for that.
But it does not include support for divide-and-conquer checkpointing.
However, there is no barrier, in principle, to supporting divide-and-conquer
checkpointing, of the sort described above, in an aggressive optimizing
compiler that implements AD via source-code transformation.
One would simply need to reformulate the source-code transformations that
implement AD, along with the aggressive compiler optimizations, in CPS instead
of direct style.
Moreover, the techniques presented here could be integrated into other
compilers for other languages that generate target code in CPS by instrumenting
the CPS conversion with step counting, step limits, and limit-check
interruptions.\footnote{We note that many optimizing compilers, for example GCC,
  use an intermediate program representation called Single Static Assignment,
  or SSA, which is formally equivalent to CPS \citep{kelsey1995a}.}
A driver can be wrapped around such code to implement $\CHECKPOINTREVERSEJ$.\@
For example, existing compilers, like \SMLNJ\ \citep{appel2006compiling}, for
functional languages like \SML, already generate target code in CPS, so it
seems feasible to adapt their techniques to the purpose of AD with divide-and-conquer
checkpointing.
In fact, the overhead of the requisite instrumentation for step counting, step
limits, and limit-check interruptions need not be onerous because the
step counting, step limits, and limit-check interruptions for basic blocks can
be factored, and those for loops can be hoisted, much as is done for the
instrumentation needed to support storage allocation and garbage collection
in implementations like \MLton\ \citep{weeks2006whole}, for languages like
\SML, that achieve very low overhead for automatic storage management.

\subsection{Advantages of Functional Languages for Interruption and Resumption}
\label{sec:advantages}

Functional languages simplify interruption and resumption, allowing these to be
much more efficient.
Two different capsules taken at two different execution points can share
common substructure, by way of pointers, without needing to copy that
substructure.
Indeed, CPS in the \checkpointVLAD\ implementation renders all program state,
including the stack and variables in the environment, as closures, possibly
nested.
Creating a capsule simply involves saving a pointer to a closure.
Resuming a capsule simply involves invoking the saved closure, a simple
function call that is passed the closure environment as its argument.
The garbage collector can traverse the pointer structure of the program state
to determine the lifetime of a capsule.
The interruption and resumption framework need not do so itself.
This simplicity and efficiency would be disrupted by structure mutation or assignment.

\subsection{Nesting of AD Operators}
\label{sec:nesting}

It has been argued that the ability to nest AD operators is important in many
practical domains \citep{christianson-1993a, agarwal2016second,
  andrychowicz-etal-2016a}.
Supporting nested use of AD operators involves many subtle issues
\citep{siskindpearlmutter2008a}.
\checkpointVLAD\ addresses these issues and fully supports nested use of AD
operators.
One can write programs of the form
\begin{equation}
  \alpha\;(\lambda x\LD{}(\ldots (\beta\;f\;\ldots)))\;\ldots
  \label{eq:nest}
\end{equation}
where each of~$\alpha$ and~$\beta$ can be any of~$\FORWARDJ$, $\REVERSEJ$,
and~$\CHECKPOINTREVERSEJ$.
\Ie\ one can apply AD to one function that, in turn, applies AD to another
function.
This allows one not only to do forward-over-reverse, reverse-over-forward,
and reverse-over-reverse, it also allows one to do things like
divide-and-conquer-reverse-over-forward,
reverse-over-divide-and-conquer-reverse, and even
divide-and-conquer-reverse-over-divide-and-conquer-reverse.

There is one catch however.
When one applies an AD operator, that application is considered to be atomic
by an application of a surrounding AD operator.
This is evident by the $n+1$
in~\cref{eq:cps-jstar,eq:cps-starj,eq:cps-checkpoint-starj},
What this means is that if~$\alpha$ in~\eqref{eq:nest}
were~$\CHECKPOINTREVERSEJ$, a split point for~$\alpha$ could not occur
inside~$f$.
While this does not affect the correctness of the result, it could affect the
space and time complexity.

The reason for this is that while the semantics of the AD operators are
functional, their internal implementation involves mutation.
In particular, to support nesting, $\FORWARDJ$, $\REVERSEJ$,
and~$\CHECKPOINTREVERSEJ$ internally maintain and update an $\epsilon$ tag
as described in \citep{siskindpearlmutter2008a}.
The~$\epsilon$ tag is incremented upon entry to an AD operator and
decremented upon exit from that invocation to keep track of the nesting level.
All computation within a level must be performed with the same~$\epsilon$ tag.
This requires that the entries to and exits from AD operator invocations obey
last-in-first-out sequencing.
If~$\alpha$ were~$\CHECKPOINTREVERSEJ$, and the computation of~$f$ were
interrupted, then situations could arise where the last-in-first-out sequencing
of~$\beta$ was violated.
Moreover, since divide-and-conquer checkpointing executes different portions of
the forward sweep different numbers of times, the number of entries into a
nested AD operator could exceed the number of exits from that operator.

Reverse mode involves a further kind of mutation.
The forward sweep creates a tape represented as a directed acyclic graph.
The nodes in this tape contain slots for the cotangent values associated with
the corresponding primal values.
The reverse sweep operates by traversing this graph to accumulate the
cotangents in these slots.
Such accumulation is done by mutation.

The above issues arise because of mutation in the implementation of AD
operators.
Conceivably, these could be addressed using methods that are well-known in the
programming languages community for supporting mutation in functional languages
(\eg\ monads and uniqueness types).
Issues arise beyond this, however.
If both~$\alpha$ and~$\beta$ were~$\CHECKPOINTREVERSEJ$,
and~$\CHECKPOINTREVERSEJ$ was not atomic, situations could arise where~$f$
could interrupt for~$\alpha$ instead of~$\beta$.
Currently, interruption is indicated by returning instead of calling a
continuation.
If~$f$ were to return instead of calling a continuation, there is no way to
indicate that that interruption was due to~$\alpha$ instead of~$\beta$.
It is unclear whether this issue could be resolved.

\section{Conclusion}

Reverse-mode AD with divide-and-conquer checkpointing is an enabling
technology, allowing gradients to be efficiently calculated even where
classical reverse mode imposes an impractical storage overhead.
We have shown that it is possible to provide an operator that implements
reverse-mode AD with divide-and-conquer checkpointing, implemented as an
interpreter, a hybrid compiler/interpreter, and a compiler, which
\begin{itemize}
\item has an identical API to the classical reverse-mode AD operator,
\item requires no user annotation,
\item takes the entire derivative calculation as the root execution interval,
  not just the execution intervals corresponding to the invocations of
  particular constructs such as DO loops,
\item takes arbitrary execution points as candidate split points, not just the
  execution points corresponding to the program points at the boundaries of
  particular constructs like the iteration boundaries of DO loops,
\item supports both an algorithm that constructs binary checkpoint trees and
  the treeverse algorithm that constructs n-ary checkpoint trees,
\item supports selection of actual split points from candidate split points
  using both a bisection and a binomial criterion, and
\item supports any of the termination criteria of fixed space overhead, fixed
  time overhead, or logarithmic space and time overhead,
\end{itemize}
yet still provides the favorable storage requirements of reverse mode with
divide-and-conquer checkpointing, guaranteeing sublinear space and time
overhead.

% Authors must also incorporate a Disclosure Statement which will acknowledge
% any financial interest or benefit they have arising from the direct
% applications of their research.

\section*{Acknowledgments}      % Disclosure

This manuscript is an expanded version of the extended abstract of a
presentation at AD~(2016), which is available as a technical report
\citep{ad2016a}.
Aspects of this work may be subject to the intellectual property policies of
Purdue University and/or Maynooth University.
Code for the implementations described above and related materials are
available at \url{https://github.com/qobi/checkpoint-VLAD/}.
%
% Please supply all details required by any funding and grant-awarding bodies
% as an Acknowledgement on the title page of the manuscript, in a separate
% paragraph, as follows:
% For single agency grants: "This work was supported by the [Funding Agency]
% under Grant [number xxxx]."
% For multiple agency grants: "This work was supported by the [Funding Agency
% 1] under Grant [number xxxx]; [Funding Agency 2] under Grant [number xxxx];
% and [Funding Agency 3] under Grant [number xxxx]."
%
%% \section*{Funding}
%
This work was supported, in part, by the US National Science Foundation under
Grants 1522954-IIS and 1734938-IIS, by the Intelligence Advanced Research
Projects Activity (IARPA) via Department of Interior/Interior Business Center
(DOI/IBC) contract number D17PC00341, and by Science Foundation Ireland under
Grant SFI 09/IN.1/I2637.
Any opinions, findings, views, and conclusions or recommendations expressed in
this material are those of the authors and do not necessarily reflect the
views, official policies, or endorsements, either expressed or implied, of the
sponsors.
The U.S. Government is authorized to reproduce and distribute reprints for
Government purposes, notwithstanding any copyright notation herein.

\bibliographystyle{gOMS}
\bibliography{oms2017a}

\providecommand{\Scheme}{\mbox{\textsc{Scheme}}}
  \providecommand{\Lisp}{\mbox{\textsc{Lisp}}}
\begin{thebibliography}{10}
\newcommand{\noopsort}[1]{}
\newcommand{\printfirst}[2]{#1}
\newcommand{\singleletter}[1]{#1}
\newcommand{\switchargs}[2]{#2#1}
\providecommand{\url}[1]{\normalfont{#1}}
\providecommand{\urlprefix}{Available at }

\bibitem{achten1993high}
P. Achten, J. Van~Groningen, and R. Plasmeijer, \emph{High level specification
  of {I/O} in functional languages}, in \emph{Functional Programming, Glasgow
  1992}, Springer,  1993, pp. 1--17.

\bibitem{agarwal2016second}
N. Agarwal, B. Bullins, and E. Hazan, \emph{Second order stochastic
  optimization in linear time} (2016),
  \urlprefix\url{https://arxiv.org/abs/1602.03943}.

\bibitem{andrychowicz-etal-2016a}
M. Andrychowicz, M. Denil, S.G. Colmenarejo, M.W. Hoffman, D. Pfau, T. Schaul,
  and N. de  Freitas, \emph{Learning to learn by gradient descent by gradient
  descent}, in \emph{Neural Information Processing Systems},
  \urlprefix\url{https://arxiv.org/abs/1606.04474}, NIPS, 2016.

\bibitem{appel2006compiling}
A.W. Appel, \emph{Compiling with continuations}, Cambridge University Press,
  2006.

\bibitem{bendtsen1996faf}
C. Bendtsen and O. Stauning, \emph{{FADBAD}, a flexible {C++} package for
  automatic differentiation}, Technical Report IMM--REP--1996--17, Department
  of Mathematical Modelling, Technical University of Denmark, Lyngby, Denmark,
  1996.

\bibitem{chen-etal-2016a}
T. Chen, B. Xu, Z. Zhang, and C. Guestrin, \emph{Training deep nets with
  sublinear memory cost} (2016),
  \urlprefix\url{https://arxiv.org/abs/1604.06174}.

\bibitem{christianson-1993a}
B. Christianson, \emph{Reverse accumulation of functions containing gradients},
  Tech. {R}ep. 278, University of Hertfordshire Numerical Optimisation Centre,
  1993, \urlprefix\url{http://hdl.handle.net/2299/4337}, presented at the
  Theory Institute Argonne National Laboratory Illinois, Procs. of the Theory
  Institute on Combinatorial Challenges in Automatic Differentiation.

\bibitem{dauvergne2006tdf}
B. Dauvergne and L. Hasco{\"e}t, \emph{The Data-Flow Equations of Checkpointing
  in Reverse Automatic Differentiation}, in \emph{Computational Science --
  {ICCS} 2006}, Lecture Notes in Computer Science, Vol. 3994, Springer,
  Heidelberg, 2006, pp. 566--573.

\bibitem{griewank1992alg}
A. Griewank, \emph{Achieving logarithmic growth of temporal and spatial
  complexity in reverse automatic differentiation}, Optimization Methods and
  Software 1 (1992), pp. 35--54.

\bibitem{griewank1996apf}
A. Griewank, D. Juedes, and J. Utke, \emph{A package for the automatic
  differentiation of algorithms written in {C/C++}. {U}ser manual}, Tech.
  {R}ep., Institute of Scientific Computing, Technical University of Dresden,
  Dresden, Germany,  1996, this version of the manual is superceded by
  \url{http://www.math.tu-dresden.de/~adol-c/adolc110.ps}.

\bibitem{gruslys-etal-2016a}
A. Gruslys, R. Munos, I. Danihelka, M. Lanctot, and A. Graves,
  \emph{Memory-Efficient Backpropagation Through Time}, in \emph{Neural
  Information Processing Systems},
  \urlprefix\url{https://arxiv.org/abs/1606.03401}, NIPS, 2016.

\bibitem{hascoet2004tug}
L. Hasco{\"e}t and V. Pascual, \emph{{TAPENADE}~2.1 user's guide}, Rapport
  technique 300, INRIA,  2004.

\bibitem{haynes1984engines}
C.T. Haynes and D.P. Friedman, \emph{Engines build process abstractions}, in
  \emph{Proceedings of the 1984 ACM Symposium on LISP and functional
  programming}, ACM, 1984, pp. 18--24.

\bibitem{heller1998checkpointing}
R. Heller, \emph{Checkpointing without operating system intervention:
  Implementing griewank's algorithm}, Master's thesis, Ohio University,  1998.

\bibitem{kang2003implementation}
Y. Kang, \emph{Implementation of forward and reverse mode automatic
  differentiation for {GNU} {Octave} applications}, Master's thesis, Ohio
  University,  2003.

\bibitem{kelsey1995a}
R.A. Kelsey, \emph{A correspondence between continuation passing style and
  static single assignment form}, ACM SIGPLAN Notices, Papers from the 1995 ACM
  SIGPLAN Workshop on Intermediate Representations 30 (1995), pp. 13--22.

\bibitem{kowarz2006ocf}
A. Kowarz and A. Walther, \emph{Optimal Checkpointing for Time-Stepping
  Procedures in {ADOL-C}}, in \emph{Computational Science -- {ICCS} 2006},
  Lecture Notes in Computer Science, Vol. 3994, Springer, Heidelberg, 2006, pp.
  566--573.

\bibitem{reynolds1993discoveries}
J.C. Reynolds, \emph{The discoveries of continuations}, Lisp and Symbolic
  Computation 6 (1993), pp. 233--247.

\bibitem{siskindpearlmutter2008a}
J.M. Siskind and B.A. Pearlmutter, \emph{Nesting forward-mode {AD} in a
  functional framework}, Higher-Order and Symbolic Computation 21 (2008), pp.
  361--376.

\bibitem{ad2016a}
J.M. Siskind and B.A. Pearlmutter, \emph{Binomial checkpointing for arbitrary
  programs with no user annotation} (2016),
  \urlprefix\url{https://arxiv.org/abs/1611.03410}.

\bibitem{ad2016b}
J.M. Siskind and B.A. Pearlmutter, \emph{Efficient implementation of a
  higher-order language with built-in {AD}} (2016),
  \urlprefix\url{https://arxiv.org/abs/1611.03416}.

\bibitem{speelpenning80}
B. Speelpenning, \emph{Compiling fast partial derivatives of functions given by
  algorithms}, Ph.D. diss., Department of Computer Science, University of
  Illinois at Urbana-Champaign,  1980.

\bibitem{stovboun2000tool}
A. Stovboun, \emph{A tool for creating high-speed, memory efficient derivative
  codes for large scale applications}, Master's thesis, Ohio University,  2000.

\bibitem{sussmanwm2001}
G.J. Sussman, J. Wisdom, and M.E. Mayer, \emph{Structure and interpretation of
  classical mechanics}, MIT Press, 2001.

\bibitem{volin1985aco}
Y.M. Volin and G.M. Ostrovskii, \emph{Automatic computation of derivatives with
  the use of the multilevel differentiating technique --- {I}: {A}lgorithmic
  basis}, Computers and Mathematics with Applications 11 (1985), pp.
  1099--1114.

\bibitem{wadler90comprehending}
P.L. Wadler, \emph{Comprehending monads}, in \emph{Proceedings of the 1990
  {ACM} Conference on {\Lisp} and Functional Programming}, ACM, 1990, pp.
  61--78.

\bibitem{weeks2006whole}
S. Weeks, \emph{Whole-program compilation in {MLton}} (2006),
  \urlprefix\url{http://www.mlton.org/References.attachments/060916-mlton.pdf},
  workshop on ML.

\end{thebibliography}

\end{document}